\documentclass[aps,twocolumn,nofootinbib,preprintnumbers,superscriptaddress]{revtex4-1}
\usepackage{amssymb,amsmath,amstext,amsfonts,textcomp,soul}
\usepackage{graphicx,multirow}
\usepackage{physics}
\usepackage{bbold}
\usepackage{hyperref}
\usepackage[capitalise]{cleveref}
\usepackage{color,xcolor,colortbl}
\usepackage{orcidlink}

\newcommand{\be}{\begin{equation}}
\newcommand{\ee}{\end{equation}}

\newcommand{\beq}{\begin{equation}}	

\newcommand{\eeq}{\end{equation}}
\newcommand{\ba}{\begin{eqnarray}}
\newcommand{\ea}{\end{eqnarray}}

\def\PP{P}

\def\PP{\text{PP}}
\def\PG{\text{PG}}

\crefformat{equation}{Eq.~(#2#1#3)}
\crefformat{table}{Tab.~(#2#1#3)}
\crefformat{figure}{Fig.~(#2#1#3)}
\crefformat{appendix}{App.~(#2#1#3)}
\crefformat{section}{Sec.~(#2#1#3)}
\crefformat{subsection}{Subsec.~(#2#1#3)}

\begin{document}
\preprint{YITP-23-53, IPMU23-0010}

\title{Global and Local Stability for Ghosts Coupled to Positive Energy Degrees of Freedom}

\author{C\'edric~Deffayet,\orcidlink{0000-0002-1907-5606}}
\email{cedric.deffayet AT ens.fr}
\affiliation{Laboratoire de Physique de l’\'Ecole normale sup\'erieure, ENS, Universit\'e PSL, CNRS, Sorbonne Universit\'e, Universit\'e Paris Cité, F-75005 Paris, France}

\author{Aaron Held\,\orcidlink{0000-0003-2701-9361}}
\email{aaron.held@uni-jena.de}
\affiliation{Theoretisch-Physikalisches Institut, Friedrich-Schiller-Universit\"at Jena,
Max-Wien-Platz 1, 07743 Jena, Germany}
\affiliation{The Princeton Gravity Initiative, Jadwin Hall, Princeton University, Princeton, New Jersey 08544, U.S.}

\author{Shinji Mukohyama} 
\email{shinji.mukohyama@yukawa.kyoto-u.ac.jp}
\affiliation{Center for Gravitational Physics and Quantum Information, Yukawa Institute for Theoretical Physics, Kyoto University, 606-8502, Kyoto, Japan}
\affiliation{Kavli Institute for the Physics and Mathematics of the Universe (WPI), The University of Tokyo Institutes for Advanced Study, The University of Tokyo, Kashiwa, Chiba 277-8583, Japan}

\author{Alexander Vikman\,\orcidlink{ 0000-0003-3957-2068}} 
\email{vikman@fzu.cz}
\affiliation{CEICO--Central European Institute for Cosmology and Fundamental Physics, FZU--Institute of Physics of the Czech Academy of Sciences, Na Slovance 1999/2, 18221 Prague 8, Czech Republic}

\begin{abstract}
Negative kinetic energies correspond to ghost degrees of freedom, which are potentially of relevance for cosmology, quantum gravity, and high energy physics. We present a novel wide class of stable mechanical systems where a positive energy degree of freedom interacts with a ghost. These theories have Hamiltonians unbounded from above and from below, are integrable, and contain free functions.
We show analytically that their classical motion is bounded for all initial data.
Moreover, we derive conditions allowing for Lyapunov stable equilibrium points. 
A subclass of these stable systems has simple polynomial potentials with stable equilibrium points entirely due to interactions with the ghost. 
All these findings are fully supported by numerical computations which we also use to gather evidence for stability in various nonintegrable systems.
\end{abstract} 

\maketitle

\section{Introduction}
\label{sec:intro}

Ghosts are dynamical degrees of freedom (DoF), described by a Hamiltonian with negative kinetic energies, as opposed to the kinetic energies of standard degrees of freedom. Such ghosts generically appear in the context of theories with higher derivatives~\cite{Ostrogradsky:1850fid} (see also~\cite{Bruneton:2007si,Ketov:2011re,Woodard:2015zca} for reviews) and have been considered in a variety of physical scenarios: 
they have been advocated as a way to regulate the ultraviolet behaviour of quantum field theories~\cite{Pais:1950za,Lee:1970iw,Stelle:1976gc,Grinstein:2007mp};
they have been suggested as a way to address the cosmological constant problem \cite{Linde:1984ir,Linde:1988ws,Kaplan:2005rr}; they have been used to obtain bouncing cosmologies~\cite{Brandenberger:2016vhg}; and they have been proposed in the context of dark energy~\cite{Caldwell:1999ew} or, more recently, as a way to ameliorate the Hubble tension~\cite{DiValentino:2021izs}.
\\

In spite of these applications, ghosts are typically disregarded since they are expected to generically lead to catastrophic instabilities, both at the classical and at the quantum level (see e.g~\cite{Cline:2003gs}) -- for the purpose of this work, we will only be concerned with the classical case. 

Whenever positive- and negative-kinetic-energy degrees of freedom do not interact, the presence of a ghost is trivially harmless. Without coupling, no energy exchange can take place and the respective ghostly and non-ghostly Hamiltonians can each be stable due to suitable self-interactions.

In contrast, it is widely expected that a generic coupling between the positive- and negative-kinetic-energy degrees of freedom, leads to an energy exchange between them that instigates an unbounded growth of absolute values of both energies. Hence, one expects a runaway evolution to larger and larger phase-space variables.
However, in previous work \cite{Deffayet:2021nnt}, coauthored by some of the authors of the current paper, a specific counterexample to such a catastrophe was presented. 
Besides completing a stability proof for a class of models introduced in~\cite{Deffayet:2021nnt}, the key physical motivation for this work can be understood in terms of the following simple question: 
Is the stable model from \cite{Deffayet:2021nnt} too specific or can stability be extended to a wider class of integrable and nonintegrable systems with ghosts?
\\

The potential for stable motion in the presence of interacting ghosts has, of course, been previously considered. (While we will be concerned only with classical theories, we refer to~\cite{Lee:1970iw, Hawking:2001yt, Grinstein:2007mp, Bender:2007wu,Garriga:2012pk, Salvio:2014soa, Smilga:2017arl, Becker:2017tcx, Anselmi:2018kgz, Donoghue:2019fcb, Gross:2020tph, Donoghue:2021eto, Platania:2022gtt} for proposals at the quantum level.)
At the classical level, it has been argued that theories with ghosts can be considered ``benign'', provided that the runaway is sufficiently mild and the motion can be extended to the infinite future~\cite{Pagani:1987ue, Smilga:2004cy, Boulanger:2018tue, Gross:2020tph, Damour:2021fva}. At the same time, indications have been found that classical motion in the presence of ghosts can exhibit so-called ``islands of stability'' (i.e., bounded motion for a restricted set of initial conditions)~\cite{Pagani:1987ue, Smilga:2004cy, Carroll:2003st, Ilhan:2013xe,Pavsic:2016ykq, Smilga:2017arl,Pavsic:2013noa, Pavsic:2020aqi, Boulanger:2018tue, Damour:2021fva}. All of these indications are numerical (and, as such, cannot be fully conclusive, as they cannot cover the whole evolution of the system to the infinite future along even a single trajectory, not to mention \emph{all} the Hamiltonian trajectories) and/or 
hold only for restricted sets of initial conditions, i.e., do not conclusively advocate for global stability.
Moreover, we still lack a physical rationale or criterion to distinguish potentially stable ghostly interactions from unstable ones.
\\

As mentioned before, the recent work~\cite{Deffayet:2021nnt} has established the first conclusive proof of globally stable motion for a positive-energy harmonic oscillator interacting through a specific non-polynomial potential with a ghost in the form of a negative-energy oscillator.
It was shown that, for arbitrary initial conditions, the motion remains bounded. The proof relies on the integrable nature of the model. We stress that this kind of global stability (also referred to as the Lagrange stability) is different from the above notions of ``benign'' ghosts and/or ``islands of stability''. We also stress that there is no direct connection of global stability to local (Lyapunov) stability~\cite{Lefschetz}. In this context, we would like to mention that~\cite{Deffayet:2021nnt} has also proven Lyapunov stability of the equilibrium configuration of the model. 
It is global (Lagrange) stability that we are mostly concerned with and we will, from here on, refer to it simply as stability, everywhere where this would not cause confusion. 
One of the reason for this concern is that quantum mechanics effectively probes the whole configuration space. Thus, regions beyond the, usually small, "islands of stability" can force the wave function and probability distributions to runaway to large values of phase-space coordinates in the quantized theory.

The purpose of this paper is the following: First, we extend the proof in~\cite{Deffayet:2021nnt} to a much larger class of integrable models with 
kinetic terms quadratic in canonical momenta and with two free functions of (particular combinations) of the
coordinates canonically conjugated to them. This class of systems encompasses, in particular, a tower of stable polynomial interactions. Our proof also covers the subclass that has been announced in~\cite{Deffayet:2021nnt}.
Second, we prove that a wide subclass of these models possesses locally stable vacua.
Third, we propose physical criteria for stable motion of more generic systems with ghosts interacting with positive energy degrees of freedom. Fourth, we investigate the above numerically, both for integrable and non integrable models.
\\

With these goals in mind, we start in \cref{sec:integrable-models} by introducing a generic procedure to relate classical mechanical systems of $N$ degrees of freedom
with and without ghosts by means of suitable complex canonical  transformations. We then consider a particular class of integrable models in \cref{sec:Liouville}, for which in \cref{subsec:proof} we provide a fully analytic proof of the boundedness of motion, identify a polynomial subclass in \cref{subsec:integrable-motion}, and discuss Lyapunov stability of equilibrium points in \cref{subsec:Lagrange_vs_Lyapunov}. 
In particular, we find that new Lyapunov stable vacua can be instigated by polynomial interactions with a ghost. 
Incidentally, we stress that the polynomial models introduced in \cref{subsec:integrable-motion} stay polynomial and integrable when the degrees of freedom all have positive kinetic energies and appear as such to be novel to this work. Finally, we extend the discussion to nonintegrable models in \cref{sec:nonintegrable}, for which we gather numerical evidence that stable (or at least longlived) motion can persist in absence of integrability. We close with our conclusions in \cref{sec:discussion}.
We provide additional technical material in Appendices. In particular, in  \cref{app:integrable-Hamiltonians} we list various classes of integrable models; in \cref{app:Integrability_Equilibrium} we derive equations defining the models with the fist integral quadratic in canonical momenta and demonstrate that Lyapunov stable equilibrium points are located in the saddle points of the potential; in \cref{app:ghoslty x} we discuss stability in alternatively  complexified Liouville models; in \cref{app:polynomial-potentials} and \cref{app:polynomial-potentials-class-3} we provide more details on integrable models with polynomial potentials.

\section{Integrable interacting ghosts via complex canonical transformations}
\label{sec:integrable-models}

We start by considering $N$ classical mechanical degrees of freedom (DoFs) with a Hamiltonian
\begin{align}
\label{eq:Ham-general-1}
	H(x_i,p_i) = 
	\sum_{i=1}^N\frac{p_{i}^2}{2 m_i}\,
	+ V(x_i,...x_N)\;,
\end{align}
where $p_i$ denotes the canonical momenta conjugate to the coordinates $x_i$. Together, these denote the phase-space variables $\xi_i = (x_i,p_i)$. For now, all $N$ DoFs have same-sign kinetic terms and thus correspond
to modes with positive kinetic energy, as the masses $m_i$ are assumed to be positive. 
We notice that a 
rotation in the complex plane to the imaginary axis of any of the momentum variables, i.e., $p_n\rightarrow \pm ip_n$, flips the sign of the respective kinetic term, i.e., $p_n^2/m_n\rightarrow - p_n^2/m_n$, and thus turns the n-th degree of freedom (DoF) with positive kinetic energy
into a DoF with negative kinetic energy. Hence, this transformation effectively flips the sign of the mass $m_n$.
This rotation in the complex plane can be completed into a complex canonical transformation by simultaneously transforming the respective $x_n\rightarrow \mp ix_n$. The resulting (univalent) transformation
\begin{align} \label{Cnpm}
	\mathfrak{C}^{\pm}_{x^n}:
	\left\{\begin{array}{lclcl}
		x_j & \rightarrow & \pm i\,x_j & \text{for} & j=n
		\\
		p_j & \rightarrow & \mp i\,p_j & \text{for} & j=n
		\\
		x_j & \rightarrow & x_j & \forall & j\neq n
		\\
		p_j & \rightarrow & p_j & \forall & j\neq n
	\end{array}\right\}\;,
\end{align}
is canonical in the sense that it preserves the Poisson bracket $\{ \}_\text{PB}$, i.e.,
\begin{align}	
	\{\mathfrak{C}^{\pm}_{x^n}(x_i),\,\mathfrak{C}^{\pm}_{x^n}(p_j)\}_\text{PB} &= 
	- \{\mathfrak{C}^{\pm}_{x^n}(p_i),\,\mathfrak{C}^{\pm}_{x^n}(x_j)\}_\text{PB} = 
	\delta_{ij}\;,
	\notag\\
	\{\mathfrak{C}^{\pm}_{x^n}(x_i),\,\mathfrak{C}^{\pm}_{x^n}(x_j)\}_\text{PB} &= 0\;,
	\notag\\
	\{\mathfrak{C}^{\pm}_{x^n}(p_i),\,\mathfrak{C}^{\pm}_{x^n}(p_j)\}_\text{PB} &= 0\;.
\end{align}
Given that $\mathfrak{C}^{\pm}_{x^n}$ is a canonical transformation, it moreover preserves the Hamilton equations, the time-independence of the Hamiltonian itself, as well as all the time-independence of any other existing constant of motion.
We note also that $\mathfrak{C}^{\pm}_{x^n} \circ \mathfrak{C}^{\mp}_{x^n} = \mathfrak{C}^{\mp}_{x^n} \circ \mathfrak{C}^{\pm}_{x^n} = \mathbb{1}$. Moreover, any function $F(\vec{\xi})$ of the phase-space variables $\vec{\xi}$ which is even under the simultaneous change of the sign of the variables $x_n$ and $p_n$, i.e.,
\begin{align}
	F(x_{i\neq n},x_n,p_{i\neq n},p_n)
	= 
	F(x_{i\neq n},-x_n,p_{i\neq n},-p_n)\;,
\end{align}
is invariant under the successive application of two identical $\mathfrak{C}^{\pm}_{x^n}$.
We will also refer to such functions as `($x_n$,$p_n$)-parity even'. On the other hand, we refer to a potential satisfying $V(x_{i\neq n},x_n)=V(x_{i\neq n},-x_n)$ as `$x_n$-parity even'. Obviously, the Hamiltonian in \cref{eq:Ham-general-1} is ($x_n$,$p_n$)-parity even if and only if the potential is $x_n$-parity even.
\\

In general, $\mathfrak{C}^{\pm}_{x^n}$ does \emph{not} preserve real-valuedness of the Hamiltonian. (The same, and all of what follows, also holds for other constants of motion.) However, real-valuedness is obviously preserved if the potential is polynomial (or Taylor expandable with respect to $x_n$) and $x_n$-parity even. This is the case, e.g., for the canonical Hamiltonian of $N$ non interacting positive kinetic-energy modes (where here and henceforth we normalize the kinetic terms just to be $\pm p_i^2 /2 $, {\it i.e.}, we set the masses to one, but keep the possibility to have a different frequency  $\omega_i$ for each mode)
\begin{equation} \label{freeN}
    H_\text{canonical} = 
    \sum_{i=1}^N \frac{1}{2}\left(p_i^2 + \omega_i^2 x_i^2 \right)\,.
\end{equation}
In this case, the canonical transformation $\mathfrak{C}^{\pm}_{x^n}$ transforms a real-valued Hamiltonian with $N$ positive-energy modes to a dual real-valued Hamiltonian with $N-1$ positive kinetic-energy DoFs and one negative kinetic-energy DoF, i.e., a ghost. This can easily be extended by performing several distinct such canonical transformations to introduce multiple ghost degrees of freedom. For the general Hamiltonian in \cref{eq:Ham-general-1}, if the potential is parity-even in all of the respective variables, then all of these Hamiltonians are real-valued as far as the original one is.
All of the above also holds if further parity-even interactions are included.
\\

In the rest of this work, we will specify to systems with two degrees of freedom $x$ and $y$ and with a Hamiltonian 
\begin{align}
\label{eq:Ham-1}
    H = \frac{1}{2}p_x^2 + \sigma\frac{1}{2}p_y^2 + V(x,y)\;,
\end{align}
where $\sigma = \pm 1$ decides between the case of two positive kinetic terms ($\sigma = +1$) and the case of two opposite-sign kinetic terms ($\sigma = -1$). We refer to $\sigma = +1$ as the ``PP'' case, denoting two positive kinetic-energy DoFs, and to $\sigma = -1$ as the ``PG'' case, where a positive kinetic-energy DoF is coupled to a negative kinetic-energy (ghost) DoF.
Starting with a PP model, and applying a complex canonical transformation (cf.~\cref{Cnpm}) on the variable $y$, we obtain a new model of PG type with Hamiltonian $\frac{1}{2}p_x^2 -\frac{1}{2}p_y^2 + V(x,\pm i y)$ which can be real depending on the $V(x,y)$ one starts out with.
\\

In the following, it will be sometimes convenient to split the potential $V(x,y)$ into three parts as follows for a PG system with $\sigma=-1$,
\begin{align}
&V_\text{P}(x)=V(x,0)\,,\quad \text{and}\quad V_\text{G}(y) =V(0,y)\,,\quad \text{while}\notag \\[0.5em]
&V_\text{int}(x,y)= V(x,y) - V_\text{P}(x) - V_\text{G}(y)\;,
\label{eq:potential-split-notation}
\end{align}
such that $V = V_\text{P} + V_\text{G} + V_\text{int}$. We refer to $V_\text{P}(x)$ and $V_\text{G}(y)$ as the decoupled potentials, to $V_\text{int}(x,y)$ as the interaction potential, and to $V(x,y)$ as the full potential.
This makes sense, in particular, if the structure of the potential $V(x,y)$ is such that $y=0$ is a solution of $y$-equation of motion for arbitrary $x(t)$ and analogously $x=0$ is a solution of the $x$-equation of motion for arbitrary $y(t)$. If this property holds\footnote{
    It may be that this property does not hold for the pair $(x, y)$ but holds for a pair $(x', y')$ that is related to $(x, y)$ by a Poincar\'{e} transformation. Then, one can redefine variables as $(x, y) \to (x', y')$. Of course, this property may not hold at all.
} one can separate self-interactions from cross-coupling interactions.
\\

More specifically, for what concerns the analytic part of this work, our starting point will be integrable PP systems
with a time independent Hamiltonian $H$ of the form in \cref{eq:Ham-1} and with an extra constant of motion $I(\xi)$ such that
\begin{align} \label{constantI}
    \frac{dI}{dt} = \left\{I,H\right\}_\text{PB} = 0 \;.
\end{align}
The study of such systems has been pioneered by Liouville~\cite{Liouville1855note,Arnold2013book} who showed that if an $N$ dimensional Hamiltonian system possesses $N$ functionally independent and Poisson-commuting constants of motion, then the motion is integrable (by quadrature).
In principle, there can be more than $N$ (i.e., up to $(2N-1)$) functionally independent constants of motion. If so, the system is referred to as superintegrable.

Some Hamiltonian systems are trivially integrable. For instance, the canonical Hamiltonian of $N$ uncoupled harmonic oscillators, see \cref{freeN}, possesses $N$ conserved quantities: the individual Hamiltonians of each oscillator.  For other Hamiltonian systems, integrability is, of course, nontrivial and takes more effort to be revealed.

One way to systematically search for constants of motion is by direct methods.
Direct methods (i) make an ansatz for $I(\xi)$ -- typically a polynomial in momenta, dressed with undetermined functions as prefactors -- and then (ii) classify solutions to the resulting (highly overdetermined) set of partial differential equations (PDEs) implied by $\{I,H\}_\text{PB}=0$. 
When restricting to two particle systems, such direct searches for constants of motion (and the corresponding integrable potentials) have been performed at fixed polynomial order in the momenta. Up to quadratic order, the classification has been pioneered in~\cite{darboux1901archives}, formalized in~\cite{Whittaker1964book}, and subsequently completed in~\cite{Frivs1967symmetry,Holt1982construction,Ankiewicz1983complete,Ramani:1983,Ramani:1983a,Thompson1984darboux,Sen1985integrable,Hietarinta:1986tw}. 
We summarize the complete set of resulting pairs of integrable potentials $V(x,y)$ and constants of motion in \cref{app:integrable-Hamiltonians}.
At cubic order in the momenta and beyond, no full classification is known, cf.~\cite{Hietarinta:1986tw} for a comprehensive review of partial results and for references to related methods.
\\

As stressed above, the on-shell constancy of some quantity $I$ is preserved by the action of $\mathfrak{C}^{\pm}_{x^n}$ as a mere consequence of the algebraic nature of equation \cref{constantI} and the canonical nature of the transformation $\mathfrak{C}^{\pm}_{x^n}$. Hence, acting on a PP integrable system with a transformation 
$\mathfrak{C}^{\pm}_{x^n}$, we obtain a PG integrable one. Upon this action, the Hamiltonian can, in general, become complex. However, an inspection of the potentials given in \cref{app:integrable-Hamiltonians} shows that there exist large classes of PP integrable systems yielding real-valued PG integrable systems upon application of the complex canonical transformation $\mathfrak{C}^{\pm}_{y}$ (or $\mathfrak{C}^{\pm}_{x}$). This is the case, e.g., for class 1 of \cref{app:integrable-Hamiltonians}. 
In the next section, we use this real-valued $\PP\Leftrightarrow\PG$ pair to analytically discuss the conditions for bounded motion in the presence of a ghost.

In passing, we also note that there can be real-valued integrable PG systems, for which the PP side is complex-valued, and which may thus have not yet received much attention. An integrable example of this kind is given by class 5 in \cref{app:integrable-Hamiltonians}. In this case the application of the complex canonical transformation  $\mathfrak{C}^{\pm}_{y}$ (or $\mathfrak{C}^{\pm}_{x}$) transforms a complex $V$ to a real one. 
\\

\begin{figure}
    \begin{centering}
        \includegraphics[width=\linewidth]{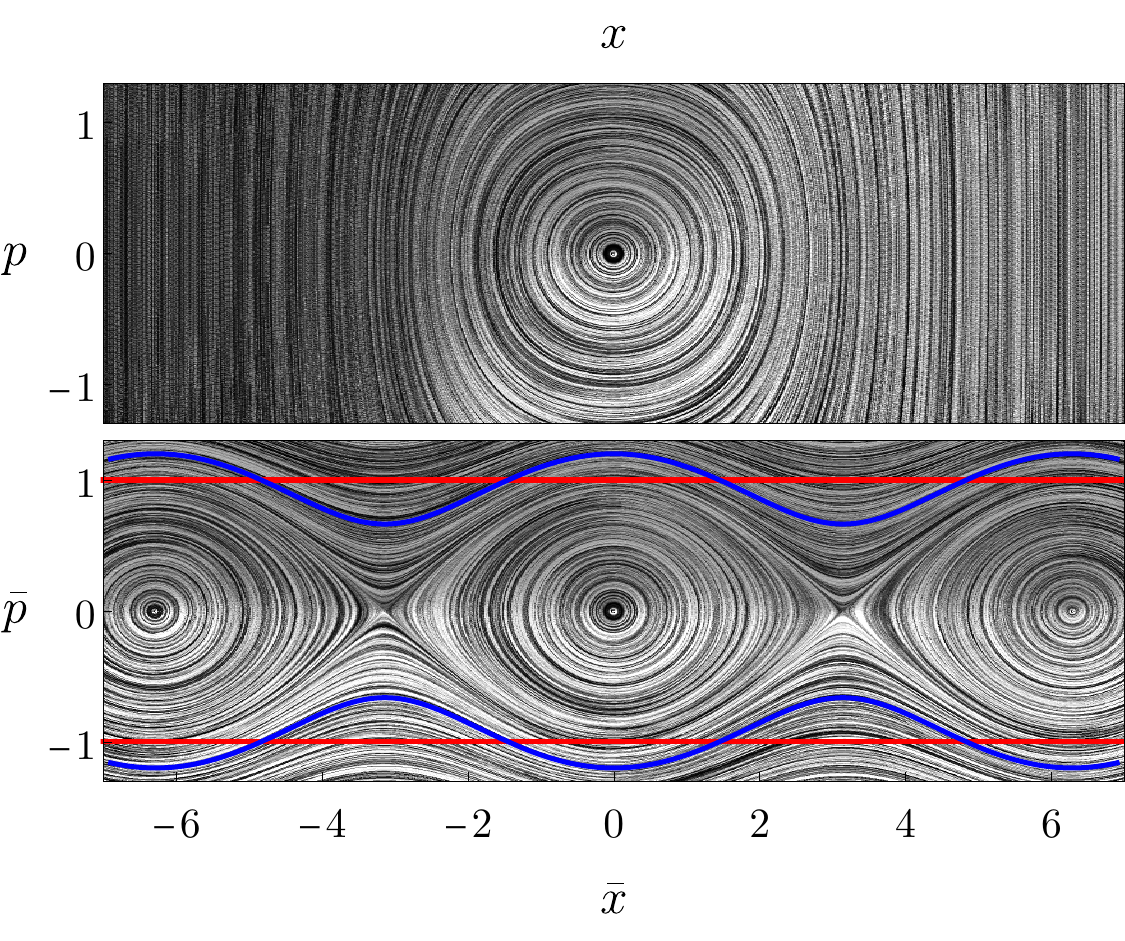}
        \par
    \end{centering}
    \caption{\label{fig:Phase_space_change}
        We show the phase space $\left(x,p\right)$ of the original
        Hamiltonian in \cref{eq:H_Cosh} (upper panel) as well as the
        phase space $\left(\bar{x},\bar{p}\right)$
        of the transformed Hamiltonian in \cref{eq:H_Cos} (lower panel).
        In the ghostified case (lower panel), all trajectories with initial momentum 
        $\left|\bar{p}_0\right|>1$ (cf. red horizontal lines)
        evolve over the potential barrier and run away to plus or minus infinity in $\bar{x}$, irrespective of the initial value of $\bar{x}_0$. Two such runaway trajectories are depicted in blue. 
        In contrast, all phase space trajectories of \cref{eq:H_Cosh} (upper panel) are closed curves.
    }
\end{figure}

To avoid confusion, we stress that complex canonical transformations $\mathfrak{C}^{\pm}_{x^n}$ preserve the physics only when accompanied by the corresponding rotations in the complex planes of the initial data. Otherwise, the transformed theory is physically different and equivalent to probing the original theory with purely imaginary initial data. Therefore, the dynamical properties of such ``ghostified'' theories can be completely different from those of the original theory when all quantities considered are real, which we will assume in our PG models.

Let us illustrate this by a simple example of a theory with one real degree of freedom $x$ (and real $p$)
\begin{equation}
H=\frac{p^{2}}{2}+\frac{1}{4}\cosh x\,.\label{eq:H_Cosh}
\end{equation}
Clearly, for arbitrary, we stress \emph{real}, initial data, the solution perpetually evolves
in a finite region of phase space bounded by the potential well $\cosh x$
and the positive kinetic energy. This solution is a closed trajectory in the phase space $\left(x,p\right)$, see the upper panel of \cref{fig:Phase_space_change}. Now let us perform the canonical transformation $\mathfrak{C}^{-}_{x}$, so that $p=i\bar{p}$ and $x=-i\bar{x}$, transforming \cref{eq:H_Cosh} to the new theory of real variables ($\bar{x}$, $\bar{p}$) and Hamiltonian, i.e.,
\begin{equation}
\bar{H}=-\frac{\bar{p}^{2}}{2}+\frac{1}{4}\cos\bar{x}\,.\label{eq:H_Cos}
\end{equation}
We could consider the solution $\left(\bar{x}\left(t\right),\bar{p}\left(t\right)\right)$
for the system in \cref{eq:H_Cos}, with purely imaginary initial data $\left(\bar{x}_0,\bar{p}_0\right)=\left(ix_{0},-ip_{0}\right)$
to obtain just the real solution $\left(x\left(t\right),p\left(t\right)\right)=\left(-i\bar{x}\left(t\right),i\bar{p}\left(t\right)\right)$
of \cref{eq:H_Cosh} with the real initial data $\left(x_{0},p_{0}\right)$. In particular, $H=\bar{H}$ for all the solutions which are related in this way. 
However, a solution of \cref{eq:H_Cos} with real initial data $\left(\bar{x}_{0},\bar{p}_{0}\right)$
corresponds to a purely imaginary solution of \cref{eq:H_Cosh} with
initial data $\left(-i\bar{x}_{0},i\bar{p}_{0}\right)$. Thus, purely
real solutions of \cref{eq:H_Cosh} are very different from purely
real solutions of \cref{eq:H_Cos} even though both systems are related
by a simple canonical transformation. In particular, for arbitrary real initial coordinates $\bar{x}_0$ and initial momenta
$\bar{p}_0>1$ (or $\bar{p}_0<-1$), the trajectory of \cref{eq:H_Cos}
evolves over the potential barrier and is unbounded, with a runaway of $\bar{x}(t)$, see the right panel of \cref{fig:Phase_space_change}. As we mentioned earlier, there are no such real-valued trajectories in \cref{eq:H_Cosh}. Clearly, in the phase space $\left(\bar{x},\bar{p}\right)$ of \cref{eq:H_Cos} such real-valued trajectories are not closed, whereas all real-valued trajectories in $\left(x,p\right)$
of \cref{eq:H_Cosh} are closed, thus even the topology of the phase space trajectories is changed.
In this regard it is worth mentioning that with real $\left(\bar{x},\bar{p}\right)$ there are not only an infinite number of new 
stable equilibrium points, but also new unstable equilibrium points in-between the former, see \cref{fig:Phase_space_change}.
While the above refers to local (Lyapunov) stability, the ``ghostified'' real theory in \cref{eq:H_Cosh} is globally (Lagrange) unstable, in contrast to the original real theory \cref{eq:H_Cosh}. Two of the respective runaway trajectories are marked as thick blue lines in \cref{fig:Phase_space_change}.

\section{The integrable Liouville Model \& a proof of bounded motion}
\label{sec:Liouville}

In the following, we consider an integrable class of theories, cf.~class 1 in \cref{app:integrable-Hamiltonians}, for which we can rigorously establish boundedness of motion. Specifically, we focus on
\begin{align}
    \label{eq:Liouville-hamiltonian}
    H_\text{LV} &= \frac{p_x^2}{2} + \sigma \frac{p_y^2}{2} + V_\text{LV}(x,y)
    \\
    \label{eq:Liouville-potential}
    \text{with}\quad
    V_\text{LV} &= \frac{f(u) - g(v)}{u^2 - v^2}\;,
\end{align}
where $f$ and $g$ are arbitrary functions of the coordinates $(u,v)$. The coordinates $(u,v)$ are related to $(x,y)$ via
\begin{align}
    u^2 &= \frac{1}{2}\left(r^2 + c +\sqrt{(r^2 + c)^2 - 4\,c\,x^2}\right)\,, 
    \label{defu}
    \\
    v^2 &= \frac{1}{2}\left(r^2 + c -\sqrt{(r^2 + c)^2 - 4\,c\,x^2}\right)\,, 
    \label{defv}
    \\[0.5em] \label{defr}
    r^2 &= x^2 + \sigma\,y^2\,.
\end{align}
where $c$ is an arbitrary constant. 
Besides the Hamiltonian, this theory exhibits a second constant of motion, i.e.,
\begin{align}
    \label{eq:Liouville-constant-of-motion_x-y}
    I_\text{LV} &= -\sigma\left(
        p_y x - \sigma\,p_x y
    \right)^2
    - c\,p_x^2
    + \mathcal{V}
    \\[0.5em]\label{defmathcalV}
    \text{with}\quad
    \mathcal{V} &=
    2\,\frac{u^2g(v) - v^2f(u)}{u^2 - v^2}\;,
\end{align}
where we have (for later purpose) defined the momentum-independent part of the constant of motion $\mathcal{V} = I_\text{LV}|_{p_{x,y}\rightarrow0}$.
Although this model was initially studied in the PP case (i.e. with $\sigma =+1$) the discussion of the previous section shows that it stays integrable also for $\sigma =-1$, i.e., in the PG case. The corresponding extra integral of motion (i.e. \cref{eq:Liouville-constant-of-motion_x-y} with $\sigma=-1$) 
is obtained from the PP one (i.e. \cref{eq:Liouville-constant-of-motion_x-y} with $\sigma=+1$), as explained in the previous section.

\begin{figure}
    \centering
    \includegraphics[width=\linewidth]{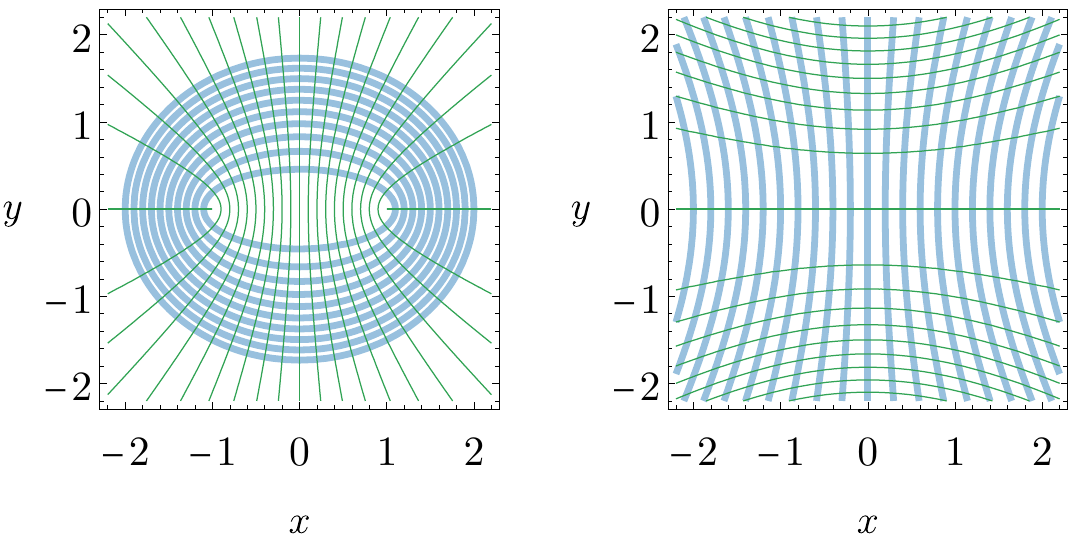}
    \caption{
    Left-hand panel: Curves of constant $u$ (thick) and $v$ (thin) in the $(x,y)$ plane for the Liouville model without ghost ($\sigma=+1$).
    Right-hand panel: Curves of constant $u$ (thick) and $-iv$ (thin) in the $(x,y)$ plane for the Liouville model with ghost ($\sigma=-1$).)
    }
    \label{fig:coords-uv-xy}
\end{figure}

For $\sigma = 1$ and $c\geqslant 0$, $u$ and $v$ define so-called elliptic coordinates, i.e., curves of constant $u\in(\sqrt{c},\infty)$ trace out ellipses and curves of constant $v\in(0,\sqrt{c})$ trace out hyperbolas in the $(x,y)$ plane. For $\sigma = -1$ and $c\leqslant 0$, curves of constant $u\in(0,\infty)$ and curves of constant $-i\,v\in(\sqrt{c},\infty)$ both trace out hyperbolas in the $(x,y)$ plane, cf.~\cref{fig:coords-uv-xy}.
\\

We note that, even though this class of integrable systems was derived, and is usually presented, with an arbitrary constant $c$, it is only the sign of this constant which influences the
dynamics. Indeed, for $c\neq0$ one can first perform a canonical transformation 
\begin{align}
x\rightarrow\left|c\right|^{1/2}\,x\,, & \qquad y\rightarrow\left|c\right|^{1/2}\,y\,,
\nonumber\\
p_{x}\rightarrow\left|c\right|^{-1/2}\,p_{x}\,, & \qquad p_{y}\rightarrow\left|c\right|^{-1/2}\,p_{x}\,.
\label{eliminate}
\end{align}
Then one can complete it with rescaling of time
\begin{align}
t\rightarrow\left|c\right|\,t\,, & \qquad H\rightarrow \left|c\right|^{-1} \, H\,.
\label{time_transform}
\end{align}
These transformations completely absorb $|c|$ into the arbitrariness of $f(u)$ and $g(v)$, except for the limit cases
$c\rightarrow\infty$ or $c\rightarrow0$. For instance, the latter limit transforms
the Liouville model to a subclass of class 2, see \cref{eq:class2}, with $f=0$. 
\\
The above integrable Hamiltonian in \cref{eq:Liouville-hamiltonian} follows as a subclass of a system first investigated by Liouville~\cite{Liouville1846} (in the PP case). Liouville starts out from a Hamiltonian with non-standard kinetic terms, i.e.,\begin{align}
    \label{eq:Liouville-hamiltonian_alpha-beta}
    H_\text{LV} = \frac{1}{F(\alpha) - G(\beta)}\left[
        \frac{p_\alpha^2}{2}
        + \frac{p_\beta^2}{2}
        +\left(
            f(\alpha) - g(\beta)
        \right)
    \right]\;,
\end{align}
where $(\alpha,\,\beta)$ are yet another set of coordinates, $(p_\alpha,\,p_\beta)$ are the respective canonical momenta, and $f(\alpha)$, $F(\alpha)$, $g(\beta)$, $G(\beta)$ are arbitrary functions. Following Liouville, we found that
\begin{align}
    I_\text{LV} = p_\alpha^2 + 2\,f(\alpha) - 2\,F(\alpha)\,H_\text{LV}
\end{align}
Poisson-commutes with the Hamiltonian $H_\text{LV}$ and thus defines a second constant of motion\footnote{
    Equivalently, $K_\text{LV} = p_\beta^2 - 2\,g(\alpha) + 2\,G(\alpha)\,H_\text{LV}$ denotes a constant of motion, but $K_\text{LV} = -I_\text{LV}$ and is therefore not independent.
}. Hence, the model is integrable, irrespective of the choice of $f(\alpha)$, $F(\alpha)$, $g(\beta)$, and $G(\beta)$. 

The coordinates $(\alpha,\,\beta)$ are related to $(u,\,v)$ via
\begin{align}
    u^2 &= c\,\sinh{\alpha}^2\;,
    \notag\\
    v^2 &= c\,\sin{\beta}^2\;.
\end{align}
In coordinates $(u,\,v)$, the Liouville Hamiltonian $H_\text{LV}$ and constant of motion $I_\text{LV}$ read
\begin{align}
    H_\text{LV} &= \frac{1}{F - G}\Bigg[
        \frac{p_u^2}{2} (u^2 - c)
        + \frac{p_v^2}{2} (c - v^2)
        +\left(
            f - g
        \right)
    \Bigg],
    \label{eq:Liouville-hamiltonian_u-v}
    \\
    I_\text{LV} &= \frac{1}{F - G}\Bigg[
        \frac{p_u^2}{2} (u^2 - c)\,G
        + \frac{p_v^2}{2} (c - v^2)\,F
    \notag\\&\quad\quad\quad\quad\quad
        -\left(
            f\,G - g\,F
        \right)
    \Bigg]\;.
    \label{eq:Liouville-constant-of-motion_u-v}
\end{align}
For the specific choice of $F(u)=u^2$ and $G(v)=v^2$, and with the transformation from $(u,v)$ to $(x,y)$, these reduce to the Hamiltonian and constant of motion previously given in \cref{eq:Liouville-hamiltonian,eq:Liouville-constant-of-motion_x-y} with $\sigma=+1$. For more general $F(u)$ and $G(v)$, the transformations to $(x,\,y)$ coordinates does not bring the Hamiltonian to a form with standard kinetic terms. In the following, we will thus restrict to the case of $F(u)=u^2$ and $G(v)=v^2$.

\subsection{Proof of boundedness of motion}
\label{subsec:proof}
We show here that the PG Liouville model with $\sigma=-1$ (in the form of \cref{eq:Liouville-hamiltonian} and with the additional constant of motion in \cref{eq:Liouville-constant-of-motion_x-y}) is such that -- under certain conditions on $f(u)$ and $g(v)$ detailed below -- the phase-space motion is always bounded, irrespective of the choice of initial conditions. This applies to the model considered in~\cite{Deffayet:2021nnt}, which is a special class of PG Liouville models, but also extends the result of ~\cite{Deffayet:2021nnt} to a much larger class of ghostly oscillator interacting with a positive energy oscillator where the motion can hence be proven analytically not to run away.

We start by defining three further helpful coordinate combinations
\begin{align}
    W &\equiv u^2+v^2 =x^2-y^2+c \notag \\
    & = r^2 +c \, ,  \label{defW}
    \\ 
    \tilde{W} &\equiv u^2-v^2 \notag \label{defV}
    \\
    &= \sqrt{r^{4}- 2 c \left(x^{2}+y^{2}\right)+c^2} \, ,\\
    \label{uvx2}
    w &\equiv -4\,u^2\,v^2 =  -4\,c\,x^2\;,
\end{align}
with which we may re-express $\tilde{W} = \sqrt{W^2 + w}$ and write
\begin{align}
    u^2 &= \frac{1}{2} \left(W + \sqrt{W^2 + w} \right) =\frac{1}{2} \left(W+\tilde{W}\right)
    \label{uWC}\\ 
    \label{vWC}
    v^2 &= \frac{1}{2} \left(W - \sqrt{W^2 + w} \right) = \frac{1}{2} \left(W-\tilde{W}\right)
\end{align}
We choose here and henceforth a negative $c < 0$  such that the momentum-dependent part of the constant of motion (cf.~the first two terms in \cref{eq:Liouville-constant-of-motion_u-v}) are positive. Furthermore, a negative $c < 0$ implies
\begin{align}
    v^2 \leqslant  0 \label{signv}\;,
\end{align}
and while the above $v^2$ (given in \cref{uWC}) is a real expression, the so defined $v$ is purely imaginary. In order to manipulate real and positive variables, we define the real and positive quantities $\tilde{v}$ and $\tilde{c}$ as 
\begin{align}
    \tilde{v} &= |v| \label{defvt} \\
    \tilde{c} &= |c| = -c
\end{align}
such that $v^2 = - \tilde{v}^2$, and thus we have 
\begin{align}
    u^2 &\geqslant 0 \label{signu}\;,\\
    \tilde{v}^2  &\geqslant 0 \label{signvtilde}\;,\\
    \tilde{W} &\geqslant 0 \;, \label{signV}\\
    \tilde{c} &> 0 \label{sign_tilde_c} \;.
\end{align}
all these quantities being real, as well as, by choice, the functions $f$ and $g$.
\\

We now prove that phase-space motion is bounded if
\begin{enumerate}
    \item[(i)]
    $f(u)$ and $g(v)$ are bounded below, i.e.,
    \begin{align}
        f(u)&\geqslant f_0
        \label{eq:bounded-f}
        \;,
        \\
        g(v)&\geqslant g_0
        \label{eq:bounded-g}
        \;,
    \end{align}
    with constants $f_0,g_0\in\mathbb{R}$; and
    \item[(ii)]
    at large $|u|$ and $|v|$,
    these lower bounds sharpen to
    \begin{align}
        f(u)&\geqslant 4 F_0\,|u|^\zeta >  0
        \label{Hyponf}
        \;,
        \\
        g(v)&\geqslant 4  G_0\,|v|^\eta >  0
        \label{Hypong}
        \;,
    \end{align}
    with positive constants $F_0,G_0\in\mathbb{R}^+$ as well as $\zeta > 2$ and $\eta >2$ (and the factor 4 is just here for later convenience).
\end{enumerate}

The first  step of our proof is the introduction of a new first integral $J_{\text{LV}}$ defined as (using  \cref{defvt,signvtilde,sign_tilde_c})  
\begin{align}
J_{\text{LV}}&=I_{\text{LV}}-\tilde{c}\,H_{\text{LV}}\notag \\&=\left(xp_{y}+yp_{x}\right)^{2}+\frac{\tilde{c}}{2}\left(p_{x}^{2}+p_{y}^{2}\right)+\mathcal{U}\,, \label{NewFirstIntegral}
\end{align}
where 
\begin{align}
\mathcal{U}\left(u,v\right)&=\frac{\left(2\,u^{2}+\tilde{c}\right)g\left(v\right)+\left(2\,\tilde{v}^{2}-\tilde{c}\right)f\left(u\right)}{u^{2}+\tilde{v}^{2}}
\notag \\
&= \gamma_+ g(v) + \gamma_- f(u) \, ,
\label{defU}
\end{align}
and $\gamma_{\pm}$ are obtained to be (using \cref{defV,uWC,vWC})
\begin{align}
\gamma_\pm =    1 \pm \frac{r^2}{\tilde{W}}\, .
\end{align}
Using \cref{defV}, we get $-1<\frac{r^{2}}{\tilde{W}}<1$
implying 
\begin{equation}
0<\gamma_{\pm}<2\,.
\end{equation}
This, using the condition (i) above implies that 
\begin{align}
& \gamma_+ g(v) \geq -2 |g_0| \label{gvg0}\,,\\
& \gamma_- f(u) \geq -2 |f_0| \label{fuf0}\,,
\end{align}
where the equality may hold if $g_0=0$ or/and $f_0=0$,
and hence, using \cref{defU}, we get that  
\begin{equation}
    \mathcal{U}\geq -2\left(\left|g_{0}\right|+\left|f_{0}\right|\right)\,, 
\label{bound_on_U}
\end{equation}
showing that $\mathcal{U}$ is bounded below. As  $J_{\text{LV}}$ defined in \cref{NewFirstIntegral} is conserved, this implies that $|x p_y + y p_x|$, $|p_x|$ and $|p_y|$ are bounded on shell. This is the first step of our proof. As it should be clear, this first step puts forward an analogy between a conserved and bounded from below Hamiltonian and the conserved $J_{\text{LV}}$, where all the ``kinetic energies'' are positive definite and the ``potential energy'' $\mathcal{U}\left(u,v\right)$ has just been shown to be bounded from below. 

To proceed further, we use again this analogy, showing that conditions (ii) imply that $\mathcal{U}\left(x,y\right)$ is growing without bound at large radius $R=\sqrt{x^{2}+y^{2}}$, which excludes runaways of the dynamical variables $x$ and $y$. To do so we introduce polar coordinates $(R,\phi)$ such that $x=R \cos \phi$ and $y = R \sin \phi$, yielding $r^2=x^{2}-y^{2}=R^{2} \kappa$ where we have defined
$\kappa = \cos 2 \phi$.
As a consequence, one has 
\begin{align} \label{gammaRkappa}
    \gamma_{\pm} = 1 \pm \frac{R^{2} \kappa}{\sqrt{R^{4}\kappa^2+2R^{2}\tilde{c}+\tilde{c}^2}}\, . 
\end{align}
As we are interested in large $R$, we can assume that 
\begin{align}
    \sqrt{2} R \gg \sqrt{\tilde{c}}\;, \label{boundRC}
\end{align}
implying, in particular, that the last term in the square-root in \cref{gammaRkappa} is much smaller than the second one.
We also have from \cref{defV}  
\begin{align}
    u^2 + \tilde{v}^2 = \sqrt{R^{4}\kappa^2+2R^{2}\tilde{c}+\tilde{c}^2} >  \sqrt{R^{4}\kappa^2+2R^{2}\tilde{c}}  \,.
\end{align}
Using this we conclude that one has either 
\begin{align}
    u^2 > \frac{1}{2}\left(\sqrt{R^{4}\kappa^2+2R^{2}\tilde{c}}\right)\,, \label{Generalboundu2}
\end{align}
or 
\begin{align}
    \tilde{v}^2 > \frac{1}{2}\left(\sqrt{R^{4}\kappa^2+2R^{2}\tilde{c}}\right)\,.\label{Generalboundv2}
\end{align}
Distinguishing the above, and moreover between
\begin{align}
R^4 \kappa^2 &< 2 R^2 \tilde{c} \; \;  \Leftrightarrow \; \; R  |\kappa| < \sqrt{2 \tilde{c}}\label{case1RKappa} \, , \\
\text{or}\quad\quad
R^4 \kappa^2 & \geq 2 R^2 \tilde{c}\; \;  \Leftrightarrow \; \; R  |\kappa| \geq \sqrt{2 \tilde{c}}\label{case2RKappa}\;,
\end{align}
we will now show that, in all the respective four cases, $\mathcal{U}$ grows at least as fast as a suitable power of $R$.

Let us first assume that \cref{case1RKappa} holds. 
In this case we have 
\begin{align}
    \sqrt{R^4 \kappa^2 + 2 R^2 \tilde{c} + \tilde{c}^2} > \sqrt{2  R^4 \kappa^2}. 
\end{align}
This implies a lower bound on $\gamma_\pm$ given by\footnote{where the $1/4$ has been put just for convenience in order to alleviate some later expressions.}
\begin{align} \label{boundgamma1over4}
\gamma_{\pm} \geq 1 - \frac{1}{\sqrt{2}} > \frac{1}{4}\;,
\end{align}
which holds for either sign of $\kappa$ (and including $\kappa=0$).
Moreover, we have from \cref{Generalboundu2,Generalboundv2} that either $u^2 \geq R \sqrt{\tilde{c}/2}$
or $\tilde{v}^2 \geq R \sqrt{\tilde{c}/2}$.
In the first subcase, we can use the bounds \cref{Hyponf,gvg0,boundgamma1over4} to get that at large 
 $R$ 
\begin{align}
\mathcal{U} \geq -2\, |g_0| + F_0\ \left(\tilde{c}/2\right)^{\zeta/4}\,R^{\zeta/2}\,. \label{Ubound1}
\end{align}
In the second subcase, we get similarly 
\begin{align}
\mathcal{U} \geq -2\, |f_0| + G_0\ \left(\tilde{c}/2\right)^{\eta/4}\,R^{\eta/2}\,. \label{Ubound2}
\end{align}
Hence, at least one of these two bounds holds whenever \cref{case1RKappa} does. 

We then turn to the second 
case in \cref{case2RKappa}, where obviously $\kappa \neq 0$, and we can hence rewrite \cref{gammaRkappa} as 
\begin{align} 
    \gamma_{\pm} = 1 \pm \frac{\kappa}{|\kappa|} \frac{1}{\sqrt{1 + \xi}}\,,
\end{align}
with $\xi$ given by 
\begin{align} 
   \xi =  \frac{2 \tilde{c} }{R^2 \kappa^2} + \frac{\tilde{c}^2 }{R^4 \kappa^2}\,.
\end{align}
The inequalities in \cref{boundRC,case2RKappa} (and the assumption $\tilde{c}>0$) imply in turn that\footnote{This upper bound can be strengthened but is enough for the present discussion.} 
\begin{align} \label{rangexi}
    0 < \xi \leq 2\,.
\end{align}
 Let us then first assume that $\kappa >0$. In this case we can obtain a useful lower bound on $\gamma_-$ noticing that, for $\xi$ in the range in \cref{rangexi}, one has 
 \begin{align}
     1-\frac{1}{\sqrt{1+ \xi}} 
     \geq \left(1-\frac{1}{\sqrt{3}}\right)\frac{\xi}{2}
     > \frac{1}{8} \xi\,.
 \end{align}
Using this, we conclude that, for positive $\kappa$ and whenever \cref{case2RKappa} holds, we have
\begin{align} \label{boundgammabis}
    \gamma_- >
    \frac{\tilde{c}}{ 4 R^2 \kappa^2} \,.
\end{align}
We also note, following the same logic, that $\gamma_+$ obeys exactly the same bound for negative $\kappa$ whenever \cref{case2RKappa} holds.
We can then, as previously, distinguish between the two subcases in \cref{Generalboundu2,Generalboundv2} which implies that 
either $u^2 \geq |\kappa| R^2 /2$
or $\tilde{v}^2 \geq |\kappa| R^2 /2$.
In the first subcase, and for positive $\kappa$, we can use the bounds in \cref{Hyponf,boundgammabis} to get that at large 
 $R$
\begin{align} 
\gamma_- f(u) &>
\tilde{c} F_0 2^{-\frac{\zeta}{2}} \left(\kappa R^2\right)^{\frac{\zeta}{2}-1}\kappa^{-1} \notag \\
&\geq \tilde{c} F_0 2^{-\frac{\zeta}{2}} \left( R \sqrt{2 \tilde{c}}\right)^{\zeta/2-1}\kappa^{-1} \notag \\
&\geq \tilde{c} F_0 2^{-\frac{\zeta}{2}} \left( R \sqrt{2 \tilde{c}}\right)^{\zeta/2-1}\,, \notag
\end{align}
where to go from the first line to the second line we used once more \cref{case2RKappa} and to go from the second line to the third line we used the fact that $ 1 \geq \kappa >0$. Hence, using this last inequality as well as \cref{gvg0}, we get that, for positive $\kappa$ and whenever \cref{case2RKappa} and $u^2 \geq \kappa R^2 /2 $ hold, we have
\begin{align}
\mathcal{U} >
-2\, |g_0| +\tilde{c}F_0 2^{-\frac{\zeta}{2}} \left( R \sqrt{2 \tilde{c}}\right)^{\zeta/2-1}\,. \label{Ubound3}
\end{align}
Let us then turn to discuss the subcase where $\tilde{v}^2 > |\kappa| R^2 /2$. In this subcase, still assuming that $\kappa>0$ we notice that $\gamma_+ \geq 1$ and hence we get simply using \cref{Hypong,fuf0} that 
\begin{align}
\mathcal{U} &\geq -2\, |f_0| + 4 G_0 2^{-\frac{\eta}{2}} \left(\kappa R^2\right)^{\frac{\eta}{2}} \notag \\
&\geq -2\, |f_0| + 4 G_0 2^{-\frac{\eta}{2}} \left(R \sqrt{2 \tilde{c}}\right)^{\frac{\eta}{2}}\,, \label{Ubound4}
\end{align}
where to go from the first line to the second line we used again \cref{case2RKappa}. 
Similar bounds can be obtained for negative $\kappa$, {\it mutatis mutandis}.
Indeed, we find in this case, following the same logic, one has either 
\begin{align}
\mathcal{U} 
&\geq -2\, |g_0| + 4 F_0 2^{-\frac{\zeta}{2}} \left(R \sqrt{2 \tilde{c}}\right)^{\frac{\zeta}{2}} \label{Ubound5}
\end{align}
or
\begin{align}
\mathcal{U} >
-2\, |f_0| +\tilde{c}G_0 2^{-\frac{\eta}{2}} \left( R \sqrt{2 \tilde{c}}\right)^{\eta/2-1}\,. \label{Ubound6}
\end{align}
Hence, we conclude from the considerations of \cref{Ubound1,Ubound2,Ubound3,Ubound4,Ubound5,Ubound6} that the hypotheses (i) and (ii) above are enough to conclude that $\forall\;  \mathcal{U}_0 >0 \;, \exists R_0 > 0$ such that $R>R_0 \Rightarrow \mathcal{U} > \mathcal{U}_0$. This concludes our proof: in all cases,  the ``potential energy'' $\mathcal{U}$ grows without bound with $R$ implying that $|x|$ and $|y|$ stay bounded on shell.
\\

To summarize, we have shown that the phase-space motion stays bounded at all times, irrespective of the initial conditions, for functions $f$ and $g$ which (i) are bounded below, and (ii) at large values of their arguments, obey respectively \cref{Hyponf} (where $F_0 >0 $ and $\zeta > 2$) and \cref{Hypong} (where $G_0 >0$ and $\eta > 2$). 
We stress that the behaviour of the functions $f$ and $g$, entering in condition (ii), only needs to be specified at large $|u|$ and $|v|$, implying, in particular, that the lower bounds $f_0$ and $g_0$ can be negative. 

The above proof can also be extended to the case where, starting from the Liouville PP model, one ghostifies $x$ 
by applying a canonical transformation $\mathfrak{C}^{\pm}_{x}$. The obtained models are in fact identical (up to a global sign in the Hamiltonian) to the previously considered PG models obtained by ghostifying $y$ from the PP model via a canonical transformation $\mathfrak{C}^{\pm}_{y}$. As a result, the two kinds of models with a ghost are stable under the same conditions (i) and (ii). This is shown explicitly in~\cref{app:ghoslty x}. 
\\

\begin{figure*}
    \centering
    \includegraphics[trim=0cm 0cm 0cm 0cm, clip, width=0.49\linewidth]{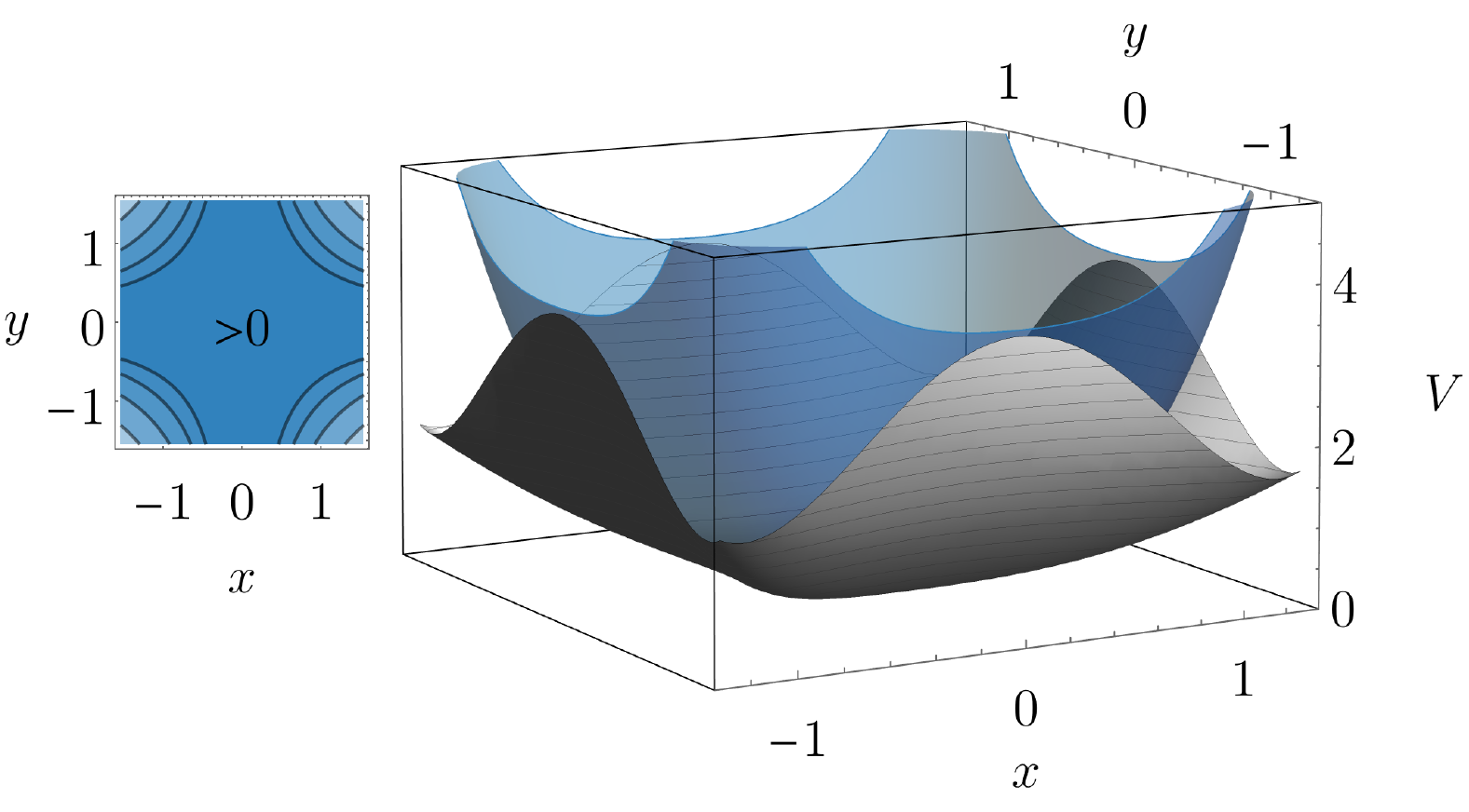}
    \hfill\vline\hfill
    \includegraphics[trim=0cm 0cm 0cm 0cm, clip, width=0.49\linewidth]{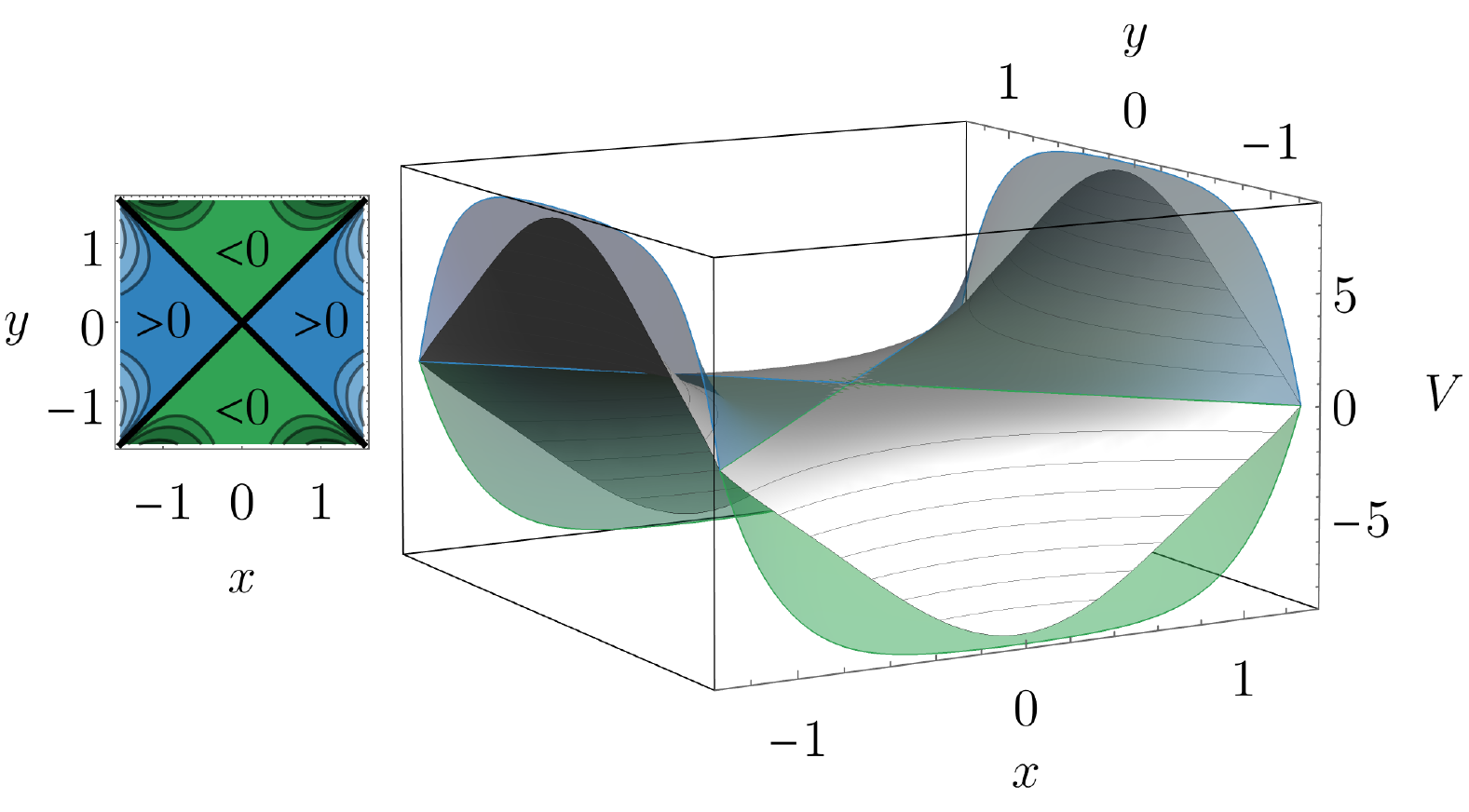}
    $\phantom{1}$
    \caption{
    We show the first two non-trivial examples of polynomial Liouville potentials, corresponding to $N=3$ (left-hand panel) and $N=4$ (right-hand panel) in \cref{defVN}.
    With the notation of \cref{V4simp}: the left-hand panel, corresponds to $\omega_x^2=1$, $\omega_y^2=-1$, $\mathcal{C}_4=0$, and $\tilde{c}=1$; the right-hand panel, corresponds to $\omega_x^2=1$, $\omega_y^2=1$, $\mathcal{C}_4=1$, and $\tilde{c}=1$. 
    With regards to \cref{eq:pre-conjecture-condition}, we also show $V_\text{P}(x) + V_\text{G}(y)$ as transparent surfaces. 
    For clarity, we add respective contour plots of $(V_P + V_G) - V$ with contours at $(V_P + V_G) - V=(-8,-4,-2,0,2,4,8)$ and with respective shading from darker to lighter colour, where we also indicate regions in which $(V_P + V_G) - V\gtrless 0$.
    }
    \label{fig:condition3D}
\end{figure*}
%

\subsection{A first application of the proof}
\label{subsec:firstapp}
As a first application of the above proof, we consider potentials $V_\text{LV}(x,y)$ generated by functions $f$ and $g$ which are polynomials in $u^2$ and $v^2$ respectively, i.e. of the form 
\begin{align} \label{fpoly-expl}
    f(u) &=
    \sum_{n=1}^{N_f}
    \mathcal{C}_{n}
    \left(u^2\right)^n
    \;,
    \\ \label{gpoly-expl}
    g(v) &= 
    \sum_{n=1}^{N_g}
    \mathcal{D}_{n}
    \left(v^2\right)^n
    \;.
\end{align}
with $N_f\in\mathbb{N}$, $N_g\in\mathbb{N}$, $\mathcal{C}_n\in\mathbb{R}$ and $\mathcal{D}_n\in\mathbb{R}$.
More specifically, choosing $N_f = N_g =2$, we get
\begin{align} \label{fpoly2}
    f(u) &= \mathcal{C}_{0} + \mathcal{C}_{1} u^2 + \mathcal{C}_{2} u^4
    \;,
    \\ \label{gpoly2}
    g(v) &= \mathcal{D}_{0} + \mathcal{D}_{1} v^2 + \mathcal{D}_{2} v^4
    \;,
\end{align}
and the potential $V_\text{LV}(x,y)$ is given by 
\begin{align} \label{VLV-PRL}
     V_\text{LV}(x,y)&=
\frac{\mathcal{C}_1+\mathcal{D}_1}{2} + \frac{\mathcal{C}_2+\mathcal{D}_2}{2} W + \frac{\mathcal{C}_0-\mathcal{D}_0}{\tilde{W}}
\notag\\&\quad + (\mathcal{C}_1-\mathcal{D}_1)\frac{W}{2\tilde{W}} \notag\\&\quad + \frac{\mathcal{C}_2-\mathcal{D}_2}{4}\left(\frac{W^2}{\tilde{W}}+\tilde{W}\right) \;,
\end{align}
where $W$, $\tilde{W}$ and $w$ are defined in the previous subsection. 
Note that, in the above, the first term on the right-hand side is a mere constant and, as such, can be omitted. Moreover, the second term proportional to $W$ is a quadratic term for $x$ and $y$ (plus an irrelevant constant proportional to $\tilde{c}$). 
Choosing 
\begin{align}
\mathcal{C}_2 &>0 \;, \label{condC2}\\
\mathcal{D}_2 &>0 \;, \label{condD2}\\
\tilde{c}&>0\;, \label{condc}
\end{align}
the functions $f$ and $g$ defined by \cref{fpoly2,gpoly2} fulfill conditions (i) in \cref{eq:bounded-f,eq:bounded-g} and (ii) in \cref{Hyponf,Hypong} of \cref{subsec:proof} and as such yield models with bounded motion. 
The further choice 
$\mathcal{C}_1 = \mathcal{D}_1=1/2$, 
$\mathcal{C}_2 = \mathcal{D}_2=1/2$, 
$\mathcal{C}_0 - \mathcal{D}_0= \lambda$, and 
$\tilde{c}=1$ 
yields exactly the main model of reference~\cite{Deffayet:2021nnt} (whose Hamiltonian is given in equation (1) of this reference). The more general choice of a strictly negative $c<0$, strictly positive and equal $\mathcal{C}_2=\mathcal{D}_2>0$, and 
$\mathcal{C}_1 \geq \mathcal{D}_1$ (and, moreover, setting $\mathcal{D}_1 = -\mathcal{C}_1$ to remove an overall constant)
gives the largest set of models introduced in~\cite{Deffayet:2021nnt} (i.e., in Eq.~(20) and~(21) of this reference). The parameters $a$, $b$, $c$ and $d$ of~\cite{Deffayet:2021nnt} can be read off from the above as
\begin{align} 
    a &= - \frac{\mathcal{C}_{2} + \mathcal{D}_{2}}{2}\;,\notag\\
    b &= \mathcal{C}_{0} - \mathcal{D}_{0}\;,\notag\\
    c &= - \frac{\mathcal{C}_{1} - \mathcal{D}_{1}}{2}\;,\notag\\
    d &= \tilde{c}
    \notag
    \;,
\end{align}
where the left hand side of the above refers to the notation of~\cite{Deffayet:2021nnt} (and we stress, in particular, that $c$ on the left hand side of the third equality is introduced in \cite{Deffayet:2021nnt} and is {\it not} the same as the $c$ of the present work). We see that the conditions \cref{condC2,condc} imply that $a<0$ and $d>0$, and the present work provides a stability proof that was not given in \cite{Deffayet:2021nnt}. Note further that the condition on $c$ given in \cite{Deffayet:2021nnt} (i.e. $c \leq 0$) can, in fact, be relaxed and still yields stable motion.

\subsection{A polynomial subclass of the Liouville model: towards more general conditions for stable motion}
\label{subsec:integrable-motion}

The Liouville model at the heart of this work, and defined in \cref{eq:Liouville-potential,defr}, possess an interesting subclass for which the potential $V_\text{LV}(x,y)$ is polynomial in $x$ and $y$. This subclass is generated by
choosing
\begin{align} \label{fpoly}
    f(u) &=
    \sum_{n=1}^{N}
    \mathcal{C}_{n}
    \left(u^2\right)^n
    \;,
    \\ \label{gpoly}
    g(v) &= 
    \sum_{n=1}^{N}
    \mathcal{C}_{n}
    \left(v^2\right)^n
    \;,
\end{align}
with $N\in\mathbb{N}$ (and still $\mathcal{C}_n\in\mathbb{R}$), i.e., the \emph{same} polynomial form for $f$ and $g$. In \cref{app:polynomial-potentials}, we provide a proof that the resulting potential $V_\text{LV}(x,y)$, given by 
\begin{align} \label{defVN}
    V_\text{LV}^{(N)} &= 
    \sum_{n=1}^N \frac{\mathcal{C}_n}{u^2-v^2} \left[
        \left(u^2\right)^n - \left(v^2\right)^n
    \right]\;,
\end{align}
is indeed polynomial in $x$ and $y$.
To our knowledge, it was not noticed before
that this particular choice leads to polynomials, and hence, for {\it both} the PP and PG cases, to integrable systems with polynomial interactions whose specific form can be found below and in the \cref{app:polynomial-potentials}.

\begin{figure}
    \includegraphics[trim=0cm 0cm 0cm 0cm, clip, height=0.68\linewidth]{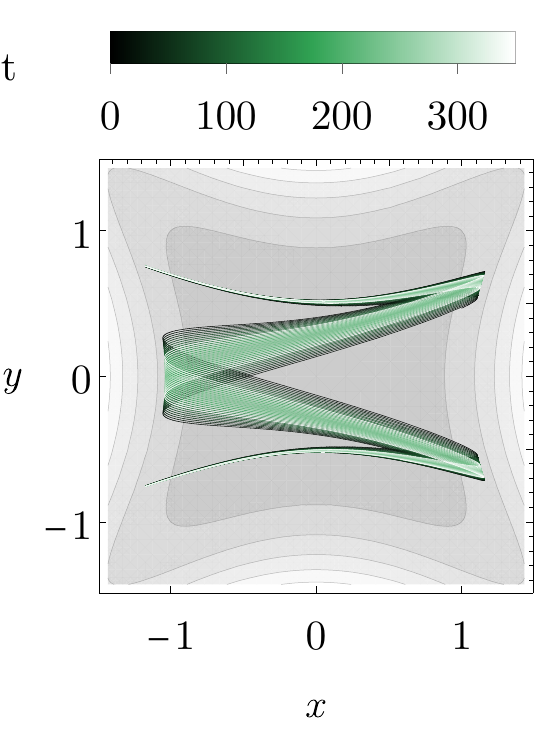}
    \hfill
    \includegraphics[trim=0cm 0cm 0cm 0cm, clip, height=0.68\linewidth]{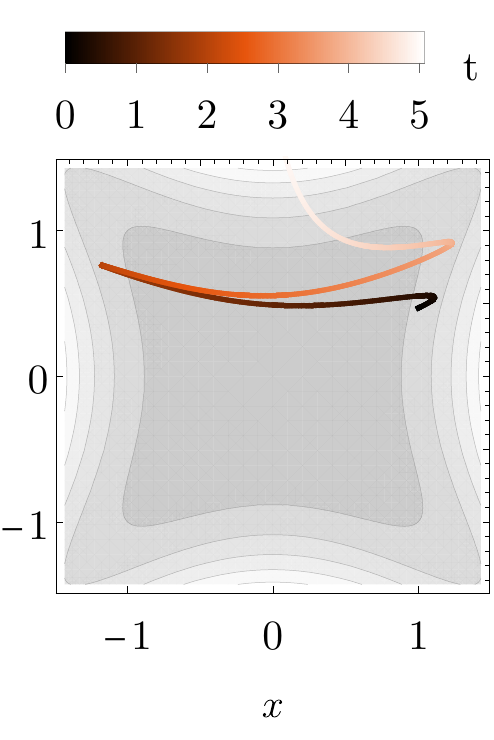}
    \caption{
    We show trajectories (with initial conditions tuned to the boundary of an apparent island of stability) for $N=3$, i.e., for $\omega_x^2=1$, $\omega_y^2=-1$, $\mathcal{C}_4=0$, and $\tilde{c}=1$, in \cref{V4simp}. We superimpose a contour plot of the potential (cf.~left-hand panel in \cref{fig:condition3D}) where increasingly light shades indicate increasingly large values of the potential. In the left-hand panel, we show a bounded trajectory. In the right-hand panel, we show an adjacent unstable trajectory. To be specific, and in case the reader wants to reproduce our results, we have chosen $x(t=0)=1$, $x'(t=0)=1$, $y(t=0)=A$, and $y'(t=0)=A$, for which the difference between bounded and unbounded motions is obtained by choosing respectively $A\approx0.468$ in the left-hand panel and $A\approx0.469$ in the right-hand panel. These values hence also locate the boundary between the two kinds of motions considered here.}
    \label{fig:N3_lissajous}
\end{figure}

\begin{figure*}
    \includegraphics[trim=0cm 0cm 0cm 0cm, clip, width=0.47\linewidth]{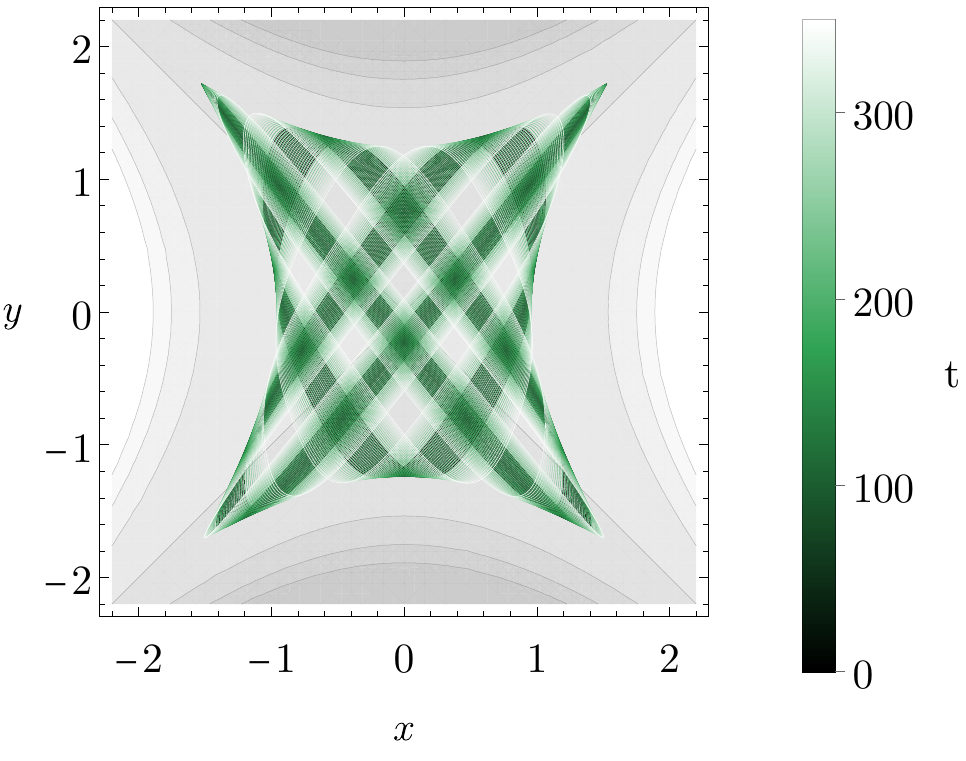}
    \hfill
    \includegraphics[trim=0cm 0cm 0cm 0cm, clip, width=0.49\linewidth]{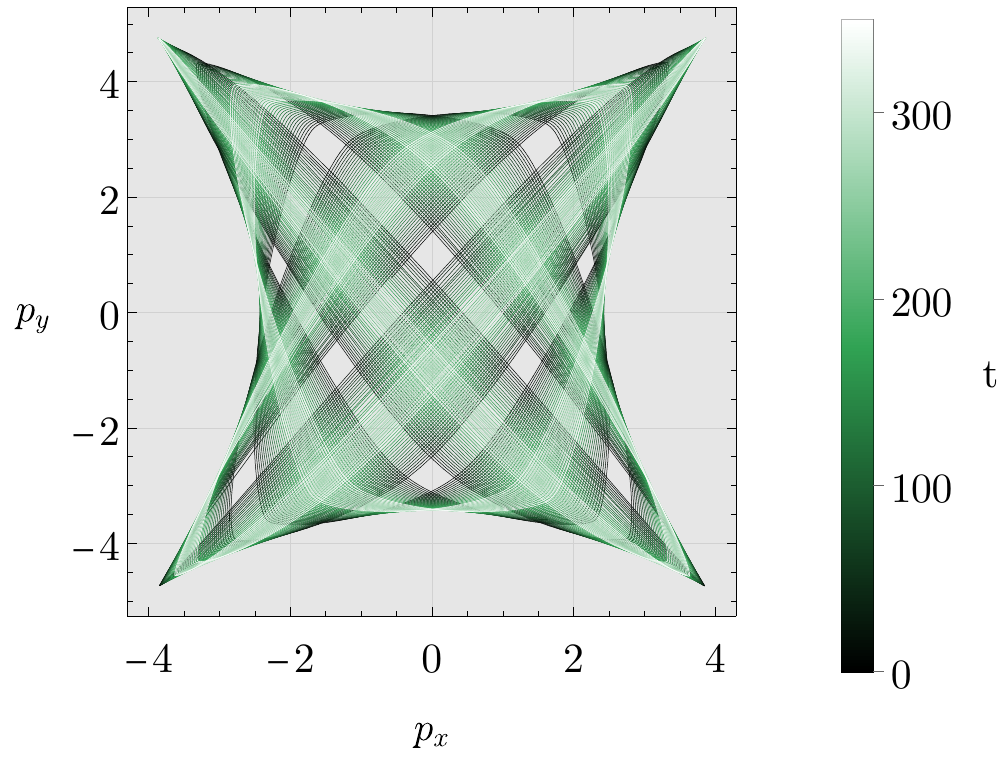}
    \caption{
    We show the trajectory (for random initial conditions drawn from a Gaussian distributions with standard deviation $\text{SD}=2$) for $N=4$, i.e., for $\omega_x^2=1$, $\omega_y^2=1$, $\mathcal{C}_4=1$, and $\tilde{c}=1$ in \cref{V4simp}. In the left-hand panel, we superimpose a contour plot of the potential (cf.~right-hand panel in \cref{fig:condition3D}) where increasingly light shades indicate increasingly large (positive) values of the potential. In the right-hand panel, we show the respective evolution of the momenta.
    }
    \label{fig:N4_lissajous}
\end{figure*}

\begin{figure*}
    \includegraphics[trim=0cm 0cm 0cm 0cm, clip, width=0.47\linewidth]{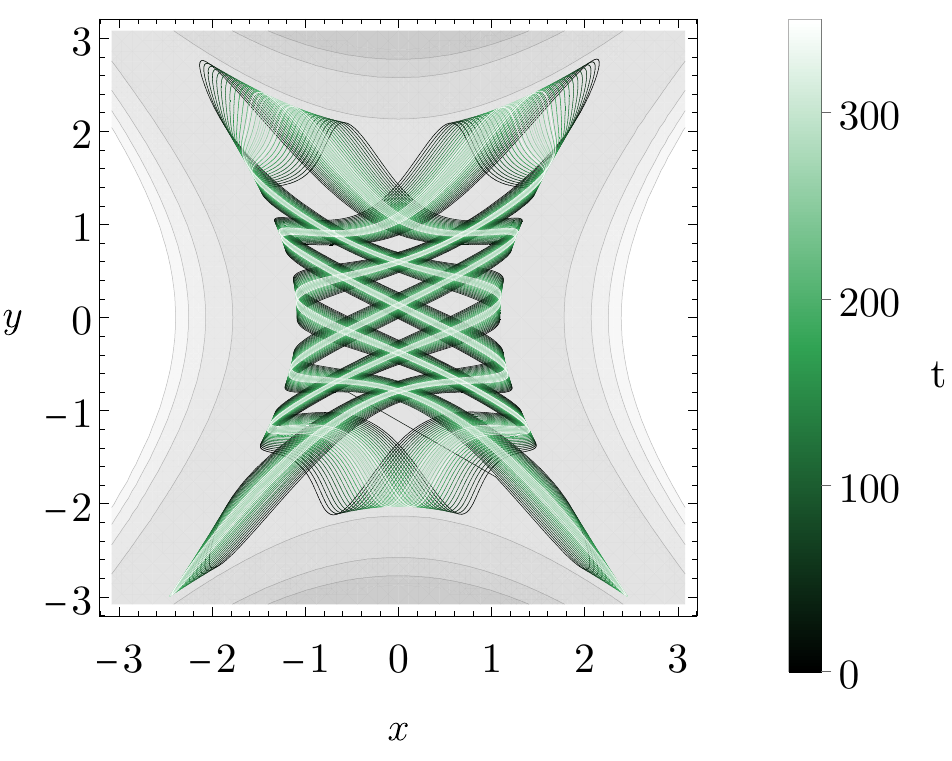}
    \hfill
    \includegraphics[trim=0cm 0cm 0cm 0cm, clip, width=0.49\linewidth]{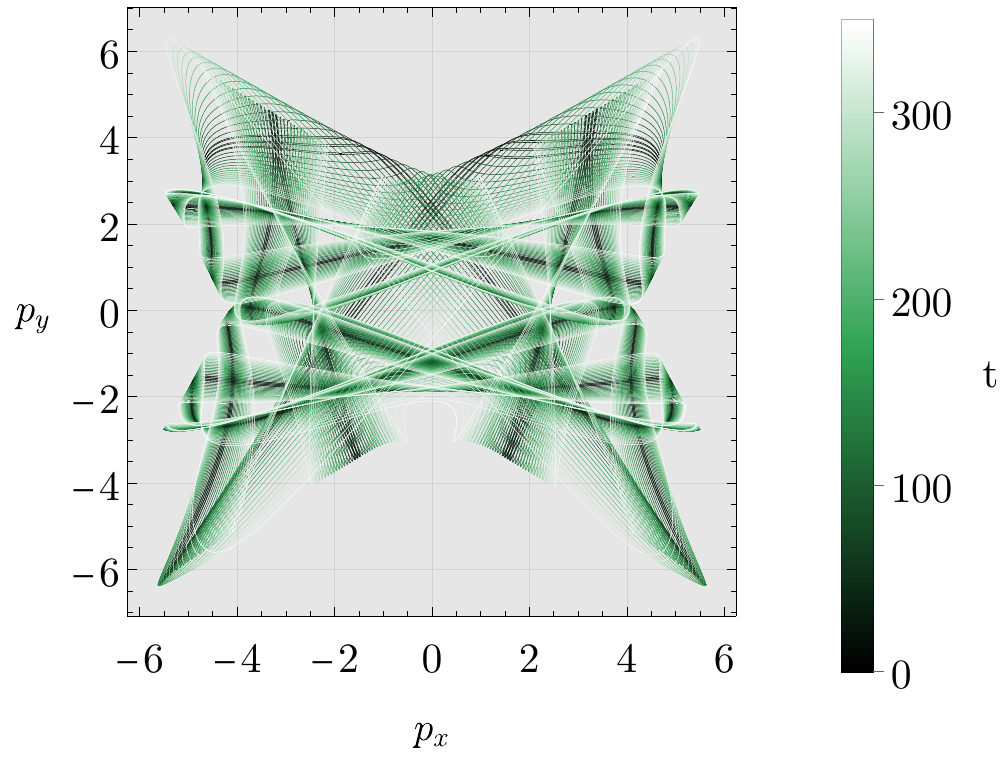}
    \caption{
    As in \cref{fig:N4_lissajous} but for a case with a ghost frequency which is tachyonic around the origin, i.e., for $\omega_x^2=5$, $\omega_y^2=-5$, $\mathcal{C}_4=1$, and $\tilde{c}=1$ in \cref{V4simp}.
    }
    \label{fig:N4_lissajous_tachyonic}
\end{figure*}

\begin{figure*}
    \includegraphics[trim=0cm 0cm 0cm 0cm, clip, width=0.47\linewidth]{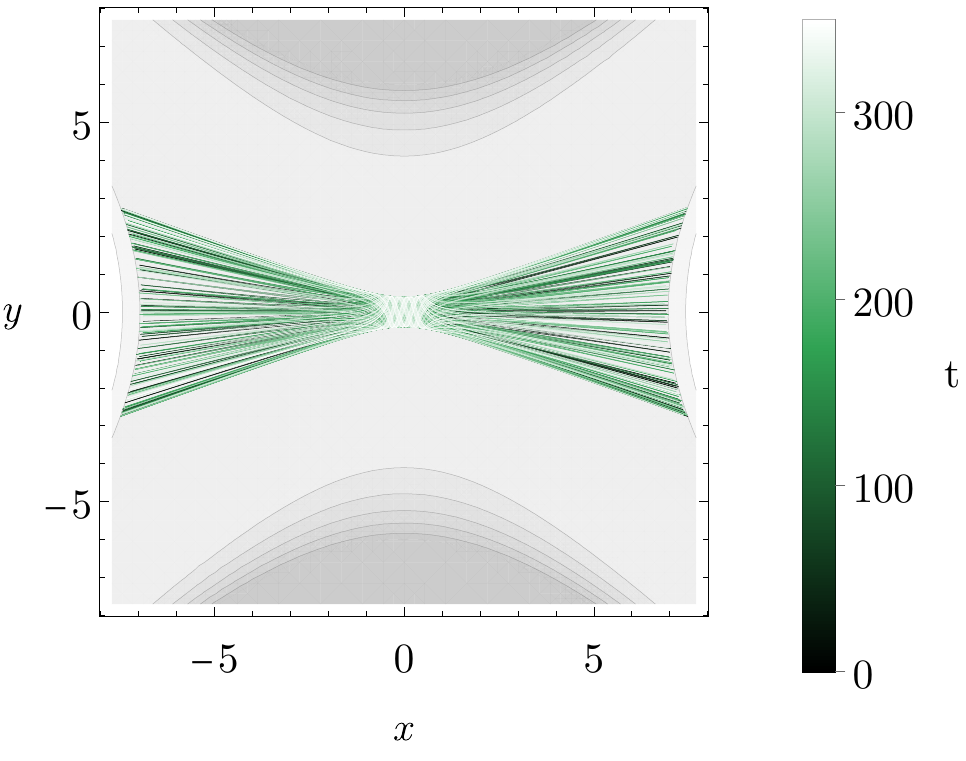}
    \hfill
    \includegraphics[trim=0cm 0cm 0cm 0cm, clip, width=0.49\linewidth]{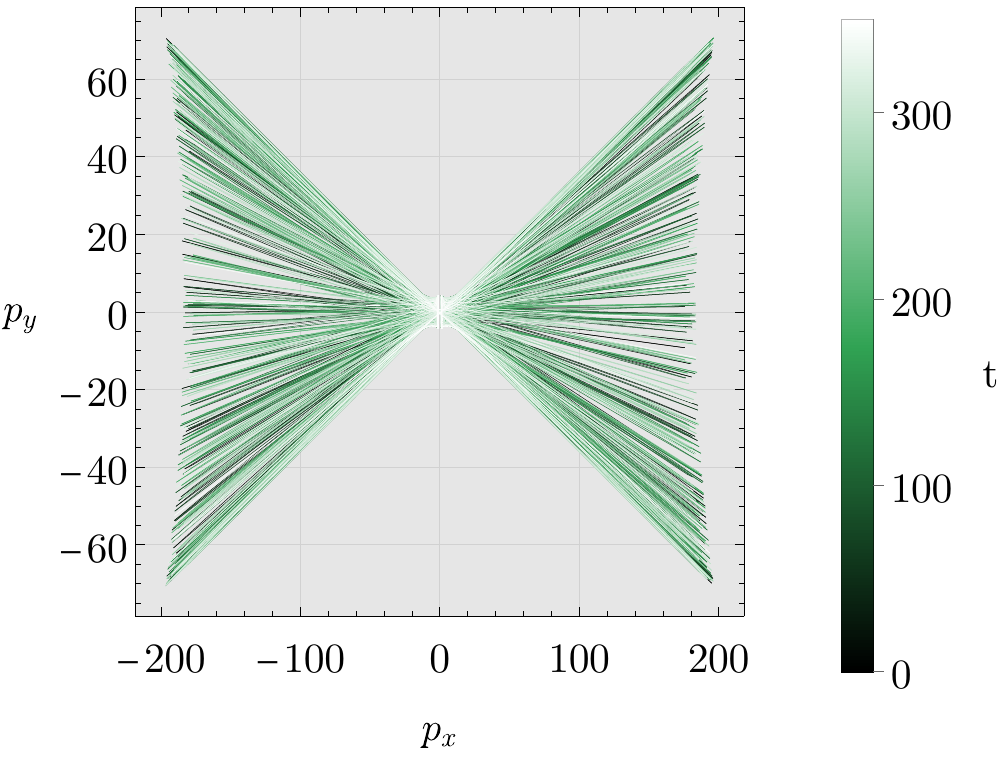}
    \caption{
    As in \cref{fig:N4_lissajous} but for a case with large frequency ratio between positive-energy and negative-energy mode, i.e., for $\omega_x^2=1$, $\omega_y^2=100$, $\mathcal{C}_4=1$, and $\tilde{c}=1$ in \cref{V4simp}.
    }
    \label{fig:N4_lissajous_large-mass-ratio}
\end{figure*}

The explicit form of $V_\text{LV}(x,y)$, in the PG case, and for the simplest non trivial case of interest here, is obtained from the expression in \cref{app:polynomial-potentials}, where one trades the constants  $\mathcal{C}_2$ and $ \mathcal{C}_3$ for quadratic terms and chooses the constant $\mathcal{C}_1$ to eliminate an irrelevant constant term. One obtains, ordering the interactions by powers of $x$ and $y$,
\begin{align} \label{V4simp}
    V_\text{LV}^{(4)}(x,y) &= 
    \frac{\omega_x^2}{2} x^2 - \frac{\omega_y^2}{2} y ^2 \notag\\&\quad
    +\frac{1}{\tilde{c}} \left( \frac{\omega_x^2}{2}- \frac{\omega_y^2}{2}\right)(x^2-y^2)^2\notag\\&\quad
      + \tilde{c}\;  \mathcal{C}_4 (x^4 - y^4)
      + \mathcal{C}_4 (x^2 - y^2)^3 \;.
\end{align}
This rewriting captures the first two non-trivial cases, i.e., $N=3$ and $N=4$, for which we visualize examples of the potential in \cref{fig:condition3D} and of the resulting phase-space motion in \cref{fig:N3_lissajous} (for $N=3$) and \cref{fig:N4_lissajous,fig:N4_lissajous_tachyonic,fig:N4_lissajous_large-mass-ratio} (for $N=4$).

Except for the degenerate choice of $\omega_x=\omega_y$ (for which the $N=3$ case reduces to the $N=2$ case), the $N=3$ case does not lead to fully stable motion, as conditions (i) in \cref{eq:bounded-f,eq:bounded-g} and (ii) in \cref{Hyponf,Hypong} of \cref{subsec:proof} are not fulfilled\footnote{
    Even though our proof in \cref{subsec:proof} only provides sufficient not necessary conditions for the boundedness of motion, on physical grounds, we expect that integrals of motion unbounded from below and from above cannot bound the motion.
}. Indeed, either one chooses $\mathcal{C}_{3}$ positive and for large negative $v^2$, $g(v)$ diverges to $-\infty$, or one chooses  $\mathcal{C}_{3}$ negative and for large positive $u^2$, $f(u)$ diverges to $-\infty$. An example of two near-by sets of initial conditions -- one stable and one unstable -- is visualized in \cref{fig:N3_lissajous}.

In contrast, the $N=4$ case leads to bounded motion since, for $\mathcal{C}_{4}>0$, conditions (i) in \cref{eq:bounded-f,eq:bounded-g} and (ii) in \cref{Hyponf,Hypong} of \cref{subsec:proof} are obeyed. We visualize three different examples:
In \cref{fig:N4_lissajous}, we show an example of the motion for equal frequencies, i.e., for $\omega_x^2=1$, $\omega_y^2=1$. In \cref{fig:N4_lissajous_tachyonic}, we show an example of the motion for which the ghost has a tachyonic term, i.e., for $\omega_x^2=5$, but $\omega_y^2=-5$. Finally, in \cref{fig:N4_lissajous_large-mass-ratio}, we show an example of the motion with a significant frequency hierarchy, i.e., for $\omega_x^2=1$ and $\omega_y^2=100$. In all cases, $\mathcal{C}_4=1$.
The ``potential'' in \cref{defU} that stabilizes the motion is bounded from below and grows without bound at large $x$ and $y$ for $\mathcal{C}_{4}>0$. It is 
given here by 
\begin{align} \notag
& \mathcal{U}_\text{LV}^{(4)}\left(x,y\right)=\frac{\tilde{c}}{2}\left(\omega_{x}^{2}x^{2}+\omega_{y}^{2}y^{2}\right)\\
 & +\frac{1}{2}\left(x^{4}-y^{4}\right)\left(\omega_{x}^{2}-\omega_{y}^{2}\right)\nonumber\\
 & +\mathcal{C}_{4}\tilde{c}\left[\left(x^{2}+y^{2}\right)\left(x^{2}-y^{2}\right)^{2}+\tilde{c}\left(x^{4}+y^{4}\right)\right]\,, \label{U_4} 
\end{align}
where we rewrote \cref{U4} using $\tilde{c}>0$. Note that both the above $\mathcal{U}_\text{LV}^{(4)}$ and $V_\text{LV}^{(4)}\left(x,y\right)$ are symmetric with respect
to separate reflections 
\begin{equation}
\label{reflection}
    x\rightarrow-x\,,\quad \text{and} \quad y\rightarrow-y\,,
\end{equation} 
as they depend
only on $x$ and $y$ via their squares $x^{2}$ and $y^{2}$. The potential $V_\text{LV}^{(4)}\left(x,y\right)$ 
has up to power six self-interactions and PG cross-couplings
\begin{equation}
\label{PG-Interaction_Poly}
\frac{\left(\omega_{y}^{2}-\omega_{x}^{2}\right)}{\tilde{c}}\,x^{2}y^{2}\,,\quad\text{and}\quad3\,\mathcal{C}_{4}\,x^{2}y^{2}\,\left(y^{2}-x^{2}\right)\,,
\end{equation}
between the usual DoF $x$ and the ghost $y$. These cross-coupling terms vanish when one
of the coordinates is zero. However, elsewhere these cross-couplings
cannot be neglected in comparison with the self-interactions. Without
cross-couplings the positive energy degree of freedom $x$ has a self-interaction
potential 
\begin{equation}
\label{V_P}
V_{\text{P}}\left(x\right)=\frac{\omega_{x}^{2}x^{2}}{2}+\left[\frac{\left(\omega_{x}^{2}-\omega_{y}^{2}\right)}{2\tilde{c}}+\mathcal{C}_{4}\tilde{c}\right]x^{4}+\mathcal{C}_{4}x^{6}\,,
\end{equation}
while the the ghost self-interaction potential is 
\begin{equation}\label{V_G}
V_{\text{G}}\left(y\right)=-\left[\frac{\omega_{y}^{2}y^{2}}{2}+\left[\frac{\left(\omega_{y}^{2}-\omega_{x}^{2}\right)}{2\tilde{c}}+\mathcal{C}_{4}\tilde{c}\right]y^{4}+\mathcal{C}_{4}y^{6}\right]\,.
\end{equation}
Without interactions one could flip the sign in front of the ghost's
Hamiltonian, so that uncoupled ghost effectively experiences
inverted self-interaction potential $\left(-V_{\text{G}}\left(y\right)\right)$. Note that $\left(-V_{\text{G}}\left(y\right)\right)$ can be obtained from $\left(V_{\text{P}}\left(x\right)\right)$ by changing $x$ to $y$ everywhere, including indices.

The next non-trivial polynomial and stable PG theory is obtained for $N=6$ (and positive $\mathcal{C}_6$) and the corresponding potential is given in \cref{app:polynomial-potentials}.
\\

More generally, all polynomial theories with even $N$ (and positive $\mathcal{C}_N$) are stable according to the above proof.
This becomes apparent when looking at large $u^2\gg|\mathcal{C}_n|$ (as well as large $-v^2\gg|\mathcal{C}_n|$), for which
\begin{align}
    f(u)
    \longrightarrow\;&
    \mathcal{C}_N\left(u^2\right)^{N}
    \;, 
    \notag\\
    g(v)
    \longrightarrow\;&
    (-1)^N\mathcal{C}_N\left(-v^2\right)^{N}
    \;.
    \label{eq:polynomial-subclass}
\end{align}
In writing the second expression, we arrange minus signs such as to account for $-v^2>0$.

For odd $N$, the proof conditions cannot be fulfilled since (for any choice of $\mathcal{C}_N$) either $f(u)$ or $g(v)$ is unbounded below. While islands of stable initial conditions are possible (cf.~left-hand panel in \cref{fig:N3_lissajous}), we expect that there are always some initial conditions for which divergent behaviour remains (cf.~right-hand panel in \cref{fig:N3_lissajous}).
\\

So, how do the proof conditions relate to the behavior at large $x^2\gg \tilde{c}$ and $x^2\gg|\mathcal{C}_n|$ (as well as large $y^2\gg \tilde{c}$ and $y^2\gg|\mathcal{C}_n|$)? Taking both limits,
\begin{align}
    \label{eq:polynomial-limit-potential}
    V_\text{LV}^{(N)}(x,y) 
    &\rightarrow
    \mathcal{C}_N\left(x^2-y^2\right)^{N-1}
    \;.
\end{align}
Together with the proof conditions, i.e., even $N$ and $\mathcal{C}_N>0$, this implies: 
\begin{itemize}
    \item The decoupled potential $V_\text{P}(x)$ (see \cref{eq:potential-split-notation,V_P})
    is bounded below and unbounded above and thus serves as a stable potential for the decoupled motion of the $x$ mode.
    \item Similarly, $V_\text{G}(y)$ (see \cref{eq:potential-split-notation,V_G})
    is bounded above and unbounded below and thus serves as a stable potential for the decoupled motion of the $y$ mode.
    \item
    At sufficiently large\footnote{For $N=4$, $|x|\gtrless|y|$ in \eqref{eq:pre-conjecture-condition} should be understood as $|x|\gtrless |y| + \frac{\omega_y^2-\omega_x^2}{3\tilde{c}\mathcal{C}_4}\frac{1}{|x|+|y|}$, which agrees with $|x|\gtrless|y|$ either in the limit $|x|+|y|\to\infty$ or for $\omega_y^2=\omega_x^2$.} $x^2$ and $y^2$, the interaction potential $V(x,y)$ is bounded 
    by the decoupled potential, i.e.,
    \begin{align}
        \label{eq:pre-conjecture-condition}
        V(x,y) \lessgtr V_\text{P}(x) + V_\text{G}(y)
        \quad \forall\;|x|\gtrless|y|
        \;.
    \end{align}
    \item As we will see, the special case of $|x|=|y|$, for which the full and the decoupled potential agree at sufficiently large $x^2$ and $y^2$, seems to require an additional criterion.
\end{itemize}
To elucidate the latter subtlety further, we take a look at another class of integrable polynomial potentials.

%

\subsection{An unstable polynomial subclass: refining more general conditions for stable motion}
\label{subsec:class-3}
%
\begin{figure}
    \centering
    \includegraphics[trim=0cm 0cm 0cm 0cm, clip, width=\linewidth]{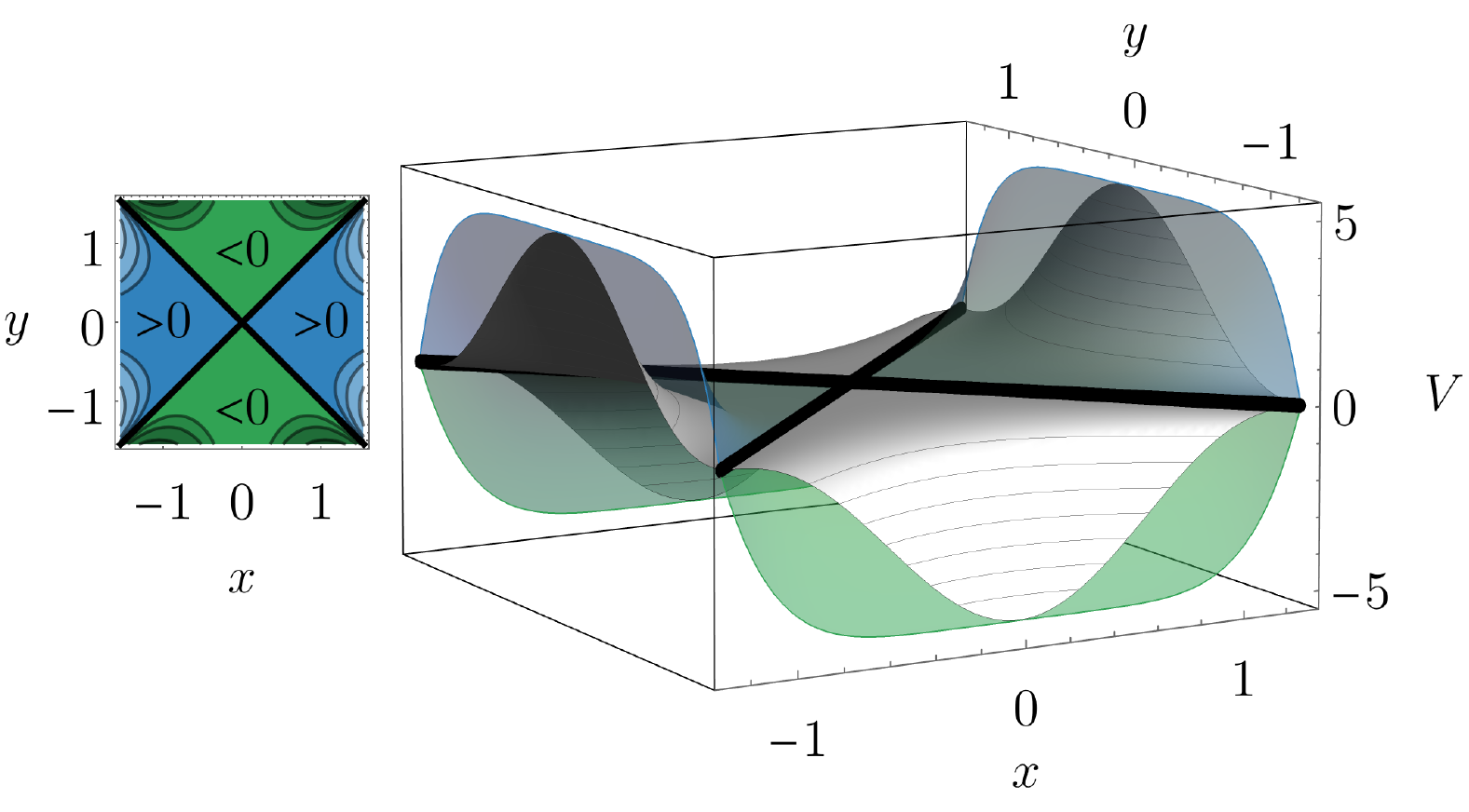}
    \caption{
    As in \cref{fig:condition3D} but for the class 3 potential in \cref{eq:class-3-polynomial-pot} with $\omega_x^2 = \omega_y^2 = 0$ and $\mathcal{C}_4=1$. The thick lines indicate the directions along which unstable motion occurs. Note that the potential exhibits sets of saddle points along flat directions extending to infinity, cf.~right-hand panel in~\cref{fig:condition3D}.
    }
    \label{fig:condition3D_class3}
\end{figure}

To elucidate the possibility of stable motion in the Liouville class further, we can compare it to the motion in another class of integrable two-particle Hamiltonians.

As explained in~\cref{sec:integrable-models} real-valued PG Hamiltonians can sometimes also be obtained from complex-valued PP Hamiltonians. This turns out to be the case for another class of integrable two-particle Hamiltonians, closely related to the Liouville model, cf.~\cref{app:integrable-Hamiltonians}. The details of this integrable model are discussed in \cref{app:polynomial-potentials-class-3} and, following~\cite{Hietarinta:1986tw}, we refer to it as class 3 from here on. 

In fact, just as the integrable Liouville model, the integrable class 3 contains a polynomial subclass which is analogously obtained, cf.~\cref{app:polynomial-potentials-class-3}. Introducing frequencies, removing a constant shift (as in \cref{V4simp}), and specifying to equal frequencies (i.e., $\omega_x=\omega_y=\omega$), the first few terms can be written as
\begin{align}
    \label{eq:class-3-polynomial-pot}
    V_\text{(class 3)}^{(4)}(x,y) &\stackrel{\omega_x=\omega_y}{=} 
    \frac{\omega^2}{2}\left(x^2-y^2\right)
    + \mathcal{C}_4\left(x^2-y^2\right)^3
    \notag\\&\quad
    - \mathcal{C}_4\,c\,(x+y)(x-y)(x\pm y)^2
    \;.
\end{align}
The general expression with unequal frequencies is given in \cref{app:polynomial-potentials-class-3}.
For $\mathcal{C}_4=0$, the potential $V_\text{(class 3)}^{(4)}$ is (depending on the other parameter choices) either bounded from below or bounded from above. More generally, just as for the Liouville polynomial subclass, each odd $N$ case is thus unstable because one of the decoupled potentials is necessarily unstable.

For $\mathcal{C}_4>0$, cf.~also~\cref{fig:condition3D_class3}, the potential $V_\text{(class 3)}^{(4)}$ fulfills all the conditions stated around \cref{eq:pre-conjecture-condition}: Both of the decoupled potentials are stable and the full potential is bounded by the decoupled potential. Nevertheless, we find that this case exhibits runaways with polynomial growth rate and is thus \emph{not} stable. We thus conclude that the conditions stated around \cref{eq:pre-conjecture-condition} are not sufficient to guarantee the absence of runaway behaviour.
\\

We first look at the case ($\omega = 0$), for which we observe that the runaway behaviour occurs either along the $x-y=0$ or along the $x+y=0$ direction. In fact, the potential in \cref{fig:condition3D_class3} exhibits sets of saddle points (marked with thick lines) extending out to infinity along flat directions of the potential. As a consequence of this, the local potential around $x\pm y=0$ is no longer dominated by the respective decoupled potentials. 

This can be seen from the Hessian matrix\footnote{
    For the case at hand, the Hessian matrix is defined by 
    $
        \text{Hess}\left(V_\text{(class 3)}^{(4)}(x,y)\right)\equiv
        \begin{pmatrix}
            \partial_x\partial_x V_\text{(class 3)}^{(4)} & \partial_x\partial_y V_\text{(class 3)}^{(4)}\\
            \partial_y\partial_x V_\text{(class 3)}^{(4)} & \partial_y\partial_y V_\text{(class 3)}^{(4)}
        \end{pmatrix}
    $.
}, which vanishes at $x=\pm y$, i.e.,
\begin{align}
    \text{Hess}\left(V_\text{(class 3)}^{(4)}(x,y)\right) \stackrel{x=\pm y}{=}
    \begin{pmatrix}
        0 & 0\\
        0 & 0
    \end{pmatrix}\;.
\end{align}
In contrast, the Hessian matrix of the decoupled potential (at $x=\pm y$) is non-vanishing, i.e.,
\begin{align}
    \text{Hess}\left(V_\text{(class 3)}^{(4)}(x,0) + V_\text{(class 3)}^{(4)}(0,y)\right) \stackrel{x=\pm y}{=} 
    30\,x^4\begin{pmatrix}
        1 & 0\\
        0 & -1
    \end{pmatrix}\;.
\end{align}
Along the saddle surfaces, the local potential $V_\text{(class 3)}^{(4)}$ does, therefore, \emph{not} resemble the decoupled potential. In fact, it does not resemble \emph{any} decoupled potential. Put differently, there exist regions in which decoupled interactions no longer dominate the potential at large phase-space variables. 
\\

For $V_\text{(class 3)}^{(4)}$ with non-vanishing but equal frequencies (and still for $\mathcal{C}_4>0$), the issue of runaways is lifted along one of the two directions. Without loss of generality, picking the `$-$' sign in the last term in \cref{eq:class-3-polynomial-pot}, the $x=-y$ direction is stable but the $x=y$ direction remains unstable. Once more, this can be understood in terms of the Hessian matrix. For $x=-y$, the Hessian matrix now reads
\begin{align}
    \text{Hess}\left(V_\text{(class 3)}^{(4)}(x,y)\right) \stackrel{x=- y}{=}
    (\omega^2+24\,\mathcal{C}_4\,y^2)\begin{pmatrix}
        1 & 0\\
        0 & -1
    \end{pmatrix}\;.
\end{align}
This agrees with the Hessian matrix of a free theory, to be specific, the free theory with $V(x,y)=\frac{\omega^2}{2}(x^2-y^2)+2\,\mathcal{C}_4(x^4-y^4)$.
In contrast, for $x=y$, we find
\begin{align}
    \text{Hess}\left(V_\text{(class 3)}^{(4)}(x,y)\right) \stackrel{x=y}{=}
    \omega^2\begin{pmatrix}
        1 & 0\\
        0 & -1
    \end{pmatrix}\;,
\end{align}
for which the behaviour at large phase-space variables still has no free-theory equivalent. In agreement with the above, we find that, with equal frequencies, polynomial runaway behaviour no longer occurs along the $x=-y$ but still occurs along the $x=y$ direction.
\\

Numerically, we observe that the runaways can be fully removed by the addition of decoupled interactions of the form $x^4-y^4$. (Numerics are necessary since the addition of these terms breaks integrability.) With this addition, the Hessian matrix has the same structure as for the Liouville potential $V_\text{LV}^{(4)}(x,y)$. In both cases, it does still \emph{not} agree with the Hessian matrix of the respective decoupled potential. However, one can identify a decoupled potential for which the Hessian matrices (at sufficiently large $x=\pm y$) agree. We conclude that it is sufficient (at least for the specific case at hand) that the Hessian matrix at large $|x|=|y|$ agrees with \emph{some} decoupled potential, not necessarily with the decoupled potential of the respective model.
\\

In passing, we mention that similar polynomial runaway behaviour also occurs in an integrable model previously obtained (from a supersymmetric construction) by Smilga {\it et al.} \cite{Smilga:2017arl,Robert:2006nj}. Its Hamiltonian is given by 
\begin{align} 
\label{Hamilt1}
H &= \frac{1}{2}\left(p_x^2 + \omega^2 x^2\right) - \frac{1}{2}\left(p_y^2 + \omega^2 y^2\right)  \notag\\&\quad
+ \frac{\lambda}{4 \omega}(x-y)(x+y)^3\,,
\end{align}
where $\lambda$ and $\omega$ are constants. It follows as a degenerate limit of \cref{eq:class-3-polynomial-pot}, i.e., when picking the $+$ sign in the last term of \cref{eq:class-3-polynomial-pot} and taking $\mathcal{C}_4\rightarrow0$ as well as $c\rightarrow\infty$ such that $\mathcal{C}_4\,c = -\lambda/4\omega$. (Alternatively, \cref{Hamilt1} can also be obtained as a specific case\footnote{
    To be specific, the Hamiltonian in~\cref{Hamilt1} can be obtained from the integrable class 8 in \cref{eq:class8} by choosing $g=0$ and $f'' = \frac{1}{2} \omega^2 + \frac{\lambda}{\omega}(x+ iy)^2$. This results in the potential (for a PP theory) given by 
    $V= (x^2 +y^2)(\frac{1}{2}\omega^2 + \frac{\lambda}{4 \omega}(x+iy)^2)$. 
    We then obtain the theory in~\cref{Hamilt1} by applying a canonical transformation $\mathfrak{C}^{-}_{y}$.
} of class 8 in \cref{app:integrable-Hamiltonians}.)
The motion of this theory does not stay bounded (even though the ghost there is argued to be ``benign'' due to the slow (polynomial) growth-rate of the runaway solutions found numerically \cite{Smilga:2017arl,Robert:2006nj}). 
In contrast to \cref{eq:class-3-polynomial-pot}, the potential of \cref{Hamilt1} also violates the condition in \cref{eq:pre-conjecture-condition}.
\\

We conclude that all of the discussed cases can be understood if the following criterion is added to the list in \cref{subsec:integrable-motion}:
\begin{itemize}
    \item For $|x|=|y|$ (or more generally, for all $(x,y)$ for which $V(x,y) =  V_\text{P}(x) + V_\text{G}(y)$), the Hessian matrix at large phase-space variables approaches the Hessian matrix of some free theory.
\end{itemize}
Having understood these more subtle cases, we conclude: By translating the proof conditions to conditions on the potential $V(x,y)$, we understand that stable motion seems to be possible because stable decoupled potentials $V_\text{P}(x)$ and $V_\text{G}(y)$ dominate all interaction terms at sufficiently large $x^2$ and $y^2$. 

In the following \cref{sec:nonintegrable}, we will use this insight to formulate general conditions for stability.
Before doing so, we comment on the relation to local Lyapunov stability.

\subsection{Lagrange vs Lyapunov Stability}
\label{subsec:Lagrange_vs_Lyapunov}
In the previous subsections, we were concerned with global boundedness of motion which corresponds to the stability in the sense of Lagrange. However, one can be also interested in local stability in the sense of Lyapunov, see, e.g., \cite{Lefschetz}. In particular, this is crucial to specify the nature of equilibrium points which are fixed points of the system of equations of motion and can be considered as vacua. The two notions of stability are not equivalent. Indeed, a Lagrange stable system may not have locally stable solutions, while a system with Lyapunov stable equilibrium points may have other solutions with runaways invalidating stability in the sense of Lagrange. Here we will show that the Liouville class of systems in \cref{eq:Liouville-hamiltonian}, describing ghosts interacting with usual degrees of freedom ($\sigma=-1$), contains plenty of theories with Lyapunov stable equilibrium points. 
More specifically, we focus here first on systems in this class with functions $f(u)$ and $g(v)$ which are respectively analytic functions of $u^2$ and $\tilde{v}^2$ around the origin $\left(x,y\right)=\left(0,0\right)$. These systems contain the polynomial case \cref{fpoly,gpoly} as a subcase.

For such theories, the origin $\left(x,y\right)=\left(0,0\right)$ is always an equilibrium point. Indeed, at the origin one has $u^{2}=0$ and $\tilde{v}^{2}=\tilde{c}$, while 
\begin{equation}
\partial_{x}V_{\text{LV}}=0\,,\,\quad\text{and}\quad\,\,\partial_{y}V_{\text{LV}}=0\,,
\end{equation}
because all partial derivatives
\begin{align}
	&\frac{\partial u^{2}}{\partial x}=\left[\frac{r^{2}+\tilde{c}}{\tilde{W}}+1\right]x\,,\quad\frac{\partial\tilde{v}^{2}}{\partial x}=\left[\frac{r^{2}+\tilde{c}}{\tilde{W}}-1\right]x\,,\\
	&\frac{\partial u^{2}}{\partial y}=\left[\frac{r^{2}-\tilde{c}}{\tilde{W}}-1\right]y\,,\quad\frac{\partial\tilde{v}^{2}}{\partial y}=\left[\frac{r^{2}-\tilde{c}}{\tilde{W}}+1\right]y\,,
\end{align}
are vanishing there. 
Now we can linearize the equations of motion\footnote{Note that the sign difference in the $y$-equation of motion appears due to the negative mass of the ghost.}
\begin{equation}
\ddot{x}+\frac{\partial V_{\text{LV}}}{\partial x}=0\,,\,\quad\text{and}\quad\,\,\ddot{y}-\frac{\partial V_{\text{LV}}}{\partial y}=0\,,
\end{equation}
around the origin and obtain corresponding frequencies
\begin{align}
\label{W_x}
&\bar{\omega}_{x}^{2}=\frac{2}{\tilde{c}^{2}}
\left[\tilde{c}f'(0)-f(0)+g\left(\tilde{c}\right)\right]\,,\\
&\bar{\omega}_{y}^{2}=\frac{2}{\tilde{c}^{2}}\left[\tilde{c}g'\left(\tilde{c}\right)-g\left(\tilde{c}\right)+f(0)\right]\,.
\label{W_y}
\end{align}
Note that $\bar{\omega}_{x}$ and $\bar{\omega}_{y}$ are just defined expanding around the origin and are hence conceptually different from unbarred frequencies defined previously, hence the different notation. Of course they could be the same depending on the potential.
Naive stability by the linearized approximation requires that both $\bar{\omega}_{x}^{2}$ and $\bar{\omega}_{y}^{2}$ are positive. 
In particular, adding the two equations above, we see that stability requires that 
\begin{equation}
g'\left(\tilde{c}\right)+f'(0)>0\,.
\label{condition_on_the_summ_of_der}
\end{equation}
Let us now confirm that the conditions $\bar{\omega}_{x}^{2}>0$ and $\bar{\omega}_{y}^{2}>0$ are sufficient for the Lyapunov stability of the equilibrium point in the origin. 
Consider the first integral $J_{\text{LV}}$ given by \cref{NewFirstIntegral}. Expanding $J_{\text{LV}}$ around the origin and around vanishing momenta one obtains
\begin{equation}
J_{\text{LV}}=g\left(\tilde{c}\right)+f\left(0\right)+\frac{\tilde{c}}{2}\left(p_{x}^{2}+p_{y}^{2}\right)+\frac{\bar{\omega}_{y}^{2}\tilde{c}}{2}y^{2}+\frac{\bar{\omega}_{x}^{2}\tilde{c}}{2}x^{2}+...,
\label{C_expansion}
\end{equation}
where ellipsis stands for higher order terms. Clearly, for $\bar{\omega}_{x}^{2}>0$ and $\bar{\omega}_{y}^{2}>0$ the first integral has a minimum at the equilibrium point in the origin. Thus, $J_{\text{LV}}-(g\left(\tilde{c}\right)+f\left(0\right))$ is a Lyapunov function. Indeed, it is conserved, vanishing at the equilibrium solution and is positive definite in an open neighborhood around it (see \cite{Lefschetz}). Thus, the origin is Lyapunov stable. Here, the conserved quantity $J_{\text{LV}}$ plays the analog of usual energy for establishing stability. Crucially, the ``kinetic energies'' are all positive definite in $J_{\text{LV}}$. Moreover, $\mathcal{U}(x,y)$ -- the ``potential part'' of the ``energy'' -- has an isolated minimum at the origin. Thus, exchanging the usual potential energy with $\mathcal{U}(x,y)$, one meets the conditions of the classical Lagrange theorem stating that the minimum of the potential energy corresponds to a stable equilibrium configuration. In contrast, it is interesting to note that this stable equilibrium point is a saddle point for the potential $V_{LV}(x,y)$ as 
\begin{equation}
V_{\text{LV}}=\frac{f(0)-g\left(\tilde{c}\right)}{\tilde{c}}\,+\,\frac{\bar{\omega}_{x}^{2}x^{2}}{2}\,-\,\frac{\bar{\omega}_{y}^{2}y^{2}}{2}+....
\end{equation}
It also appears above that $\bar{\omega}_{x}^{2}>0$ and $\bar{\omega}_{y}^{2}>0$ do guarantee Lyapunov stability for the usual PP systems due to the Lagrange theorem.
The Lyapunov stability of the origin, implies, by continuity, the same stability for solutions in a small neighborhood.
It is important to repeat that the conditions on $f(u)$ and $g(v)$ required for the Lyapunov
stability of the origin (positivity of \eqref{W_x} and \eqref{W_y}) can be easily violated in systems which are globally stable according to Lagrange. And vice versa: a Lagrange unstable system may have a Lyapunov stable origin. 
This can be well illustrated by the case of $V_\text{LV}^{(3)}$ PG system given by \cref{defVN}, see also \cref{eq:polynomial-subclass_explicit-potential}. As we discussed before, this system is not stable from the point of view of Lagrange, however the origin is Lyapunov stable provided ${\omega}_{x}^{2}$ and ${\omega}_{y}^{2}$ chosen in 
\cref{C1C2C3} are positive. 
On the other hand, as we discussed before, $V_\text{LV}^{(4)}$ from \cref{V4simp} is Lagrange stable for $\mathcal{C}_{4}>0$ and $\tilde{c}>0$, but can have a Lyapunov unstable origin if one of ${\omega}_{x}^{2}$, ${\omega}_{y}^{2}$ chosen in 
\cref{C1C2C3} is negative.
To finish, let us here mention that the Lyapunov stability of the origin equilibrium point for a distinct class of interacting ghosty systems has previously been demonstrated in \cite{Pagani:1987ue} using different methods.
Moreover, in \cite{Deffayet:2021nnt} the Lyapunov stability of the origin has been proven for a particular PG system from the same Liouville class in \cref{eq:Liouville-hamiltonian} using methods similar to those employed in the current paper.
\\

\begin{table*}[]
    {\setlength{\tabcolsep}{4pt}\renewcommand{\arraystretch}{2}
    \begin{tabular}{l||c|c|c}
         & $x^2$
         & $y^2$
         & conditions 
         \\\hline\hline
         origin
         & $0$
         & $0$
         & $\Omega_{x/y}^{2}>0$,
         \\\hline
         decoupled-$x$
         & $\frac{\tilde{c}}{3}\left[\left(\Omega_{y}^{2}-\Omega_{x}^{2}-1\right)+\sqrt{\left(\Omega_{y}^{2}-\Omega_{x}^{2}-1\right)^{2}-3\Omega_{x}^{2}}\right]$
         & $0$
         & \eqref{eq:interval}
         \\\hline
         decoupled-$y$
         & $0$
         & $\frac{\tilde{c}}{3}\left[\left(\Omega_{x}^{2}-\Omega_{y}^{2}-1\right)+\sqrt{\left(\Omega_{x}^{2}-\Omega_{y}^{2}-1\right)^{2}-3\Omega_{y}^{2}}\right]$
         & \eqref{eq:interval}\, $x\longleftrightarrow y$
         \\\hline
         coupled
         & $\frac{\tilde{c}}{8}\left[\left(\Omega_{x}^{2}-\Omega_{y}^{2}\right)^{2}-4\Omega_{x}^{2}\right]$
         & $\frac{\tilde{c}}{8}\left[\left(\Omega_{x}^{2}-\Omega_{y}^{2}\right)^{2}-4\Omega_{y}^{2}\right]$
         & \eqref{eq:interacting_bound}
    \end{tabular}}
    \caption{
    \label{tab:extrema-minimal-polynomial-model}
    We list all possible minima 
    of $\mathcal{U}^{(4)}_\text{LV}(x,y)$, from \cref{U_4} where we assumed that $\tilde{c}>0$, $\mathcal{C}_{4}>0$ and $\Omega_{x/y}^{2}>0$. In the last column, we list conditions for the existence of such minima. 
    These minima are Lyapunov stable equilibrium configurations or stable vacua of the theory. The origin respects the reflection symmetry in \cref{reflection}. The ``decoupled-$x$'' minima are
    given by \cref{eq:Nontrivial_solution} and break the reflection symmetry $x\rightarrow-x$. The ``decoupled-$y$'' minima are given by \cref{eq:Nontrivial_solution}, with exchange $x\longleftrightarrow y$, and break the reflection symmetry $y\rightarrow-y$.
    The ``coupled'' minima are given by \cref{eq:x_I,eq:y_I} and break both of the reflection symmetries in \cref{reflection}.
    }
\end{table*}
\begin{figure*}
    \includegraphics[trim=0cm 0cm 0cm 0cm, clip, width=\linewidth]
    {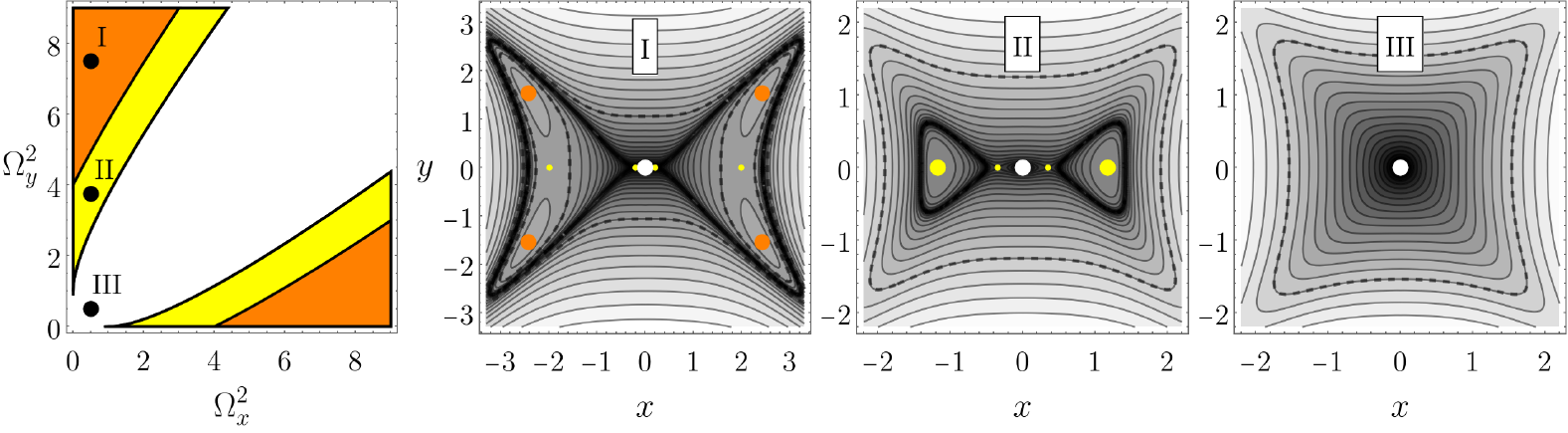}
    \caption{
    In the left-hand panel, using the rescaled frequencies in \cref{Omega}, we visualize the conditions for the existence of different stable vacua listed in \cref{tab:extrema-minimal-polynomial-model}. 
    In the white region, the origin is the only Lyapunov-stable vacuum.
    In the yellow stripe above (below), the stable origin is supplemented by the two Lyapunov-stable ``decoupled-x'' (``decoupled-y'') vacua.
    Further up and/or to the right, in the orange region, the stable origin is supplemented by the four Lyapunov-stable coupled vacua.
    In the other three panels, we specify to $\mathcal{C}_4=1$ as well as $\tilde{c}=1$ and show the three indicated examples for the respective possible cases. We super-impose the latter on contour plots of $\text{log}(|\mathcal{U}^{(4)}_\text{LV}(x,y)|$. 
    (The contours are spaced by half-integers with the thick-dashed contour indicating
    $\text{log}(|\mathcal{U}^{(4)}_\text{LV}(x,y)|=3$
    .)
    We omit further configurations which arise from exchanging $(x\leftrightarrow y)$. 
    From left to right, we show $(\Omega_x^2,\Omega_y^2)=(0.5,7.5)$ (i.e., the point labeled ``I'' in the orange region of the left-hand panel), $(\Omega_x^2,\Omega_y^2)=(0.5,3.75)$ (i.e., the point labeled ``II'' in the yellow region of the left-hand panel) and  $(\Omega_x^2,\Omega_y^2)=(0.5,0.5)$ (i.e., the point labeled ``III'' in the white region of the left-hand panel). Larger (smaller) dots indicate minima (other real-valued extrema) of $\mathcal{U}^{(4)}_\text{LV}(x,y)$ and the colours white, yellow, and orange indicate origin, decoupled extrema, and coupled extrema, respectively. 
    Even though, for illustrative purposes, we have not chosen the parameters to preserve the ratio of $\Omega$'s but rather to fit the minima at the same region of $(x,y)$ space, it is clear that, for fixed values of $\tilde{c}$ and $\omega_y$, $\omega_x$, the vacuum configurations will change from left to right with increasing strength of $\mathcal{C}_4$.
    }
    \label{fig:Lyapunov-extrema}
\end{figure*}
%

\subsubsection*{All Vacua and their Lyapunov Stability for the Polynomial Subclass of the Liouville Model}
%
\begin{figure}
    \begin{centering}
        \includegraphics[trim=1cm 0cm 1cm 0cm, clip, width=\linewidth]{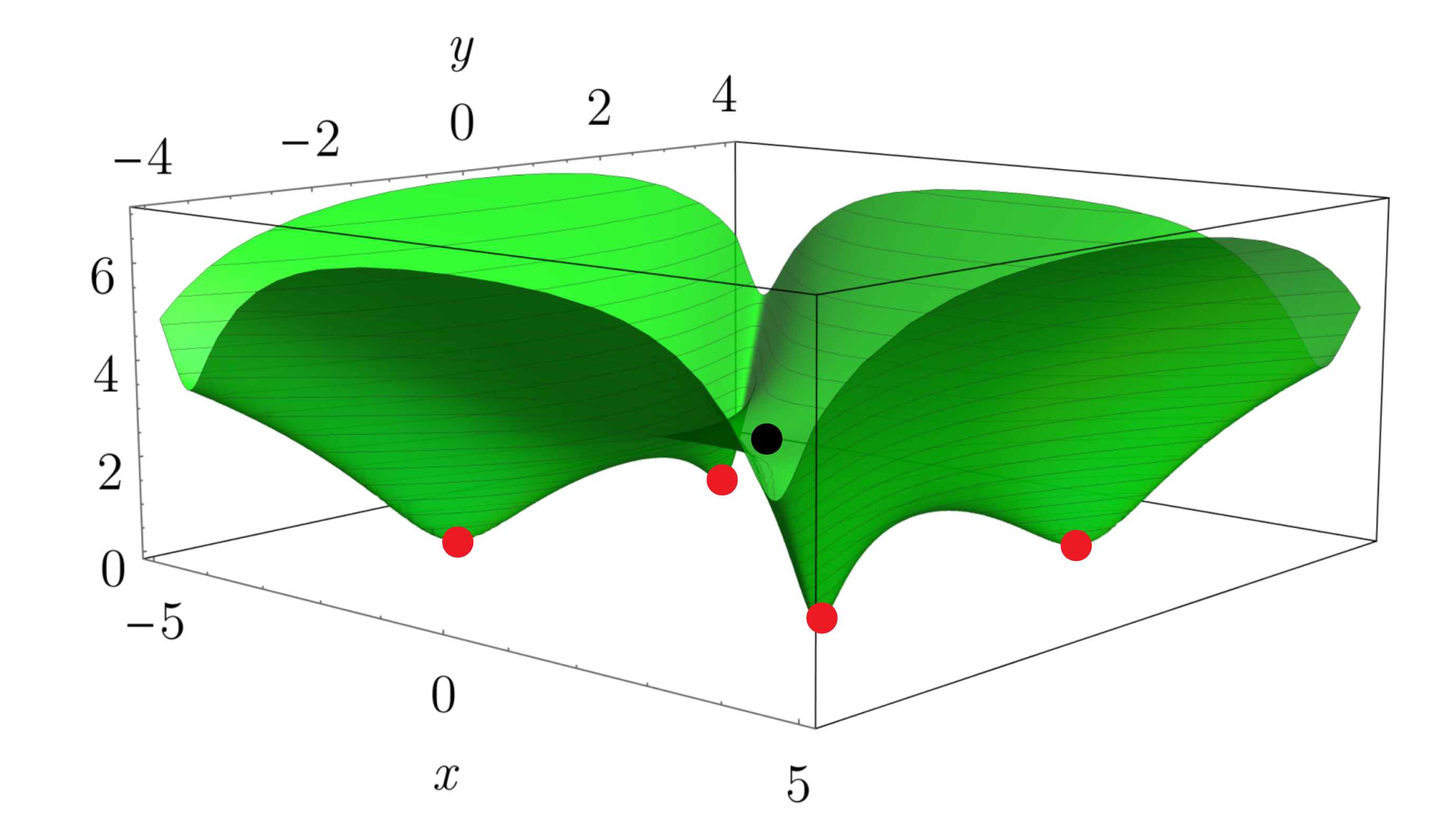}
        \par
    \end{centering}
    \caption{\label{fig:U_low_LLog}
        For values of $\mathcal{C}_{4}$ satisfying \cref{eq:interacting_bound}, we show a logarithmic plot of the ``uplifted" effective ``potential'' $\mathcal{U}^{(4)}_\text{LV}\left(x,y\right)$, i.e., the plot of $\log(1+\mathcal{U}^{(4)}_\text{LV}\left(x,y\right)-\mathcal{U}^{(4)}_\text{LV}\left(x_{I},y_{I}\right))$, such that we take the logarithm of a strictly positive quantity. The red points indicate the four global vacua $\left(x_{I},y_{I}\right)$ given by \cref{eq:x_I,eq:y_I} with all combinations of signs for the square roots. The black point indicates the vacuum at the origin which, however, is only a local minimum of $\mathcal{U}^{(4)}_\text{LV}$. The parameters are chosen as $\tilde{c}=1$,
        $\omega_{x}=1$, and 
        $\omega_{y}=1.7$, while $\mathcal{C}_{4}$ is 1.5 times smaller than the lower bound in \cref{eq:interacting_bound} so that $\mathcal{C}_{4}\approx0.103$. This case has different numerical values of constants, but its structure of vacua corresponds to case I in \cref{fig:Lyapunov-extrema}.
    }
\end{figure}
\begin{figure*}
    \begin{centering}
    \includegraphics[trim=0cm 0cm 0cm 0cm, clip, width=\linewidth]
    {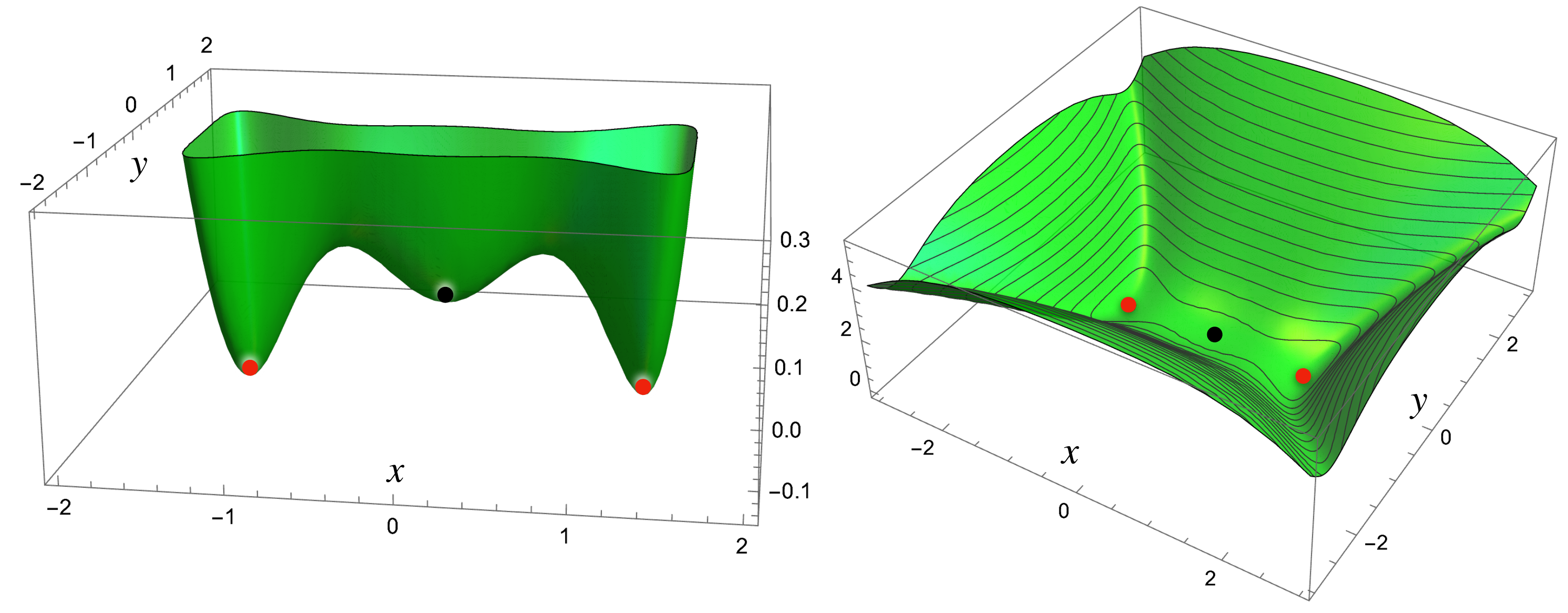}
        \par
    \end{centering}
    \caption{\label{fig:U_middle_large}
        On the left, we show the effective ``potential'' $\mathcal{U}^{(4)}_\text{LV}\left(x,y\right)$ around vacua, for $\mathcal{C}_{4}$ located in the interval \eqref{eq:interval}, while on the right, we plot, in the logarithmic scales, the same ``uplifted'' effective ``potential": $\log(1+\mathcal{U}^{(4)}_\text{LV}\left(x,y\right)-\mathcal{U}^{(4)}_\text{LV}\left(x_{b\left(+\right)},0\right))$, on larger scales. The red points indicate the two global vacua $\left(\pm x_{b\left(+\right)},0\right)$ from  \cref{eq:Nontrivial_solution}. The black point indicates the trivial vacuum at the origin, which however is only a local minimum of $\mathcal{U}^{(4)}_\text{LV}$. The parameters are set to $\tilde{c}=1$, $\omega_{x}=1$, and $\omega_{y}=1.7$, as in \cref{fig:U_low_LLog}, while  $\mathcal{C}_{4}$ is higher than  that of \cref{fig:U_low_LLog} and is in the middle of the interval \eqref{eq:interval}:  $\mathcal{C}_{4}\approx0.221$. This case has different numerical values of constants, but its structure of vacua corresponds to case II in \cref{fig:Lyapunov-extrema}.
    } 
\end{figure*}
The PG interacting system with polynomial potential in \cref{V4simp,VLV4} belongs to the class discussed above and deserves to be analyzed in more detail.
In \cref{fig:N4_lissajous},  \cref{fig:N4_lissajous_tachyonic} and \cref{fig:N4_lissajous_large-mass-ratio} we have already presented numerical solutions of this system for various parameters. 
In this subsubsection, we find \emph{all} equilibrium points, or fixed points of the evolution $(x_e,y_e)$, i.e.,
\begin{equation}
\label{def_equlibrium}
\partial_x V_\text{LV}^{(4)}\left(x_e,y_e\right)=\partial_y V_\text{LV}^{(4)}\left(x_e,y_e\right)=0\;.
\end{equation}
We then determine whether these vacua are Lyapunov stable. 
Assuming non vanishing $\mathcal{C}_{4}$ and $\tilde{c}$, the above two equations are equivalent to the vanishing of 
\begin{align*}
 & x_e\left[x_e^{2}+\frac{\left(x_e^{2}-y_e^{2}\right)}{2 \tilde{c}}\left[\frac{\left(\omega_{x}^{2}-\omega_{y}^{2}\right)}{\mathcal{C}_{4}\tilde{c}}+3 \left(x_e^{2}-y_e^{2}\right)\right]+\frac{\omega_{x}^{2}}{4\mathcal{C}_{4}\tilde{c}}\right]\,,\\
& y_e\left[y_e^{2}+\frac{\left(x_e^{2}-y_e^{2}\right)}{2 \tilde{c}}\left[\frac{\left(\omega_{x}^{2}-\omega_{y}^{2}\right)}{\mathcal{C}_{4}\tilde{c}}+3 \left(x_e^{2}-y_e^{2}\right)\right]+\frac{\omega_{y}^{2}}{4\mathcal{C}_{4}\tilde{c}}\right]\,,
\end{align*}
whose solutions will be discussed below.
We then shall demonstrate how PG interactions which are supposed to destabilize the system, instead introduce new stable vacua. Interestingly, the appearance of these new vacua leads to spontaneous breaking of the reflection symmetry in \cref{reflection}.
To establish Lyapunov stability, we will use the result of \cref{app:Integrability_Equilibrium}: the extrema of $V_\text{LV}^{(4)}$ are also extrema of $\mathcal{U}_\text{LV}^{(4)}$, see \cref{eq:U_x_vs_V_x}. Then, similarly to the discussion above on the stability of the origin, the minimum of $\mathcal{U}_\text{LV}^{(4)}$ implies the Lyapunov stability of the vacuum, as $J_{\text{LV}}$ from \cref{NewFirstIntegral} can be taken again as a Lyapunov function.

The first equilibrium point is the origin $\left(x_e,y_e\right)=\left(0,0\right)$,
which was discussed at the beginning of this section, and the Lyapunov stability just requires
that $\omega_{x}^{2}>0$ and $\omega_{y}^{2}>0$ which we shall also assume to hold below.
Note that this vacuum respects
the reflection symmetry in \cref{reflection} for both coordinates and is such that (see \cref{U_4})
\begin{align}
\mathcal{U}_\text{LV}^{(4)}\left(x_e,y_e\right)= 0\,. \label{U4xeye}
\end{align}

One can find a second type of equilibrium points where only one of the coordinates vanishes but not the other one. Indeed, $x_e=0$, or $y_e=0$, is obviously a solution to \cref{def_equlibrium} (see the two above equations). When either $x_e=0$, or $y_e=0$, the stability along the direction where the equilibrium coordinate does not vanish can be analyzed looking at the potential in \cref{V_G} or \cref{V_P}. It is seen there that, for sufficiently small $\mathcal{C}_{4}$, either the coefficient in front of $x^{4}$ in $V_P(x)$ or in front of $y^{4}$ in $\left(-V_{\text{G}}\left(y\right)\right)$ can become
negative, indicating that the origin is not the only minimum of $V_{\text{P}}\left(x\right)$
or of $\left(-V_{\text{G}}\left(y\right)\right)$. In that case, the vacuum spontaneously breaks the reflection symmetry in \cref{reflection} with respect to this coordinate. 
Let us assume for concreteness that $\omega_{x}<\omega_{y}$, so that we take $y_e=0$ as then implied by the positivity of the coefficient in front of $y^4$ in \cref{V_G}. The opposite case is obtained from all formulas with the replacement $x\leftrightarrow y$. Looking for non-vanishing solutions of $\partial_{x}V_{\text{P}}\left(x\right)=0$, yields the following quartic equation
\begin{equation}
\omega_{x}^{2}+2\left[\frac{\omega_{x}^{2}-\omega_{y}^{2}}{\tilde{c}}+2\mathcal{C}_{4}\tilde{c}\right]x^{2}+6\mathcal{C}_{4}\,x^{4}=0\,,\label{eq:quartic_equation_x}
\end{equation}
which has real non-vanishing solutions $x^{2}$ for 
\begin{equation}
\left[\frac{\omega_{x}^{2}-\omega_{y}^{2}}{\tilde{c}}+2\mathcal{C}_{4}\tilde{c}\right]^{2}>6\mathcal{C}_{4}\omega_{x}^{2}\,.\label{eq:determinant}
\end{equation}
The last condition is easy to fulfill for a sufficiently small $\mathcal{C}_{4}$,
i.e., for (see also \cref{footnote12})
\begin{equation}
4\tilde{c}^{2}\mathcal{C}_{4}<\omega_{x}^{2}+2\omega_{y}^{2}-\omega_{x}\sqrt{3\left(4\omega_{y}^{2}-\omega_{x}^{2}\right)}\,.\label{eq:upper_bound}
\end{equation}
The solution of \cref{eq:quartic_equation_x} is
\begin{equation}
\label{eq:Nontrivial_solution}
\frac{x_{b\left(\pm\right)}^{2}}{\tilde{c}}=\frac{\Omega_{y}^{2}-\Omega_{x}^{2}-1}{3}\pm\sqrt{\left(\frac{\Omega_{y}^{2}-\Omega_{x}^{2}-1}{3}\right)^{2}-\frac{\Omega_{x}^{2}}{3}}\,,
\end{equation}
where we denoted\footnote{This form reveals the essence of rescalings in \cref{eliminate}.}
\begin{equation}
\label{Omega}
\Omega_{x}^{2}=\frac{\omega_{x}^{2}}{2\mathcal{C}_{4}\tilde{c}^{2}}\,,\quad
\text{and}\quad
\Omega_{y}^{2}=\frac{\omega_{y}^{2}}{2\mathcal{C}_{4}\tilde{c}^{2}}\,.
\end{equation}
To obtain a real $x_{b(\pm)}$, one has to require that the first term in the right hand side of \cref{eq:Nontrivial_solution} is positive\footnote{
    Note that this inequality forbids $4\tilde{c}^{2}\mathcal{C}_{4}>\omega_{x}^{2}+2\omega_{y}^{2}+\omega_{x}\sqrt{3\left(4\omega_{y}^{2}-\omega_{x}^{2}\right)}$ which would be another way to satisfy \cref{eq:determinant}. \label{footnote12} 
} 
which translates to
\begin{equation}
2\mathcal{C}_{4}\tilde{c}^{2}<\omega_{y}^{2}-\omega_{x}^{2}\,.\label{eq:Condition_on_C_4}
\end{equation}
Note that this inequality in \cref{eq:Condition_on_C_4} is weaker
than \cref{eq:upper_bound}, so that, when the latter holds, both extrema $x_{b\left(\pm\right)}^{2}$ are
real and in total there are four solutions. Clearly, $x_{b\left(-\right)}^{2}$
corresponds to a local maximum of $V_{\text{P}}\left(x\right)$ while
$x_{b\left(+\right)}^{2}$ corresponds to a local minimum and hence a stable direction along the $x \neq 0$ direction.

Now we can investigate when such points $\left(\pm x_{b\left(+\right)},0\right)$ are minima of $\mathcal{U}_\text{LV}^{(4)}$ given by \cref{U_4}. First one notices, that at these points $\partial_{x}\partial_{y}\mathcal{U}_\text{LV}^{(4)}\left(\pm x_{b\left(+\right)},0\right)=0$, while
the other second derivatives are
\begin{align*}
 & \partial_{x}^{2}\mathcal{U}_\text{LV}^{(4)}=\frac{8}{3}\mathcal{C}_{4}\tilde{c}^{3}-\frac{4}{3}\tilde{c}\left(\omega_{x}^{2}+2\omega_{y}^{2}\right)+\frac{2\left(\omega_{x}^{2}-\omega_{y}^{2}\right)^{2}}{3\mathcal{C}_{4}\tilde{c}}-\\
 & -\frac{2}{3}\mathcal{R}\left[2\tilde{c}+\frac{\omega_{x}^{2}-\omega_{y}^{2}}{\mathcal{C}_{4}\tilde{c}}\right]\,,
\end{align*}
and 
\begin{align*}
 & \partial_{y}^{2}\mathcal{U}_\text{LV}^{(4)}=\frac{1}{2}\tilde{c}\left[2\omega_{y}^{2}-\frac{\left(\mathcal{R}-2\mathcal{C}_{4}\tilde{c}^{2}-\omega_{x}^{2}+\omega_{y}^{2}\right){}^{2}}{9\mathcal{C}_{4}\tilde{c}^{2}}\right]\,,
\end{align*}
where we denoted 
\[
\mathcal{R}=\sqrt{4\mathcal{C}_{4}^{2}\tilde{c}^{4}+\left(\omega_{x}^{2}-\omega_{y}^{2}\right)^{2}-2\mathcal{C}_{4}\left(\omega_{x}^{2}+2\omega_{y}^{2}\right)\tilde{c}^{2}}\,.
\]
One can deduce that both second derivatives are positive provided 
\begin{equation}
\frac{\left(\omega_{x}^{2}-\omega_{y}^{2}\right)^{2}}{2\omega_{y}^{2}}<4\mathcal{C}_{4}\tilde{c}^{2}<\omega_{x}^{2}+2\omega_{y}^{2}-\omega_{x}\sqrt{3\left(4\omega_{y}^{2}-\omega_{x}^{2}\right)}\,.\label{eq:interval}
\end{equation}
Note that for the assumed $\omega_{y}>\omega_{x}$ the lower bound above is always smaller
than the upper bound. Thus there is always a space for $\mathcal{C}_{4}$
between them\footnote{
    Note that one can reinterpret this inequality as $\left(\Omega_{x}^{2}-\Omega_{y}^{2}\right)^{2}/\left(2\Omega_{y}^{2}\right)<2<\Omega_{x}^{2}+2\Omega_{y}^{2}-\Omega_{x}\sqrt{3\left(4\Omega_{y}^{2}-\Omega_{x}^{2}\right)}$ for rescaled frequencies in \cref{Omega}, see \cref{fig:Lyapunov-extrema}.
}. 
Moreover, the upper bound is just what is required for the
very existence of the equilibrium point, see \cref{eq:upper_bound}. 
Hence, we have just shown that, for $\mathcal{C}_{4}$ satisfying the bound above, the stable points
$\left(\pm x_{b\left(+\right)},0\right)$ correspond to local minima
of $\mathcal{U}_\text{LV}^{(4)}$ (see \cref{fig:U_middle_large}) and these vacua are Lyapunov stable, as they actually also are turning off interactions with the ghost $y$.
For the opposite hierarchy between frequencies, analogous equilibrium
points are $\left(0,\pm y_{b\left(+\right)}\right)$ and the whole
analysis and formulas are applicable after exchange $x\longleftrightarrow y$. Interestingly,
for $\mathcal{C}_{4}$ smaller than the lower bound in \cref{eq:interval},
the equilibrium points $\left(\pm x_{b\left(+\right)},0\right)$ are not local minima of $\mathcal{U}_\text{LV}^{(4)}$ so that one cannot warrant stability.
This is a strong hint that these equilibrium configurations then become unstable. On the other hand, one can check that $\left(\pm x_{b\left(-\right)},0\right)$ never become minima of $\mathcal{U}_\text{LV}^{(4)}$ showing that these equilibrium configurations remain always unstable. Of course the same logic is applicable after exchange $x\longleftrightarrow y$.
\\
\\
Finally, the last category of equilibrium points appears for such small coupling constant $\mathcal{C}_{4}$ that the PG--interactions in \cref{PG-Interaction_Poly} introduce new nontrivial equilibrium
points. These vacua spontaneously break both reflections in \cref{reflection} and are solutions of the system in \cref{def_equlibrium}
\begin{align*}
 & x^{2}+\frac{\left(\omega_{x}^{2}-\omega_{y}^{2}\right)\left(x^{2}-y^{2}\right)}{2\mathcal{C}_{4}\tilde{c}^{2}}+3\frac{\left(x^{2}-y^{2}\right)^{2}}{2\tilde{c}}+\frac{\omega_{x}^{2}}{4\mathcal{C}_{4}\tilde{c}}=0\,,\\
 & y^{2}+\frac{\left(\omega_{x}^{2}-\omega_{y}^{2}\right)\left(x^{2}-y^{2}\right)}{2\mathcal{C}_{4}\tilde{c}^{2}}+3\frac{\left(x^{2}-y^{2}\right)^{2}}{2\tilde{c}}+\frac{\omega_{y}^{2}}{4\mathcal{C}_{4}\tilde{c}}=0\,.
\end{align*}
By subtracting these equations from each other one obtains 
\[
4\mathcal{C}_{4}\tilde{c}\left(y^{2}-x^{2}\right)=\omega_{x}^{2}-\omega_{y}^{2}\,,
\]
so that now one can substitute $y^{2}-x^{2}$ back into both equations
to obtain, using the same definitions as in \cref{Omega},
\begin{equation}
\frac{x_{I}^{2}}{\tilde{c}}=\frac{\left(\Omega_{x}^{2}-\Omega_{y}^{2}\right)^{2}-4\Omega_{x}^{2}}{8}\,,\label{eq:x_I}
\end{equation}
and similarly
\begin{equation}
\frac{y_{I}^{2}}{\tilde{c}}=\frac{\left(\Omega_{x}^{2}-\Omega_{y}^{2}\right)^{2}-4\Omega_{y}^{2}}{8}\,.\label{eq:y_I}
\end{equation}
The right hand side of both equations are strictly positive provided that\footnote{
    These inequalities trivially transform to inequalities on rescaled frequencies, e.g.,
    $\left(\Omega_{x}^{2}-\Omega_{y}^{2}\right)^{2}>4\Omega_{y}^{2}$, 
    see \cref{fig:Lyapunov-extrema}. 
}
(as easily seen using \cref{Omega})
\begin{equation}
\mathcal{C}_{4}<{\rm min}\left(\frac{\left(\omega_{x}^{2}-\omega_{y}^{2}\right)^{2}}{8\omega_{x}^{2}\tilde{c}^{2}}\,,\frac{\left(\omega_{x}^{2}-\omega_{y}^{2}\right)^{2}}{8\omega_{y}^{2}\tilde{c}^{2}}\right)\,,
\label{eq:interacting_bound}
\end{equation}
which guarantees that these nontrivial equilibrium
points exist. Further, this inequality also implies that $\partial_{x}^{2}\mathcal{U}_\text{LV}^{(4)}>0$
along with the determinant of the Hessian $\mathfrak{Hess}(\mathcal{U}_\text{LV}^{(4)})>0$, see \cref{Hessian_Det}.
Thus, all the four
equilibrium points\footnote{Here we assumed that the square root is taken as a principal square root returning non-negative numbers.} $\left(x_{I},y_{I}\right)$ , $\left(-x_{I},y_{I}\right)$,
$\left(x_{I},-y_{I}\right)$, $\left(-x_{I},-y_{I}\right)$ are Lyapunov
stable, as the ``potential'' $\mathcal{U}_\text{LV}^{(4)}$ has local minima
there. One can check that 
\begin{align*}
 & 12\mathcal{C}_{4}^{3}\tilde{c}^{4}\,\,\,\mathcal{U}_\text{LV}^{(4)}\left(x_{I},y_{I}\right)=-\left[32\mathcal{C}_{4}^{2}\left(\omega_{x}^{4}+\omega_{y}^{4}\right)\tilde{c}^{4}+\right.\\
 & \left.-8\mathcal{C}_{4}\left(\omega_{x}^{2}-\omega_{y}^{2}\right)^{2}\left(\omega_{x}^{2}+\omega_{y}^{2}\right)\tilde{c}^{2}+\left(\omega_{x}^{2}-\omega_{y}^{2}\right)^{4}\right]<0\,.
\end{align*}
Thus, comparing the above to \cref{U4xeye}, we see that the equilibium points above are in a sense true vacua as $\mathcal{U}_\text{LV}^{(4)}\left(x_{I},y_{I}\right)<\mathcal{U}_\text{LV}^{(4)}\left(x_e,y_e\right)$, see also \cref{fig:U_low_LLog}.

We would like to stress that we found all possible equilibrium points, i.e. solutions to \cref{def_equlibrium}.
Moreover, comparing, \cref{eq:interval} with \cref{eq:interacting_bound}
one infers that the existence of stable equilibrium points $\left(\pm x_{I},\pm y_{I}\right)$
and $\left(\pm x_{b\left(+\right)},0\right)$ (or $\left(0,\pm y_{b\left(+\right)}\right)$) is mutually exclusive. Hence, PG-interactions in \cref{PG-Interaction_Poly} can actually introduce new Lyapunov-stable equilibrium
points which would be impossible if the ghost $y$ had not interacted with the
positive-energy DoF $x$. Thus, at least for discrete dynamical degrees
of freedom, interactions with ghost can create new stable vacua.
Finally, we would like to mention that it is the strength of the coupling constant $\mathcal{C}_{4}$ which controls the symmetry breaking pattern for fixed $\tilde{c}$ and $\omega_{x/y}$. The larger is the constant the more symmetry is there, see \cref{fig:Lyapunov-extrema}.
\\

\section{Numerical investigation of nonintegrable Models}
\label{sec:nonintegrable}

We now relax the technical assumption of integrability. 
In~\cref{subsec:integrable-motion,subsec:class-3}, we have rephrased the conditions for stability of the proof of \cref{subsec:proof} (which are stated in terms of free functions, specific to the integrable model) to conditions in terms of the shape of the potential $V(x,y)$.

This motivates us to investigate whether similar conditions lead more generally to stable motion and, in particular, when there is no  integrability.

We find numerically that: $(\aleph)$,  a sufficient condition for the stable motion of a Hamiltonian point-particle system with coupled positive- and negative-sign kinetic terms for the respective coordinates $x$ and $y$ is that its full potential $V(x,y)$, at large coordinates $x$ and $y$, is sufficiently dominated everywhere by a potential of the form $\tilde{V}_P(x)+\tilde{V}_G(y)$ where both $\tilde{V}_P$ and $-\tilde{V}_G$ are yielding by themselves stable motions along $x$ and $y$. {\it I.e.} we need the interactions to be sufficiently suppressed with respect to a stable ``decoupled'' part at large values of the dynamical variables.

Note that these conditions are reminiscent of the Kolmogorov–Arnold–Moser (KAM)  theorem (for a pedagogical discussion see~\cite[Appendix 8]{Arnold2013book}), as the theory with the Hamiltonian just given by $\tilde{V}_P(x)+\tilde{V}_G(y)$, which dominates at large distance, would be integrable (as it has two separately conserved quantities: the respective energies of $x$ and $y$), our statement is just that, it is enough for the interactions between $x$ and $y$ to be sufficiently subdominant at large $x$ and $y$ to ensure stability.
The conditions stated in \cref{subsec:integrable-motion,subsec:class-3} are an explicit expression of the above statement for the specific integrable polynomial subclasses at hand. Beyond integrability, we still find that, whenever the above statement $(\aleph)$ holds, \emph{all} (at least, all numerically investigated) initial conditions avoid runaways (at least, up to the finite numerical evolution time).

Naturally, the lack of integrability also means that the analytical proof strategy in \cref{sec:Liouville} will no longer apply. Hence, we need to resort to numerical evidence. The latter comes with the difficulty that -- in absence of rigorous analytical proof -- runaways may only be encountered at very late evolution times and/or only for a small subset of the initial conditions. 
In \cref{subsec:numerics-setup}, we introduce a specific numerical setup to alleviate these issues, at least to some degree. In \cref{subsec:Liouville-numerics}, we demonstrate that the numerics are able to confidently distinguish between stable and unstable potentials in the presented subclass of integrable and polynomial Liouville models.
With this benchmark at hand, we apply the same numerical analysis to several nonintegrable models in \cref{subsec:evidence}, i.e., to two classes of non-polynomial and to one class of polynomial potentials. This provides evidence that the above statement $(\aleph)$ holds true, even in the absence of integrability.
\\

We emphasize that all of our results suggest that the statement $(\aleph)$ only needs to hold outside of a finite region in phase space.
As a special case, it thus encompasses all interaction potentials which are localised in a finite phase-space region and supplemented by standard quadratic terms $V_\text{P} = \frac{1}{2}\,\omega_x^2 x^2$ and $V_\text{G} = -\frac{1}{2}\,\omega_y^2 y^2$ with $\omega_x,\,\omega_y\in\mathbb{R}_+$. Two explicit examples of this are presented in~\cref{subsec:evidence}.
Note that this contrasts with the usual conditions used in the KAM theorem which uses perturbations of an integrable system which are everywhere small with respect to the integrable part, and not only at large values of the dynamical variables.

We also highlight that, if the above holds also for nonintegrable systems, it would imply that the addition of sufficiently strong self-interactions can stabilize any PG system. Indeed, all of our numerics support this conclusion: An explicit example is discussed in \cref{subsec:evidence}.
\\

At the same time, we caution that there is a -- potentially crucial -- distinction between integrable and nonintegrable models. Integrability ensures that whenever the motion returns to the initial position $(x_0,\,y_0)$, also the momenta must return to their initial values. For nonintegrable systems, this need not be the case. Nonintegrable systems may therefore be distinct in that they can potentially extract a pair of positive and negative kinetic energy from the finite region in phase space in which the above conditions are violated. This may lead to a slow growth of the explored phase-space region. While we do not observe any evidence for such slow growth in our simulations, we cannot exclude it. In order to avoid such a growth, one could for example require that the Hamiltonian in the finite region be sufficiently close to integrable so that when the system returns to the initial position $(x_0,\,y_0)$, the momenta return to values sufficiently close to their initial values. Intriguingly, this does not a priori forbid arbitrarily strong interactions between a ghost and a positive energy degree of freedom in the finite region.
Whether it is possible to make this kind of statement more quantitative and precise is an interesting question which could be investigated in the future.

\subsection{Numerical Setup}
\label{subsec:numerics-setup}

For our numerical experiments, we draw random initial conditions $x(t_i)$, $y(t_i)$, $x'(t_i)$, $y'(t_i)$ at $t_i=0$ from a normal distribution with standard deviation $\text{SD}$.
We then evolve the respective initial value problem with the Runge-Kutta method\footnote{We use the native NDSolve routine in \texttt{Mathematica}, specifying to an embedded Runge-Kutta 5(4) pair~\cite{Mathematica:RK} following~\cite{Shampine:1994numerical}. Further, we specify $\text{MaxSteps}\rightarrow\infty$ such that the integration is terminated either at $t=t_\text{max}$ or if NDSolve detects stiffness.} up to either $t_f=t_\text{max}$ or until the numerical routine detects a runaway (see conditions below), in which case $t_f<t_\text{max}$. We make sure that $t_\text{max}$ is sufficiently large such as to differentiate between cases with and without runaways.
We then repeat the experiment $N_\text{evol}$ times and determine the mean
evolution time $t_\text{mean}=\frac{1}{N_\text{evol}}\sum_{j=1}^{N_\text{evol}}t_f^{(j)}$. Assuming that the criteria below confidently detect a runaway, a value of $t_\text{mean}=t_\text{max}$ signals the absence of any runaway solutions in the numerical experiment.

In the above numerical experiment, we distinguish between two criteria to detect runaways:
\begin{enumerate}
    \item[(A)]
    We use stiffness detection of our numerical evolution scheme as a proxy for catastrophic (i.e., faster than exponential) runaways. Stiff numerical evolution typically occurs whenever very different scales enter the evolution. In practice, given that we evolve with a Runge-Kutta method, stiffness is detected by the dominant eigenvalue of the Jacobian, see~\cite{Shampine:1994numerical, Mathematica:Stiffness} and references therein.
    \item[(B)]
    We find that polynomial growth is harder to detect since the eigenvalues of the Jacobian can remain bounded throughout the runaway. In particular, this also includes asymptotically flat potentials in which trajectories can escape to infinity once they escape the potential in a confined phase-space region.
    To detect such cases, we perform the same type of experiment but we add an additional abort criterion that detects if $x(t)>C$ or $y(t)>C$ outgrow a specified bound $C$. This criterion thereby detects whenever the evolution escapes a specified phase-space region. When we pick $C\gg \text{SD}$, we expect that only rare cases of stable motion are falsely detected as unstable.
\end{enumerate}
We will highlight below in which of the cases it makes a difference whether criterion (B) is applied in addition to criterion (A).

\subsection{Benchmark: The Liouville Model Revisited}
\label{subsec:Liouville-numerics}
%
\begin{figure}
    \centering
    \includegraphics[width=\linewidth]{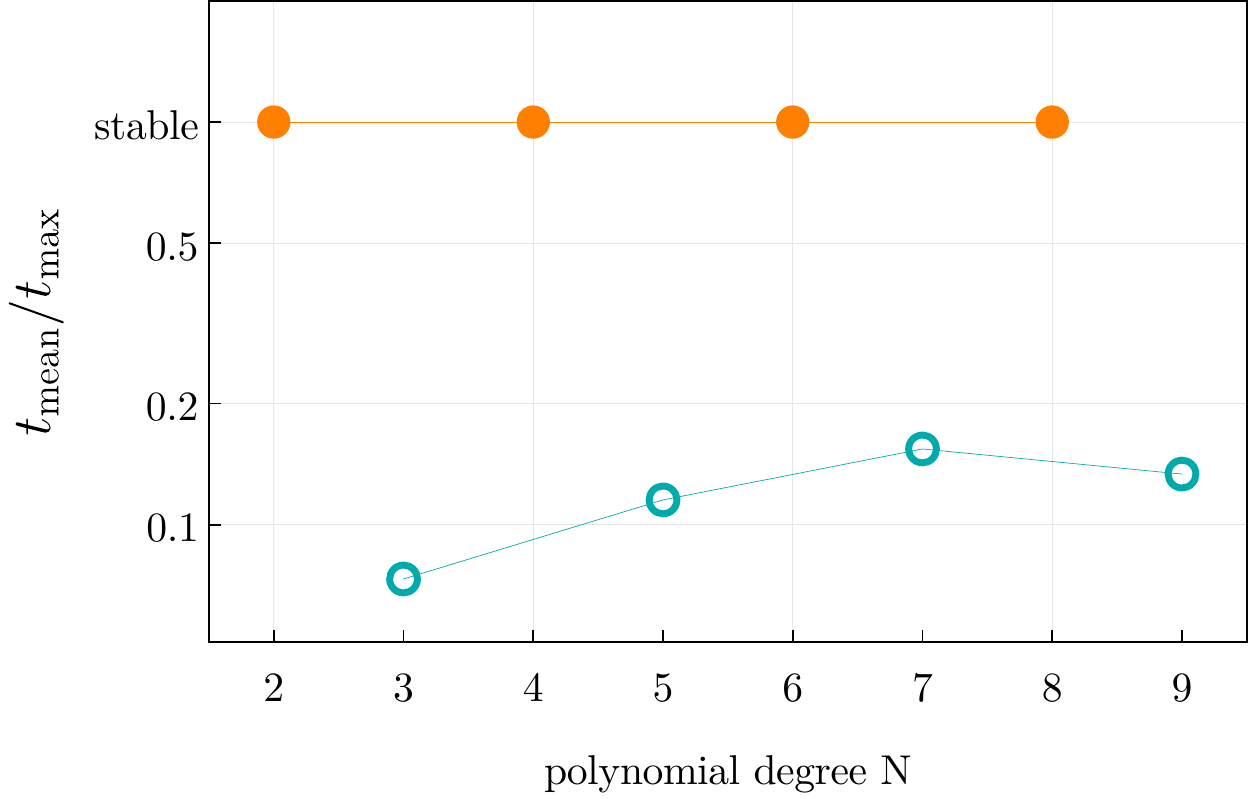}
    \caption{
    We plot the ratio of the average stable evolution time $t_\text{mean}$ and the maximal chosen evolution time of $t_\text{max}=10^2$ as a function of the polynomial degree $N$ in the polynomial subclass of integrable Liouville models, cf.~\cref{defVN}. For the coefficients, we choose $\mathcal{C}_N = 1$ and $\mathcal{C}_{n<N} = 0$ such that all even cases are stable according to \cref{subsec:integrable-motion}.
    Each point in the plot represents an average of the evolution of $N_\text{evol}=100$ random initial conditions drawn from a normal distribution with standard deviation $\text{SD} =1$.
    }
    \label{fig:numericalExperiment_polynomial_subclass}
\end{figure}

Here, we use the previously discussed integrable Liouville model, cf. \cref{sec:Liouville}, to demonstrate that the numerical experiment meaningfully distinguishes between the presence and the absence of runaways. 
\\

We first focus on the polynomial subclass, cf.~\cref{defVN,eq:polynomial-subclass},
for which the resulting potential $V(x,y)$ reduces to a polynomial of degree $N$. As discussed in \cref{subsec:integrable-motion}, the proof conditions imply stability at even degree $N$ (assuming also that $\mathcal{C}_N>0$) and, moreover, align with
conditions stated in in terms of the potential.

For each polynomial degree $2\leqslant N\leqslant 9$, we perform our numerical experiment $N_\text{evol}=10^2$ times. We choose a maximal evolution time $t_\text{max}=10^2$ and draw random initial conditions from a normal distribution with standard deviation $\text{SD}=1$. 

The result is shown in \cref{fig:numericalExperiment_polynomial_subclass} and confirms stability (instability) at even (odd) degree $N$. It turns out to be irrelevant whether or not we apply criterion (B). We take this as an indication that the runaways occur with faster than exponential growth.
\\

\begin{figure}
    \centering
    \includegraphics[height=0.48\linewidth]{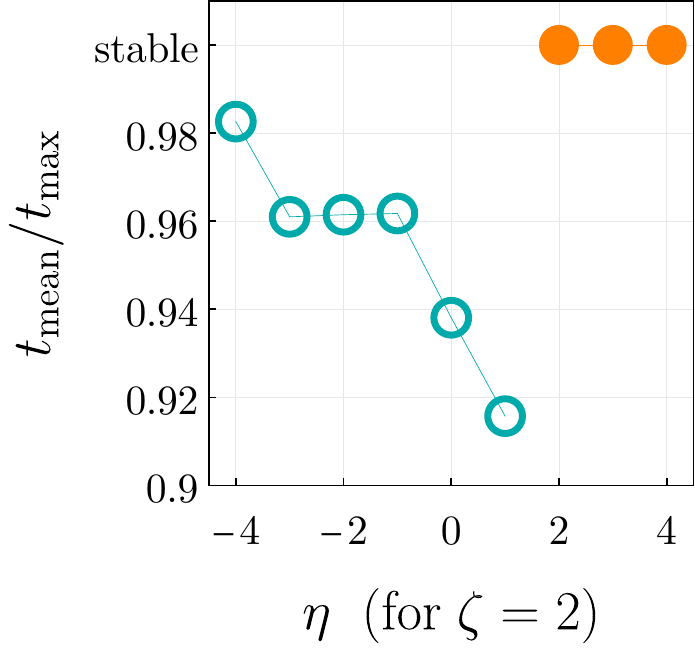}
    \includegraphics[trim=0.8cm 0cm 0cm 0cm, clip, height=0.48\linewidth]{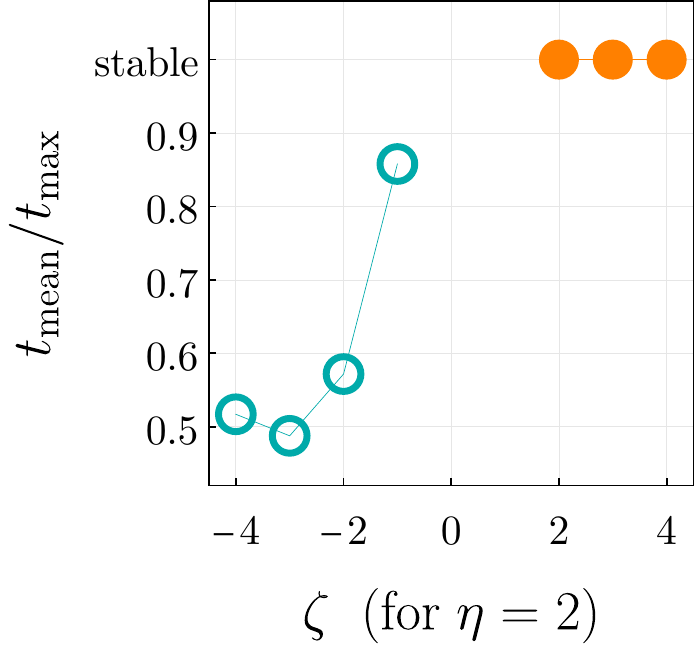}
    \caption{
    We plot the ratio of the average stable evolution time $t_\text{mean}$ and the maximal chosen evolution time of $t_\text{max}=10^3$ as a function of (i) the exponent $\eta$ (at fixed $\zeta=2$) in the left-hand panel, and for (ii) the exponent $\zeta$ (at fixed $\eta=2$) in the right-hand panel, cf.~\cref{eq:fg-proof}.
    Each point in the plot represents an average of the evolution of $N_\text{evol}=100$ random initial conditions drawn from a normal distribution with standard deviation $\text{SD} =1$. The numerical evolution is terminated if stiffness is detected or if $x(t)>C$ or $y(t)>C$, with $C=10^3$.
    }
    \label{fig:numericalExperiment_proofConditions}
\end{figure}

Going beyond the polynomial subclass, we also perform a numerical experiment with
\begin{align}
    \label{eq:fg-proof}
    f(u) = |u|^\zeta
    \;,\quad
    g(v) = |v|^\eta
    \;.
\end{align}
It turns out that, for either $\eta\leqslant2$ (irrespective of $\zeta$) or $\zeta\leqslant2$ (irrespective of $\eta$), the potential exhibits flat directions. This seems to allow for non-catastrophic runaway behaviour, the detection of which requires criterion (B). Without criterion (B), we do not detect any runaway behaviour for $\eta\leqslant2$ or $\zeta\leqslant2$ which indicates that the runaways occur with polynomial growth rate. 
We vary $\eta$ at fixed $\zeta=2$ and vice versa: The results are shown in \cref{fig:numericalExperiment_proofConditions} and reproduce the conditions required in the proof of \cref{subsec:proof}. We note that close to the critical case, the potential becomes almost flat and the runaways are, therefore, increasingly hard to confidently detect numerically.

\subsection{Numerical Evidence in Absence of Integrability}
\label{subsec:evidence}

%
\begin{figure}
    \centering
    \includegraphics[width=0.8\linewidth]{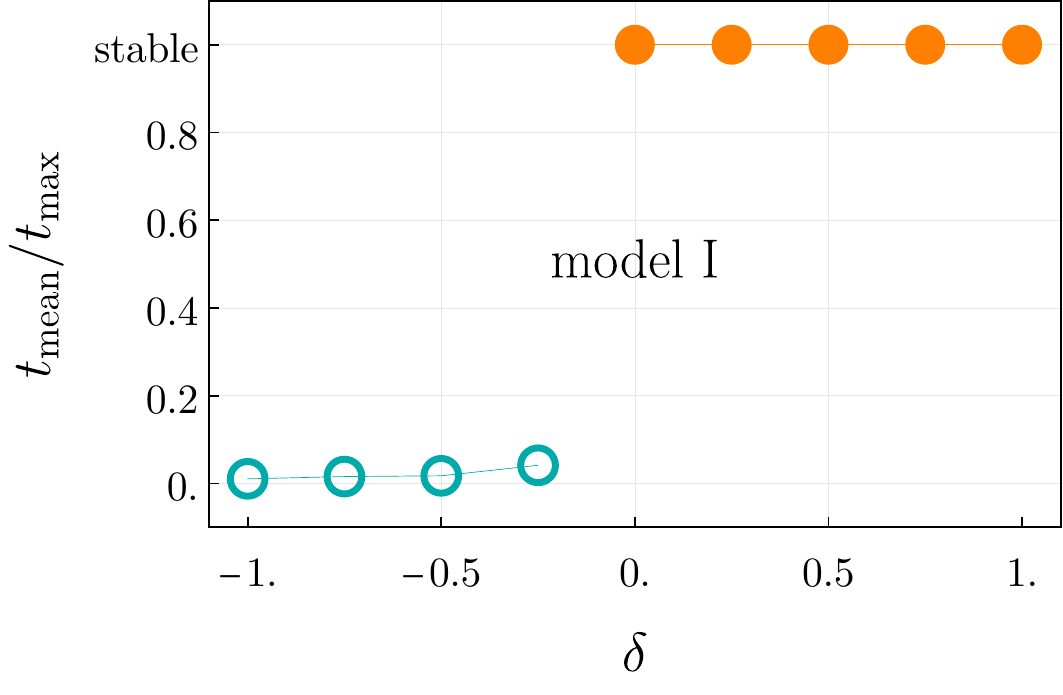}
    \\
    \includegraphics[width=0.8\linewidth]{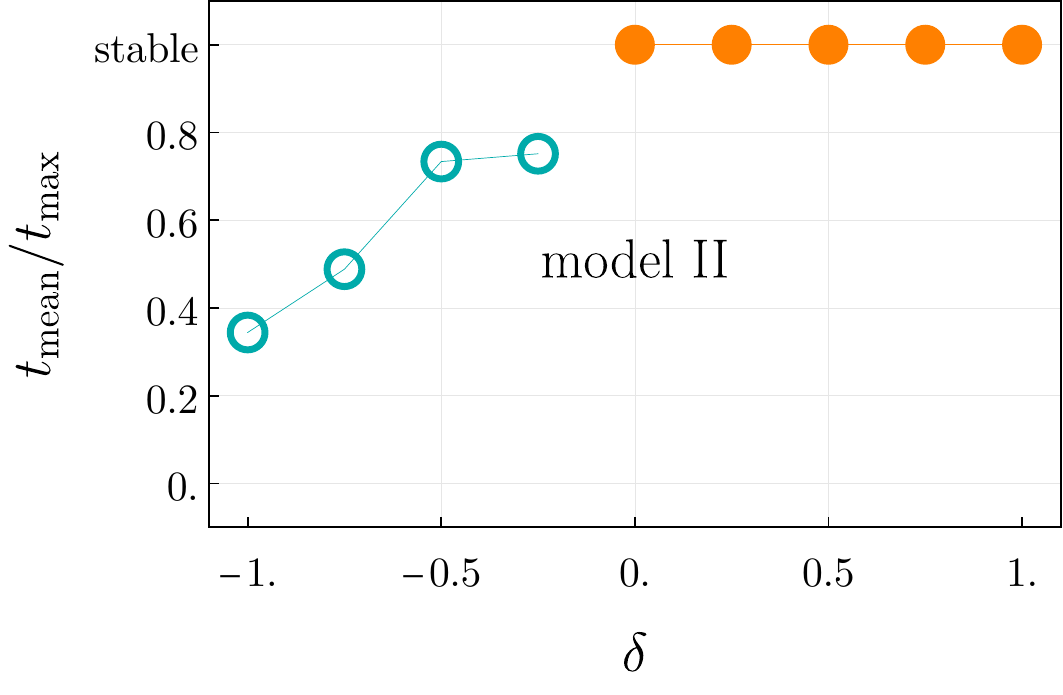}
    \\
    \includegraphics[width=0.8\linewidth]{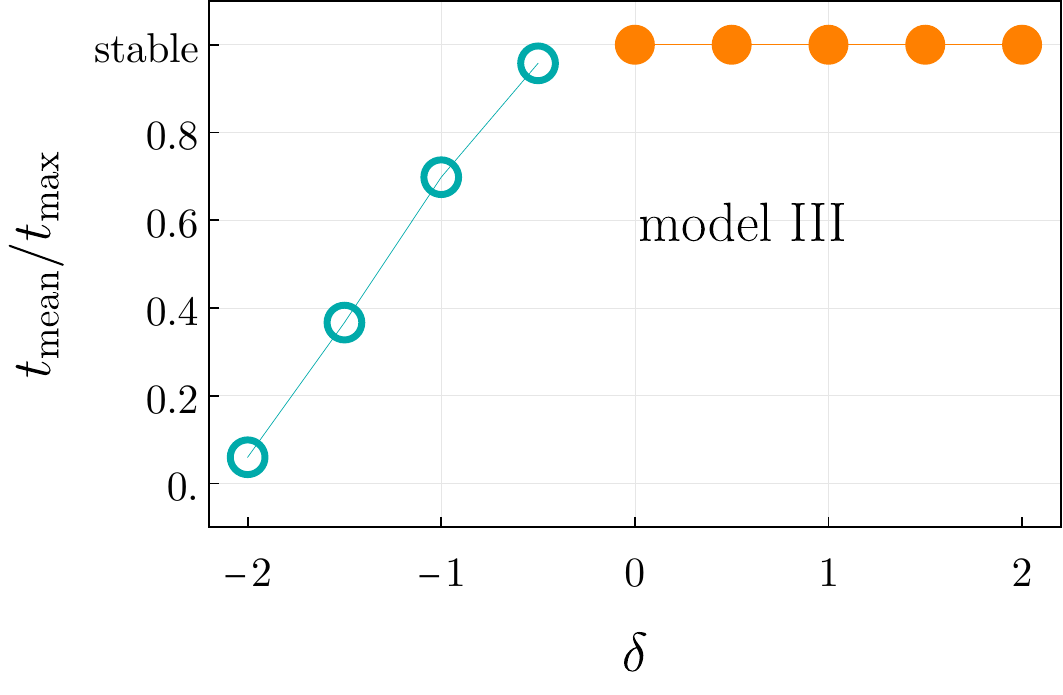}
    \caption{
	We plot the ratio of the average stable evolution time $t_\text{mean}$ and the maximal chosen evolution time of $t_\text{max}=100$ as a function of the power-law exponent $\delta$ for the three classes of potentials $V_{I,\delta}$ (upper panel, cf.~\cref{eq:V_I}), $V_{II,\delta}$ (middle panel, cf.~\cref{eq:V_II}), and $V_{III,\delta}$ (lower panel, cf.~\cref{eq:V_III}). Each point in the plot represents an average of the evolution of $N_\text{evol}=100$ random initial conditions drawn from a normal distribution with standard deviation $\text{SD} =1$.
    }
    \label{fig:numericalExperiment_nonintegrable-models}
\end{figure}

Going beyond integrability, we specify the following two classes of interaction potentials, i.e.,
\begin{align}
    \widetilde{V}_{I,\delta} &=
	\left(1+x^2+y^2\right)^{1-\delta}
	\;,
	\\
    \widetilde{V}_{II,\delta} &= 
	\left(1+x^2y^2\right)^{\frac{1}{2}-\delta}
	\;,
\end{align}
both supplemented with standard quadratic terms such that 
\begin{align}
\label{eq:V_I}
	V_{I,\delta} &= \widetilde{V}_{I,\delta} + \frac{1}{2}\,x^2 -\frac{1}{2}\,y^2\;,
	\\
\label{eq:V_II}
	V_{II,\delta} &= \widetilde{V}_{II,\delta} + \frac{1}{2}\,x^2 -\frac{1}{2}\,y^2\;.
\end{align}
(For simplicity, we only consider equal frequencies for $x$ and $y$.)
In both cases, the exponent $\delta$ is chosen such that, as expected from the statement $(\aleph)$,  
$\delta\lessgtr 0$ determine
the presence/absence of runaways. This is also what we find numerically (see below).
\\

The above two example potentials involve non-polynomial interactions (except for special choices of $\delta$). Our third choice of example potential starts with a strictly polynomial interaction, i.e.,
\begin{align}
	V_\text{int}(x,y)\equiv \widetilde{V}_{III,\delta} &=
	x^2\,y^2
	\;.
\end{align}
Supplemented only with standard kinetic terms, this interaction is unstable. However, we can generalize the decoupled potentials to
\begin{align}
	V_\text{P} &= \left(x^2\right)^{2+\delta}\;,
	\hspace*{0.15\linewidth}
	V_\text{G} = - \left(y^2\right)^{2+\delta}\;,
\end{align}
in which case, we find that for the full potential, i.e.,
\begin{align}
\label{eq:V_III}
    V_{III,\delta} = \widetilde{V}_{III,\delta} + \left(x^2\right)^{2+\delta} - \left(y^2\right)^{2+\delta}\;,
\end{align}
the exponent $\delta\lessgtr 0$, once more, seems to determine
the presence/absence of runaways (at least within out finite evolution time).
\\

We show the result of the respective numerical experiments in \cref{fig:numericalExperiment_nonintegrable-models} for $V = V_{I,\delta}$ (upper panel), for $V = V_{II,\delta}$ (middle panel), and for $V = V_{III,\delta}$ (lower panel). In all of these cases, the inclusion of criterion (B) does not affect the results. The latter suggests that, for $\delta<0$, the runaways are catastrophic, i.e., occur with exponentially fast growth rate.
\\

\begin{figure*}
    \includegraphics[trim=0cm 0cm 0cm 0cm, clip, height=0.295\linewidth]{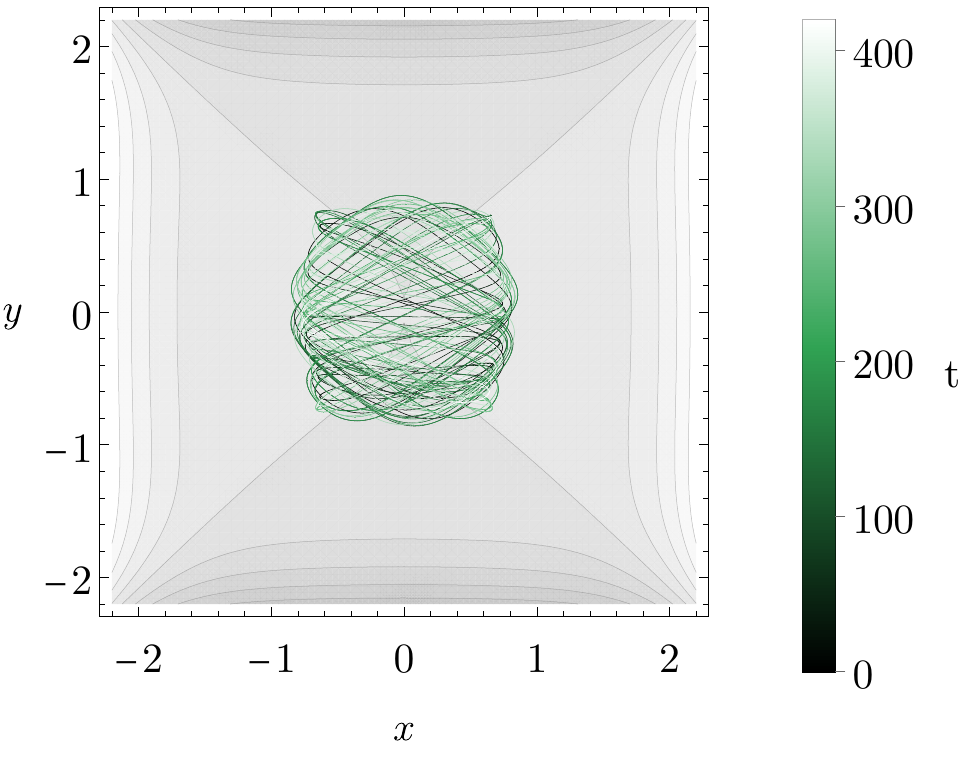}
    \hfill
    \includegraphics[trim=0cm 0cm 0cm 0cm, clip, height=0.29\linewidth]{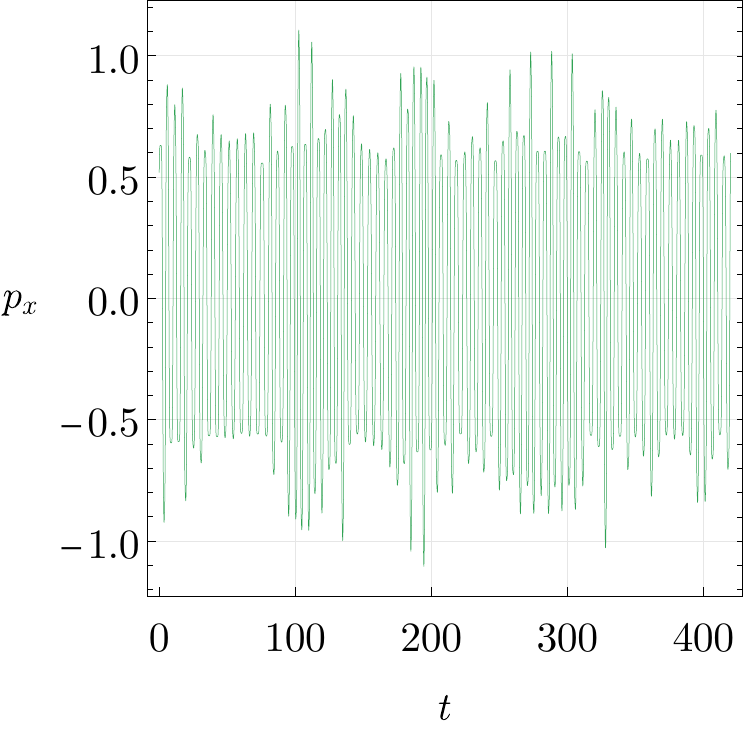}
    \hfill
    \includegraphics[trim=0cm 0cm 0cm 0cm, clip, height=0.29\linewidth]{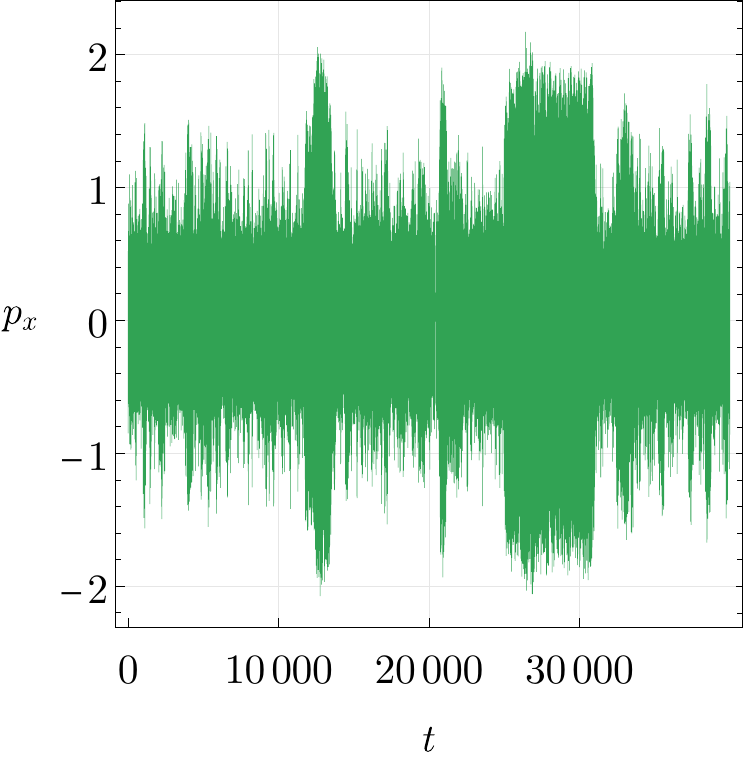}
    \\[1em]
    \includegraphics[trim=0cm 0cm 0cm 0cm, clip, height=0.295\linewidth]{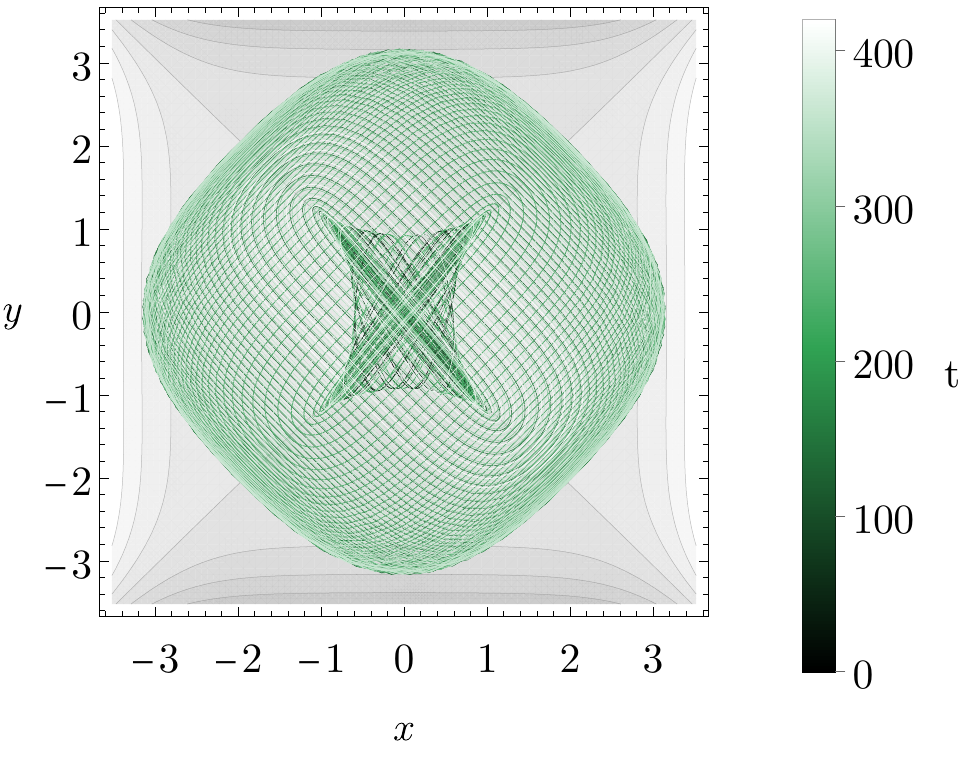}
    \hfill
    \includegraphics[trim=0cm 0cm 0cm 0cm, clip, height=0.29\linewidth]{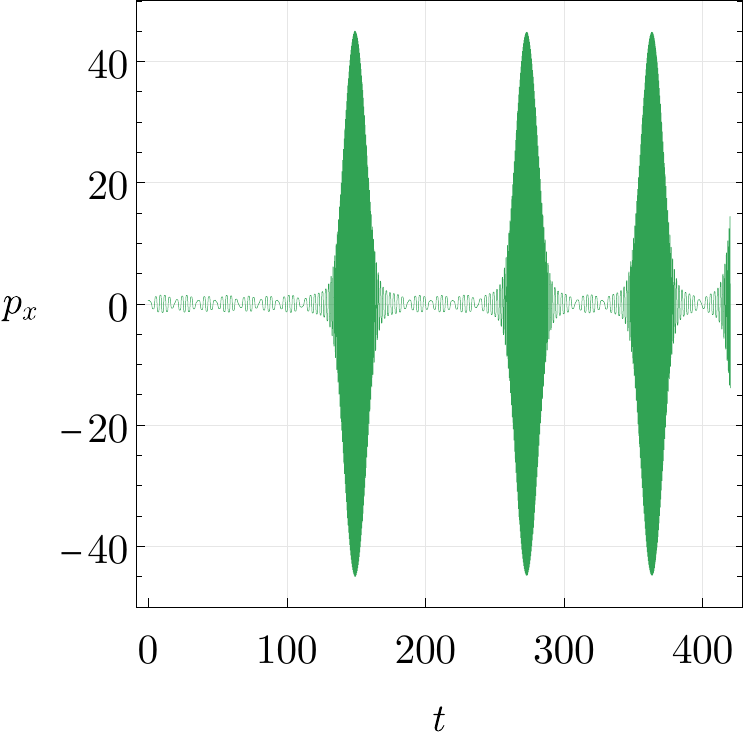}
    \hfill
    \includegraphics[trim=0cm 0cm 0cm 0cm, clip, height=0.29\linewidth]{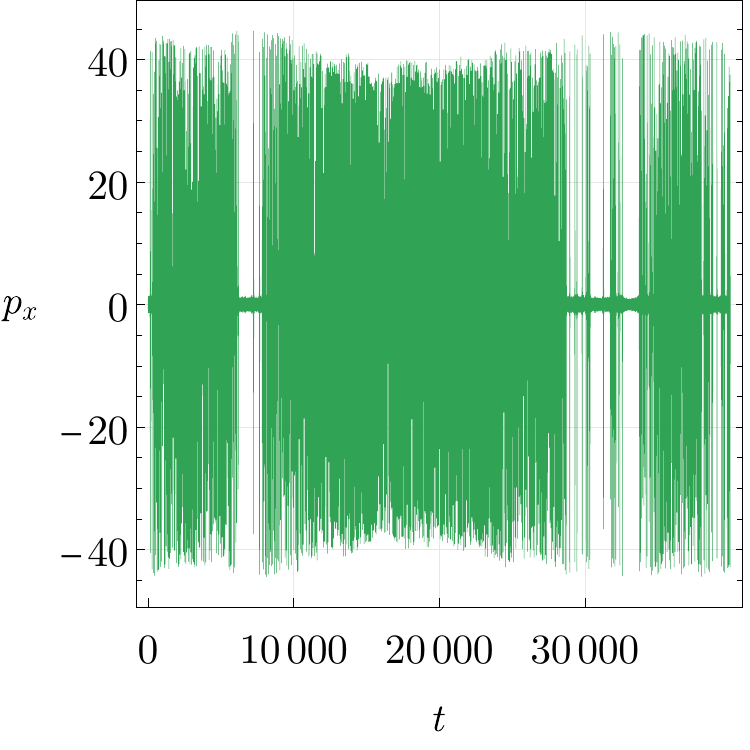}
    \caption{
    We show the evolution of two exemplary cases of random initial conditions (drawn from a Gaussian distribution with $\text{SD}=1$ but selected with regards to the discussion in the main text) for model III with $\delta=1$. 
    In the upper panels, the motion remains in the local region in which non-integrable interactions dominate: hence the motion is apparently chaotic on all timescales.
    In the lower panels, the motion transitions to larger phase-space variables for which the decoupled -- and thus approximately integrable -- potential dominates: hence the motion exhibits seemingly non-chaotic phases (cf.~lower left-hand panel).
    At very long timescales, both initial conditions lead to chaotic motion.
    }
    \label{fig:non-integrable-evolution}
\end{figure*}

Model III demonstrates in what sense added self-interactions can apparently stabilize the motion of nonintegrable systems or, at least, dramatically increase the longevity of stable motion. In the following, we give further detail on the apparent absence of runaways and the respective characterization of the motion as longlived.

We start by noting that the numerical experiment summarized in the lower panel of \cref{fig:numericalExperiment_nonintegrable-models} shows that for $\delta<0$, a significant fraction of initial conditions leads to a catastrophic runaway before $t_\text{max}=100$.

This is in stark contrast to all cases with $\delta>0$, for which we observe no sign of instabilities at all. The specific case of $\delta=1$ is further exemplified in \cref{fig:non-integrable-evolution} where we evolve two distinct and random initial conditions up to $t_\text{max}=4\times10^4$. 
We generically observe two different types of evolution, depending on the specified initial conditions. 

On the one hand, initial conditions which are sufficiently small, lead to motion which remains within the central region, throughout the entire evolution. For these cases, the nonintegrable character is always apparent. An example is shown in the upper panels of \cref{fig:non-integrable-evolution}.

On the other hand, initial conditions which are sufficiently large, lead to motion which is dominated by the decoupled -- and thus approximately integrable -- self-interactions. For these cases, the motion is close to integrable. In general we find that, the larger the initial conditions, the harder it is to reveal the nonintegrable character of the motion.
We refrain from explicitly showing one of the cases for which we cannot resolve the nonintegrable character up to $t_\text{max}=4\times10^4$. 

We also observe that, for some intermediate cases of initial conditions, the motion can transition between both types of behaviour. An example of this is shown in the lower panels of \cref{fig:non-integrable-evolution}. At apparently random evolution times, the motion transitions back and forth between the nonintegrable character in the central region and the approximately integrable character at larger phase-space values.

We emphasize once more that, in all of these cases, the motion remains bounded in a finite phase-space region, at least up to the investigated evolution time.

\section{Discussion}
\label{sec:discussion}

Two point particles with opposite-sign kinetic energies can interact and yet exhibit stable motion.
We prove this statement for a large class of integrable models, for which we show boundedness of motion, irrespective of the initial conditions (Lagrange stability).
We gather substantial numerical evidence that nonintegrable models can be stable as well or at least longlived: within the necessarily finite numerical evolution time, we detect
bounded motion and no indication of growth modes. We observe that our integrable and nonintegrable results seem to be connected by physical conditions on the interaction potential,
which can be summarized as follows (as expressed in $(\aleph)$ of \cref{sec:nonintegrable}): suppose a coupled Hamiltonian point-particle system with opposite-sign kinetic terms can be decoupled, in such a way that uncoupled ghosts and positive energy degrees of freedom are both separately stable due to their own self-interactions. Then, the coupled system avoids runaways, or at least exhibits very long-lived motion, if the full potential at sufficiently large coordinates is sufficiently dominated everywhere by its decoupled counterpart.
\\

Our analytical results are contained in \cref{sec:integrable-models,sec:Liouville} and concern integrable two-particle Hamiltonians with an additional first integral
which is quadratic in the momenta. (All the corresponding known cases with positive energies are collected in \cref{app:integrable-Hamiltonians}.) The analytical results are organized as follows.

First, we show how one can relate ghostly and non-ghostly theories by way of complex (univalent) canonical transformations and use this to build an integrable ghostly theory from a one with no ghost, {\it i.e.} to ``ghostify'' a positive kinetic energy theory while maintaining its integrability. We note that such ghostifications are \emph{not} a mere recasting of variables as we do \emph{not} complexify the initial conditions accordingly. Rather, real initial conditions on one side of the pair of theories correspond to purely complex initial conditions on the other side, and vice versa, while we work only with real variables when considering a given theory in isolation. Nevertheless, the ghostification procedure preserves integrability and thus serves as a generic construction principle for integrable models with ghostly interactions. Depending on the specific case, the Hamiltonian on either side of the pair of ghostly and non-ghostly theories can be complex-valued (and thus of little physical interest) while the other side may or may not be real-valued. With regards to the systematic construction of integrable ghostly Hamiltonians, two cases are of particular interest: First, the case in which both sides of the pair are real-valued. Second, the case in which a real-valued ghostly Hamiltonian can be obtained from a complex-valued non-ghostly Hamiltonian. We discuss examples of both these cases.

Second, we specify to a class of models with two free functions of (specific combinations) of the position variables, first obtained by Liouville. We delineate conditions on the free functions of the Liouville class and -- assuming that said conditions hold -- provide a rigorous proof that the resulting motion is bounded in a finite region in phase space, irrespective of the initial conditions. 
As in~\cite{Deffayet:2021nnt}, the key to the proof is the additional integral of motion which, in combination with the Hamiltonian, can be used to bound the absolute value of all phase-space variables. 
This integral of motion is constructed from a positive definite quadratic form of the canonical momenta, similar to the usual kinetic energy, and a ``potential'' part which is bounded from below and is unbounded from above. The conditions enunciated on the free functions ensure a sufficiently fast growth rate of the ``potential'' at large phase-space variables. These properties are in direct analogy with the usual positive definite conserved energy usually employed to establish stability. Note that for this question it is irrelevant which of the two conserved quantities drives the evolution.
We thus expect that other integrable models can be analysed in a similar manner.

Third, we systematically identify an exhaustive polynomial subclass of the Liouville model. This polynomial subclass is real-valued both for the ghostly and the non-ghostly case. To the best of our knowledge, it has not been previously identified in the literature. As for stable motion with ghostly interactions, this polynomial subclass contains, in particular, a tower of polynomial potentials for which the proof conditions hold and for which we have thus established boundedness of motion in the presence of a ghost. The lowest non-trivial order features a sextic interaction term and quadratic terms which can be chosen at will. This provides a strikingly simple model of ghostly interactions which is polynomial, integrable, and globally stable. 
Moreover, this system exhibits spontaneous symmetry breaking entirely due to the interactions with the ghost. Remarkably, the very same interactions which are often thought to catastrophically destabilize the system, instead introduce novel (Lyapunov) stable vacua.

We highlight that all the above three parts of our analytical work are systematically presented and can, presumably, be extended to other integrable systems. Our analysis is therefore far from exhaustive: Even among the known set of all integrable two-particle systems with a constant of motion, which is quadratic in the momenta (see \cref{app:integrable-Hamiltonians}), it is thus possible that further stable ghostly systems can be obtained in the same way.
\\

Our numerical results are contained in \cref{sec:nonintegrable} and concern the extension to nonintegrable models. The analytical proof for boundedness of motion in the integrable Liouville class is quite striking. It is, however, equally striking that we find no indication that integrability is necessary. All of the investigated nonintegrable models seem to behave like the integrable ones. If the interactions at large phase-space variables do not obey the observed conditions $(\aleph)$, we quickly detect runaway solutions which either lead to a breakdown of the evolution in finite time or to a less severe runaway. In contrast, whenever the interactions at large phase-space variables obey the observed conditions $(\aleph)$, we find no indication for the onset of runaways.
\\

The present work calls for various extensions in obvious directions. 
First, it is clear that our results should allow for an extension to the multiparticle case, i.e., to one or several ghost DoFs interacting with several positive-kinetic-energy DoFs. 
Second, an obvious question is that of quantization. A similar issue has been studied for some specific analogous PG system where stability could only be studied numerically and/or where the ghost still caused an instability, albeit an arguably ``benign'', i.e., a slowly-growing one (see e.g. \cite{Smilga:2004cy}). In this context the simple sextic polynomial potential which we have established to be integrable and stable presents a starting point to investigate whether stability persists, as 
we actually expect, at the quantum level.
Last, we are planning to extend our work
to field theories. In particular, one can construct a theory of two interacting scalar fields $\phi$ and $\psi$, one with
positive and one with negative kinetic energy, for which
the interaction potential is equivalent to the one investigated
here. Once more, the simple sextic polynomial potential found here, appears to be especially amenable for such a study.
Finally, an extension of our analysis to gravitational interactions remains an important question for future work.
\\

\textit{Acknowledgments.}
AH would like to thank Claudia de Rham for discussion.
CD thanks Thibault Damour and Robert Conte for discussions. We would like to thank Antonios Mitsopoulos for his helpful correspondence regarding~\cref{app:integrable-Hamiltonians}.
AH acknowledges: The work leading to this publication was supported by the PRIME programme of the German Academic Exchange Service (DAAD) with funds from the German Federal Ministry of Education and Research (BMBF). AH also acknowledges support by the Deutsche Forschungsgemeinschaft (DFG) under Grant No 406116891 within the Research Training Group RTG 2522/1. AV is supported in part by the European Regional Development Fund (ESIF/ERDF) and the Czech Ministry of Education, Youth and Sports (M\v SMT) through the Project CoGraDS–CZ.02.1.01/0.0/0.0/15 003/0000437. SM and AV collaboration is also supported by the Bilateral Czech-Japanese Mobility Plus Project JSPS-21-12 (JPJSBP120212502).

\appendix

\section{Integrable two-particle Hamiltonians with polynomial constants of motion up to quadratic order in the momenta}
\label{app:integrable-Hamiltonians}

Here, we collect all the known pairs of a potential $V(x,y)$ and the respective constant of motion $I(x,y,p_x,p_y)$ which correspond to integrable Hamiltonian two-particle systems (cf.~\cref{eq:Ham-1}) and for which $I$ is either linear or quadratic in the momenta. A systematic search for these integrable models has been pioneered~\cite{darboux1901archives}, formalized in~\cite{Whittaker1964book}, and subsequently completed in~\cite{Frivs1967symmetry,Holt1982construction,Ankiewicz1983complete,Ramani:1983,Ramani:1983a,Thompson1984darboux,Sen1985integrable,Hietarinta:1986tw}. All results stated here can also be found throughout~\cite{Hietarinta:1986tw}. For the convenience of the reader (and correcting some typos), we collect them here in compact form. In order to match with the literature, we provide the integrable classes for the PP case. Using the ``ghostification" procedure outlined in \cref{sec:integrable-models}, one can obtain the corresponding PG classes.

At linear order, only one class of integrable systems is found, i.e.,
\begin{align}
    \label{eq:class0}
    \text{class 0:}\quad
    V &= f\left(\frac{a}{2}(x^2 + y^2) + c\,x - b\,y\right)\;,
    \\\notag
    I &= a\,\left(y\,p_x - x\,p_y\right) - b\,p_x - c\,p_y\;,
\end{align}
with $a,b,c\in\mathbb{C}$ and $f$ an arbitrary function.

At quadratic order, we follow~\cite{Hietarinta:1986tw} and introduce shorthand variables (repeating some of the definitions given in the main text)
\begin{align}
    r^2 &= x^2 + y^2\;,
    \\[0.5em]
    z_\pm &= x \pm iy\;,
    \label{eq:def-zpm}
    \\[1em]
    u^2 &= \frac{1}{2}\left(r^2 + c +\sqrt{(r^2 + c)^2 - 4\,c\,x^2}\right)\,, 
    \\[0.5em]
    v^2 &= \frac{1}{2}\left(r^2 + c -\sqrt{(r^2 + c)^2 - 4\,c\,x^2}\right)\,,
    \\[1em]
    \label{eq:def-uTilde}
    \hat{u}^2 &= \frac{1}{2}\left(r^2 + \sqrt{r^4 - 2\,c\,z_\pm^2}\right)\,, 
    \\[0.5em]
    \label{eq:def-vTilde}
    \hat{v}^2 &= \frac{1}{2}\left(r^2 - \sqrt{r^4 - 2\,c\,z_\pm^2}\right)\,.
\end{align}
With these definitions the eight distinct classes of integrable systems can be written as
\begin{flalign}
    \label{eq:class1}
  \text{class 1:} \quad
&V = \frac{f(u) - g(v)}{u^2 - v^2}\;,
     \\\notag
 & I = -\left(
        xp_y - yp_x
    \right)^2
    - c\,p_x^2
    & \\
   \notag 
  & \quad  + 2\,\frac{u^2g(v) - v^2f(u)}{u^2 - v^2}\;,&
\end{flalign}
\begin{flalign}
    \label{eq:class2}
    \text{class 2:}\quad
  &  V = g(r) + \frac{f(x/y)}{r^2}
    \;,
    &\\\notag
    &I = (x p_y-y p_x)^2 + 2\,f(x/y)
    \;,&
\end{flalign}
\begin{flalign}
    \label{eq:class3}
    \text{class 3:}\quad
  &  V = \frac{f(\hat{u}) - g(\hat{v})}{\hat{u}^2 - \hat{v}^2}\;,
    &\\\notag
   & I = -\left(
        xp_y - yp_x
    \right)^2
    - \frac{c}{2}\,\left(
        p_x\pm i\,p_y
    \right)^2
    &\\\notag&\quad
    + 2\,\frac{\hat{u}^2g(\hat{v}) - \hat{v}^2f(\hat{u})}{\hat{u}^2 - \hat{v}^2}\;,&
\end{flalign}
\begin{flalign}
    \label{eq:class4}
    \text{class 4:}\quad
    & V = \frac{f(r+y) + g(r-y)}{r}
    \;,
    &\\\notag
    &I = (y p_x - x p_y)p_x 
    &\\\notag&\quad
    + \frac{(r+y)g(r-y) - (r-y)f(r+y)}{r}
    \;,&
\end{flalign}
\begin{flalign}
    \label{eq:class5}
    \text{class 5:}\quad
    &V = \frac{f(z_+ + \sqrt{z_-}) + g(z_+ - \sqrt{z_-})}{\sqrt{z_-}}
    \;,
    &\\\notag
    &I = 
    (y p_x - x p_y)(p_x + i\,p_y)
    + i\frac{(p_x - i\,p_y)^2}{8}
    &\\\notag&\quad
    - i\,\left(\frac{z_+ - \sqrt{z_-}}{\sqrt{z_-}}\right)f(z_+ + \sqrt{z_-})&
    \\\notag&\quad
    - i\,\left(\frac{z_+ + \sqrt{z_-}}{\sqrt{z_-}}\right)g(z_+ - \sqrt{z_-})
    \;,&
\end{flalign}
\begin{flalign}
    \label{eq:class6}
    \text{class 6:}\quad
    &V = 
    \frac{f(z_\pm )}{r} + g'(z_\pm )
    \;,
   & \\\notag
   & I = 
    (y p_x - x p_y)(p_x \pm i\,p_y)
    \pm i g(z_\pm )
    &\\\notag&\quad
    \mp i\,z_\pm\,\left(\frac{f(z_\pm )}{r} + g'(z_\pm )\right)
    \;,&
\end{flalign}
\begin{flalign}
    \label{eq:class7}
    \text{class 7:}\quad
   & V = f(x) + g(y)
    \;,
    &\\\notag
    &I = p_x^2 + 2\,f(x)
    \;,&
\end{flalign}
\begin{flalign}
    \label{eq:class8}
      \text{class 8:}\quad
   & V = 
    r^2 f''(z_\pm ) 
    + g(z_\pm )
    \;,
    &\\\notag
    &I = p_x(p_x \pm i p_y) + r^2 f''(z_\pm ) + g(z_\pm ) 
    &\\\notag&\quad
    + 2 (z_\pm )f'(z_\pm ) - 2 f(z_\pm )
    \;, &
\end{flalign}
where $c\in\mathbb{C}$ and $f$ and $g$ denote arbitrary functions. A prime denotes a first derivative and a double prime denotes a second derivative with respect to the argument. In comparison to~\cite{Hietarinta:1986tw}, we have corrected a typo in the integral of motion $I$ of class 3 (given here 
in \cref{eq:class3}) which differs from \cite[Eq.~(3.2.13)]{Hietarinta:1986tw} but agrees with \cite[Eq.~(41)]{Mitsopoulos_2020} (given there in terms of $A=c/2$, $F_1 = 2f$, and $F_2=2g$).
Invariants which include linear and quadratic terms in the momenta are prohibited by time reflection symmetry, cf.~\cite{Hietarinta:1986tw}.
\\

The above functional classes are known to contain several polynomial subclasses, cf.~\cite{Hietarinta:1986tw} for a collection of those. In \cref{app:polynomial-potentials,app:polynomial-potentials-class-3}, we identify the polynomial subclasses for class 1 (in the main text also referred to as Liouville class) and class 3. To the best of our knowledge these polynomial subclasses have not been previously identified.
\section{Integrability and Vacua}
\label{app:Integrability_Equilibrium}
For self-completeness of the paper, as well as to obtain some intermediate results useful for the discussion of \cref{subsec:Lagrange_vs_Lyapunov}, let us briefly review when a system with Hamiltonian 
\begin{equation}
H=\frac{1}{2}p_{x}^{2}+\frac{\sigma}{2}p_{y}^{2}+V\left(x,y\right)\,,\label{eq:H(xy)}
\end{equation}
possess a first integral that is quadratic in momenta 
\begin{equation}
I=K^{ik}\left(x,y\right)p_{i}p_{k}+U\left(x,y\right)\,,\label{eq:I_K}
\end{equation}
where $i,k$ run over $\left\{ x,y\right\} $ and $\sigma=\pm1$. The vanishing of the Poisson bracket $\left\{ H,I\right\}_{\text{PB}} $ implies that
\begin{align}
 & p_{x}\left[\partial_{x}U-2\left(K^{xx}\partial_{x}V+K^{xy}\partial_{y}V\right)\right]+\label{eq:Vanishing_Poisson_br}\\
 & \qquad+p_{y}\left[\sigma\partial_{y}U-2\left(K^{yx}\partial_{x}V+K^{yy}\partial_{y}V\right)\right]+\nonumber \\
 & \qquad\qquad +p_{i}p_{k}\left(p_{x}\partial_{x}K^{ik}+\sigma p_{y}\partial_{y}K^{ik}\right)=0\,.\nonumber 
\end{align}
 This equation can only be satisfied for \emph{arbitrary} $p_{x}$
and $p_{y}$ provided the following conditions hold 
\begin{align}
 & \partial_{x}U=2\left(K^{xx}\,\partial_{x}V+K^{xy}\,\partial_{y}V\right)\,,\label{eq:U_x_vs_V_x}\\
 & \partial_{y}U=2\sigma\left(K^{yx}\,\partial_{x}V+K^{yy}\,\partial_{y}V\right)\,,\nonumber \\
 & p_{i}p_{k}\left(p_{x}\partial_{x}K^{ik}+\sigma p_{y}\partial_{y}K^{ik}\right)=0\,.\nonumber 
\end{align}
As we assume that $K^{ik}\left(x,y\right)$ is a smooth function of
$x$ and $y$, the first two equations of this system in \cref{eq:U_x_vs_V_x}
imply that extrema of $V\left(x,y\right)$ are also extrema of $U\left(x,y\right)$. In particular,
this means that equilibrium points of the system in \cref{eq:H(xy)} are
always located at the extrema of $U$. The last equality of the system in \cref{eq:U_x_vs_V_x}
can be written as 
\begin{align}
 & p_{x}^{3}\,\partial_{x}K^{xx}+p_{x}^{2}p_{y}\,\left(\sigma\partial_{y}K^{xx}+2\partial_{x}K^{xy}\right)+\label{eq:condition_on_K}\\
 & \,\qquad+p_{y}^{2}p_{x}\,\left(\partial_{x}K^{yy}+2\sigma\partial_{y}K^{xy}\right)+\sigma p_{y}^{3}\,\partial_{y}K^{yy}=0\,.\nonumber 
\end{align}
Again, this equation should hold for arbitrary $p_{x}$ and $p_{y}$, which results in the following conditions
\begin{equation}
\partial_{x}K^{xx}=\partial_{y}K^{yy}=0\,,\label{eq:dyagonal_K_condition}
\end{equation}
\begin{equation}
\partial_{y}K^{xx}=-2\sigma\partial_{x}K^{xy}\,,\quad\text{and}\quad\partial_{x}K^{yy}=-2\sigma\partial_{y}K^{xy}\,,\label{eq:K_xx_condition}
\end{equation}
where we used that $\sigma^2 =1$ to go from \cref{eq:condition_on_K} to \cref{eq:K_xx_condition}.
Differentiating the first of the conditions in \cref{eq:K_xx_condition} with $x$ and second
with $y$ and using \cref{eq:dyagonal_K_condition} we obtain 
\begin{equation}
\partial_{x}^{2}K^{xy}=\partial_{y}^{2}K^{xy}=0\,,\label{eq:linearity_for_Kxy}
\end{equation}
which yields  
\begin{equation}
K^{xy}=\varkappa_{0}+\varkappa_{x}\,x+\varkappa_{y}\,y+\varkappa_{xy}\,xy\,,\label{eq:K_xy_function}
\end{equation}
where $\varkappa_{0},$ $\varkappa_{x}$,$\varkappa_{y}$ and $\varkappa_{xy}$
are constants. Substituting this expression into the conditions in \cref{eq:K_xx_condition}
provides 
\[
\partial_{y}K^{xx}=-2\sigma\left(\varkappa_{x}+\varkappa_{xy}y\right)\,,
\]
and consequently (using also \cref{eq:dyagonal_K_condition}). 
\begin{equation}
K^{xx}=\varkappa_{xx}-2\sigma\left(\varkappa_{x}y+\frac{1}{2}\varkappa_{xy}y^{2}\right)\,,\label{eq:K_xx_function}
\end{equation}
and analogously 
\begin{equation}
K^{yy}=\varkappa_{yy}-2\sigma\left(\varkappa_{y}x+\frac{1}{2}\varkappa_{xy}x^{2}\right)\,\label{eq:K_yy_function}, 
\end{equation}
where $\varkappa_{xx}$ and $\varkappa_{yy}$ are constants.
Hence, the kinetic part can be also written as 
\begin{align}
\label{eq:Form_of_K}
&K^{ik}\,p_{k}p_{i}=2\left[\varkappa_{0}+\varkappa_{x}x+\varkappa_{y}y\right]p_{x}p_{y}+\\
&+\left[\varkappa_{xx}-2\sigma\varkappa_{x}y\right]p_{x}^{2}+\left[\varkappa_{yy}-2\sigma\varkappa_{y}x\right]p_{y}^{2}-\nonumber\\
 &
 -\sigma\varkappa_{xy}\left[yp_{x}-\sigma xp_{y}\right]^{2}\,.\nonumber 
\end{align}
The integral of motion $I$ is useful to establish stability when the kinetic part of $I$ is positive definite. This can be achieved when 
the terms linear in coordinates are absent i.e. when $\varkappa_{x}=\varkappa_{y}=0$. 
Let us compare the expression in \cref{eq:Form_of_K} with the class 1 integral of motion $J_{\text{LV}}$ from \cref{NewFirstIntegral} with $\sigma=-1$ where 
\[
K_{\text{LV}}^{ik}\,p_{k}p_{i}=\left(xp_{y}+yp_{x}\right)^{2}+\frac{\tilde{c}}{2}\left(p_{x}^{2}+p_{y}^{2}\right)\,,
\]
we have $\varkappa_{0}=\varkappa_{x}=\varkappa_{y}=0$ and the non-vanishing
coefficients are 
\[
\varkappa_{xx}=\varkappa_{yy}=\frac{\tilde{c}}{2}\,,\quad\text{and}\quad\varkappa_{xy}=1\,.
\]
After the form of $K^{ik}$ is fixed, one can solve the system of first-order PDEs in \cref{eq:U_x_vs_V_x} to find $V\left(x,y\right)$ and $U\left(x,y\right)$. This task is beyond the scope of the paper. The corresponding solutions of this system are reviewed in e.g \cite{Hietarinta:1986tw} see also~\cref{app:integrable-Hamiltonians}. 

The main point of this appendix was to demonstrate that equilibrium configurations ($\partial_{x}V=\partial_{y}V=0$) of integrable system in \cref{eq:H(xy)} are all located at extrema of $U\left(x,y\right)$. However, the opposite is not true in general. Indeed, as the structure of $K^{ik}$ in \cref{eq:Form_of_K} clearly shows, the matrix of the system of the first two equations in \cref{eq:U_x_vs_V_x} may not be invertible. In particular, one can choose $\varkappa_{xy}=\varkappa_{y}=\varkappa_{yy}$ so that $K^{yy}=0$ and one cannot invert \cref{eq:U_x_vs_V_x} to uniquely express  $\partial_{i}V$ through $\partial_{k}U$. This happens, e.g., for the models in \cref{eq:class4,eq:class8}.

Finally let us discuss a useful relation between Hessian determinant of the potential $U$,
\begin{equation}
\label{Hessian_Det}
\mathfrak{Hess}\,U=\partial_{x}^{2}U\,\partial_{y}^{2}U-\partial_{x}\partial_{y}U\,\,\partial_{y}\partial_{x}U\,,
\end{equation}
and that of the potential $V$ at an equilibrium point. The vanishing of the derivatives of $V$ at such points implies, using  
\cref{eq:U_x_vs_V_x},
that the derivatives of $K^{ik}$ will make no contribution to second derivatives of $U$ there. Using then the first equation in 
\eqref{eq:U_x_vs_V_x}
to compute $\partial_{y}\partial_{x}U$ and the second equation in
\eqref{eq:U_x_vs_V_x}
to compute $\partial_{x}\partial_{y}U$, which facilitates the computation at equilibrium points, one finds after some algebra, that one has there
\begin{equation}
\label{Hessian_relation}
\mathfrak{Hess}\,U=4\sigma\,(\text{det}\,K)\,\mathfrak{Hess}\,V\,. \end{equation}
If the the first integral is useful to establish stability, then $K^{ik}$ is positive definite and not degenerate, so that $\text{det}\,K>0$. In that case, for PP case $\sigma=1$ so that minimum of $U$ corresponds to the minimum of $V$. However, for PG case $\sigma=-1$ and $\mathfrak{Hess}\,U>0$ corresponds to $\mathfrak{Hess}\,V<0$. Hence the minimum of $U$ can only be achieved at saddle points of $V$. Thus, stable equilibrium points in systems where a ghost $y$ interacts with a usual DoF $x$ are located in saddle points of the potential $V$. 

\section{Stable motion with a ghostly $x$}
\label{app:ghoslty x}

The proof of \cref{subsec:proof} can easily be used to show that stable motions also exist in another class of PG Liouville models: the one obtained from the PP where one now ghostifies $x$ by a complex canonical transformation $\mathfrak{C}^{\pm}_{x}$. I.e. starting from the PP Liouville model \cref{eq:Liouville-hamiltonian} with $\sigma=+1$, we change $x$ to $i x$ and $p_x$ to $-i p_x$. This yields a model, dubbed here a GP model,  with a Hamiltonian and an extra constant of motion respectively given by 
\begin{align}
    \label{eq:Liouville-hamiltianxghost}
    H_\text{LV,GP} &= -\frac{p_x^2}{2} + \frac{p_y^2}{2} + V_\text{LV}(x,y)
    \\
\label{eq:Liouville-constant-of-motion_x-ynegsigma}
    I_\text{LV,GP} &=\left(
        p_y x + p_x y
    \right)^2 +
    c\,p_x^2
    + \mathcal{V}
\end{align}
where $V_\text{LV}$, $\mathcal{V}$, $u$ and $v$ are defined respectively as in \cref{eq:Liouville-potential,defmathcalV,defu,defv}, and where
$r^2$ now reads 
\begin{align}
r^2 = -x^2 + y^2\;.
\end{align}
Note in particular that the above $I_\text{LV}$ is just identical to \cref{eq:Liouville-constant-of-motion_x-ynegsigma} upon the replacement of the positive 
$\tilde{c}$ by a positive $c$. 
The analog of the constant of motion \cref{NewFirstIntegral} is now given by the linear combination 
$I_\text{LV,GP} + c H_\text{LV,GP}$. 
In fact, one can show the mere equivalence between the PG models (i.e. the one obtained from the Liouville PP model by applying $\mathfrak{C}^{\pm}_{y}$) and the GP models (i.e. the one obtained from the Liouville PP model by applying $\mathfrak{C}^{\pm}_{x}$).
Indeed, considering the expressions for $u$ and $v$ with $\sigma =+1$ given in \cref{defu,defv} and reminded below
\begin{align}
    u^2 &= \frac{1}{2}\left(r^2 + c +\sqrt{(r^2 + c)^2 - 4\,c\,x^2}\right)\,, 
    \label{defuPP}
    \\
    v^2 &= \frac{1}{2}\left(r^2 + c -\sqrt{(r^2 + c)^2 - 4\,c\,x^2}\right)\,, 
    \label{defvPP}
    \\[0.5em] \label{defrPP}
    r^2 &= x^2 + \,y^2\,,
\end{align}
we see that the application of $\mathfrak{C}^{\pm}_{y}$ to these expressions just changes $r^2=x^2+y^2$ to $r^2=x^2-y^2$ and hence 
transforms $u^2$ and $v^2$ to 
\begin{align}
    u^2 &= \frac{1}{2}\left(x^2-y^2 + c +\sqrt{(x^2-y^2 + c)^2 - 4\,c\,x^2}\right)\,, \notag \\
    &\quad \equiv u^2_\text{PG}(c)\,, \label{defuPG}
    \\
    v^2 &= \frac{1}{2}\left(x^2-y^2 + c -\sqrt{(x^2-y^2 + c)^2 - 4\,c\,x^2}\right)\,, \notag \\
    &\quad \equiv v^2_\text{PG}(c) \,.\label{defvPG}
\end{align}
Conversely, the application of $\mathfrak{C}^{\pm}_{x}$ to \cref{defuPP,defvPP} leads to 
\begin{align}
    u^2 &= \frac{1}{2}\left(-x^2+y^2 + c +\sqrt{(-x^2+y^2 + c)^2 + 4\,c\,x^2}\right)\,, \notag \\
    &\quad \equiv u^2_\text{GP}(c) = -v^2_\text{PG}(-c)\,, \label{defuGP}
    \\
    v^2 &= \frac{1}{2}\left(-x^2+y^2 + c -\sqrt{(-x^2+y^2 + c)^2 + 4\,c\,x^2}\right)\,, \notag \\
    &\quad \equiv v^2_\text{GP}(c) = -u^2_\text{PG}(-c)\,.\label{defvGP}
\end{align}
This implies that, starting from a given PP Hamiltonian of the form 
\begin{align}   
    H_\text{LV,PP} &= \frac{p_x^2}{2} + \frac{p_y^2}{2} + \frac{\Phi(u^2)-\Psi(v^2)}{u^2-v^2}\;,
\end{align}
where for simplicity, we have assumed the functions $f$ and $g$ of 
\cref{eq:Liouville-potential} to be respectively function $\Phi$ and $\Psi$ of $u^2$ and $v^2$, i.e. 
$f(u) = \Phi(u^2)$ and $g(v)=\Psi(v^2)$, we get by applying $\mathfrak{C}^{\pm}_{y}$ the PG theory of Hamiltonian 
\begin{align}   
    H_\text{LV,PG} &= \frac{p_x^2}{2} - \frac{p_y^2}{2} + \frac{\Phi(u^2_\text{PG}(c))-
    \Psi(v^2_\text{PG}(c))}{u^2_\text{PG}(c)-v^2_\text{PG}(c)} \,. \notag \\
    &\equiv  H_\text{LV,PG}(c,\Phi,\Psi) \label{defHPGapp}
    \end{align}
where $H_\text{LV,PG}$ is a function of $c$ but a functional of $\Phi: u^2 \mapsto \Phi(u^2)$  and $\Psi: u^2 \mapsto \Psi(u^2)$. 
On the other hand, the application of $\mathfrak{C}^{\pm}_{x}$ on the {\it same} PP theory yields an Hamiltonian  
\begin{align}   
    H_\text{LV,GP} &= -\frac{p_x^2}{2} + \frac{p_y^2}{2}  + \frac{\Phi(u^2_\text{GP}(c))-\Psi(v^2_\text{GP}(c))}{u^2_\text{GP}(c)-v^2_\text{GP}(c)} \notag \,,\\
    &\equiv  H_\text{LV,GP}(c,\Phi,\Psi)\, ,  \label{defHGPapp} \\
    &=   -\frac{p_x^2}{2} + \frac{p_y^2}{2}+\frac{-\Psi(-u^2_\text{PG}(-c))+\Phi(-v^2_\text{PG}(-c))}{u^2_\text{PG}(-c)-v^2_\text{PG}(-c)} \,,\notag 
    \end{align} 
where as above, 
$H_\text{LV,GP}$ is a function of $c$ but a functional of $\Phi: u^2 \mapsto \Phi(u^2)$  and $\Psi: u^2 \mapsto \Psi(u^2)$.
So we have that 
\begin{align}
    H_\text{LV,GP}(c,\Phi,\Psi) =  - H_\text{LV,PG}(-c,u^2\mapsto \Psi(u^2),v^2 \mapsto  \Phi(v^2))
\end{align}
showing that the PG and GP theories just differ by a global sign in the Hamiltonian, a proper redefinition of the function $\Phi$ and $\Psi$ (or $f$ and $g$), and a change of sign of $c$; hence these theories are equivalent. In particular the conditions enunciated in (i) and (ii)  on the functions $f$ an $g$ of the proof of \cref{subsec:proof} yield a stable theory PG \cref{defHPGapp} for $c<0$ and  a stable theory GP \cref{defHPGapp} for $c>0$.

\section{Polynomial Liouville potentials}
\label{app:polynomial-potentials}
We have systematically constructed all potentials which are polynomials in $x$ and $y$ of degree $2N$ (with $N\in\mathbb{N}$) and belong to the class of integrable Liouville potentials. (In the following, we present them for the PG case, which is of relevance for this paper. With reference to \cref{sec:integrable-models}, these can straightforwardly be transformed to the PP case.) 

At fixed degree $2N$, the polynomial potential can be found by (i) making an ansatz with arbitrary coefficients, (ii) transforming from $(x,y)$ to $(u,v)$ coordinates, (iii) multiplying by $u^2 - v^2$, and (iv) demanding that all mixed coefficients vanish, i.e., that the potential is of the integrable form in \cref{eq:Liouville-potential}. We solve this procedure up to order $N=8$ and infer that the general form of these potentials corresponds to choosing 
\begin{align}
    \label{eq:polynomial-f}
    f(u) &=
    \sum_{n=1}^{N}
    \mathcal{C}_{n}
    \left(u^2\right)^n
    \;,
    \\
    \label{eq:polynomial-g}
    g(v) &= 
    \sum_{n=1}^{N}
    \mathcal{C}_{n}
    \left(v^2\right)^n
    \;.
\end{align}
Proof that this leads to polynomial $V(x,y)$ is given as follows. We first note that $u^2 = \frac{1}{2}(r^2 + c + \tilde{W})$ while $v^2 = \frac{1}{2}(r^2 + c - \tilde{W})$ where $\tilde{W}$ is defined in \cref{defV} With this definition, we may rewrite the potential in \cref{defVN}  as follows
\begin{align}
    V_\text{LV}^{(N)}&= 
    \sum_{n=1}^N \frac{\mathcal{C}_n}{\tilde{W}} \left[
        \left(u^2\right)^n - \left(v^2\right)^n
    \right] \label{VLVCnW}
    \\&=
    \sum_{n=1}^N \frac{\mathcal{C}_n}{2^n \; \tilde{W}} \left[
        \left((r^2 + c) +\tilde{W}\right)^n - \left((r^2 + c) -\tilde{W}\right)^n
    \right]
    \notag\\&=
    \sum_{n=1}^N \sum_{k=1}^n \frac{\mathcal{C}_n}{2^n}\left[
        \binom{n}{k}\,(r^2 + c)^{n-k}\,\tilde{W}^{k-1}\,\left(1 - (-1)^k\right)
    \right]\;, \notag
\end{align}
where we have expanded and collected the binomial coefficients in the last step. Each summand with even $k$ vanishes due to the last factor. Each summand with odd $k$ gives a contribution in which $\tilde{W}$ is raised to an even power. Hence, all of the summands are indeed polynomials in $x$ and $y$.
\\

Given that all summands with even $k$ vanish, we may also shift the sum: In terms of the variables $(x,y)$ the polynomial potential thus reads
\begin{align}
    V_\text{LV}^{(N)}(x,y) = 
    &\sum_{n=1}^N \sum_{k=1}^{\lceil \frac{n}{2}\rceil-1}\Bigg[
        \frac{\mathcal{C}_n}{2^{n-1}}
        \binom{n}{2k+1}
        \times\notag\\&
        \times\left(r(x,y)^2 + c\right)^{n-(2k+1)}
        \left(\tilde{W}(x,y)^2\right)^k
    \Bigg]\;,
    \label{eq:polynomial-subclass_explicit-potential}
\end{align}
where the $\lceil \; \rceil$ designate the ceiling function which returns the lowest integer larger or equal to its argument. 
The explicit first few terms of the potential read
\begin{align}
    V_\text{LV}^{(N)}(x,y) = &
    + \mathcal{C}_1 
    + \mathcal{C}_2\left(
        x^2 - y^2 + c
    \right)
    \notag\\&
    + \mathcal{C}_3\left[
        \left(x^2 - y^2 + c\right)^2 - c\,x^2
    \right]
    \notag\\&
    + \mathcal{C}_4\left[
        \left(x^2 - y^2 + c\right)^3 
        - 2\,c\,x^2\left(x^2 - y^2 + c\right)
    \right]
    \notag\\&
    + \dots\;, \label{genericVLVN}
\end{align}
from which we can see that 
the $n=1$ term corresponds to a constant shift of the potential, 
the $n=2$ term reproduces equal (but opposite-sign) quadratic terms for the $x$ and the $y$ mode, and 
higher-order $n\geqslant3$ terms correspond to specific polynomials. 

Generally, the $n^\text{th}$-order polynomial also includes (some of) the terms appearing at lower order. In particular, all $\mathcal{C}_n$ contribute to the quadratic term for the $x$ mode. At fixed order, the quadratic terms can thus be adjusted freely. For instance, at order $n=4$, we may trade $\mathcal{C}_2$ and $\mathcal{C}_3$ for the frequencies $\omega_x^2$ and $\omega_y^2$, as well as $\mathcal{C}_1$ to remove the constant shift. More specifically, we define 
\begin{align}
    \mathcal{C}_1 &= 
    -c\,\frac{\omega_x^2}{2} 
    \notag \; ,\\
    \mathcal{C}_2& = 
    \omega_x^2 - \frac{\omega_y^2}{2}
    + c^2\,\mathcal{C}_4 
    \notag \; ,\\
    \mathcal{C}_3&= 
    -\frac{\omega_x^2 -\omega_y^2}{2 c} 
    - 2\,c\,\mathcal{C}_4
    \; . 
    \label{C1C2C3}
\end{align}
The $4^\text{th}$-order potential potential $V_4(x,y)$ can then be rewritten as
\begin{align}
    V_\text{LV}^{(4)}(x,y) &= 
    \frac{\omega_x^2}{2}\left[x^2 - \frac{(x^2 - y^2)^2}{c}\right]
    - \frac{\omega_y^2}{2}\left[y^2 - \frac{(x^2 - y^2)^2}{c}\right]
    \notag\\&\quad
    + \mathcal{C}_4 \left[(x^2 - y^2)^3 - c (x^4 - y^4)\right]\;.\label{VLV4}
\end{align}
which can be rewritten as in \cref{V4simp}. 

For $N=6$ we find (reabsorbing an additional constant shift into a redefinition of $\mathcal{C}_1$)
\begin{align}
    V_\text{LV}^{(6)}(x,y) &= V_4(x,y)   \notag\\&\quad
+ \mathcal{C}_5 (x^2-y^2)^4 \notag\\&\quad
+ c \; \mathcal{C}_5 (x^2-y^2)^2(x^2- 4 y^2) \notag\\&\quad
+ c^2 \mathcal{C}_5 (3 y^4 - 2 x^4)  \notag\\&\quad
+ \mathcal{C}_6 (x^2 -y^2)^5 \notag\\&\quad
+ c \; \mathcal{C}_6 (x^2-y^2)^3 (x^2 - 5 y^2) \notag\\&\quad
+ c^2 \mathcal{C}_6 (x^2-y^2)(x^4 - 8 x^2 y^2 + 10 y^4)  \notag\\&\quad
+ c^3 \mathcal{C}_6 (6 y^4 - 3 x^4)
\end{align}

We note that the finite-order construction from which we started out suggests that \cref{eq:polynomial-subclass_explicit-potential} is exhaustive and captures all possible polynomial potentials in the integrable Liouville model.

Finally, it is also instructive to note that, with the choice in \cref{eq:polynomial-f,eq:polynomial-g} (leading to the polynomial potentials \cref{eq:polynomial-subclass_explicit-potential} with some fixed $N$), the quantity $\mathcal{V}$ defined in \cref{defmathcalV} and entering into the constant of motion $I_{LV}$ is also polynomial and reads 
\begin{align}
    \mathcal{V}^{(N)} = 
    - 2\,u^2 v^2 \sum_{n=1}^{N-1} \frac{\mathcal{C}_{n+1}}{\tilde{W}} \left[
        \left(u^2\right)^n - \left(v^2\right)^n
    \right]
\end{align}
where, considering e.g. \cref{VLVCnW},  the second term on the right hand side above is just the same as $-2 u^2 v^2 \times V_{LV}^{(N-1)}$ where the $\mathcal{C}_n$ are replaced by $\mathcal{C}_{n+1}$. For example, for $N=4$, we get, using \cref{uvx2} as well as \cref{genericVLVN}, 
\begin{align}
    \mathcal{V}^{(4)} =& 
    - 2\,c\,x^2 \Bigg[
        \mathcal{C}_2 
        + \mathcal{C}_3\left(
            x^2 - y^2 + c
        \right) 
    \notag \\&
        + \mathcal{C}_4\left(
            \left(x^2 - y^2 + c\right)^2 - c\,x^2
        \right)
    \Bigg]\;,
\end{align}
where the coefficients $\mathcal{C}_1, \mathcal{C}_2, \mathcal{C}_3$ can be replaced as in \cref{C1C2C3} to yield the expression corresponding to the form in \cref{VLV4}, i.e.,
\begin{align}
    \mathcal{V}^{(4)} =&
    - 2\,c\,x^2 \Bigg[
        \frac{\omega_x^2}{2} 
        - \frac{\omega_x^2 - \omega_y^2}{2c}\left(x^2 - y^2\right) 
    \notag \\&
        + \mathcal{C}_4\left(
            \left(x^2 - y^2\right)^2 - c\,x^2
        \right)
    \Bigg]\;.
\end{align}
Clearly, this also implies that the alternative constant of motion $J_\text{LV}$ used in the proof and defined in \cref{NewFirstIntegral} is polynomial since $J_\text{LV} = I_\text{LV} + c\,H_\text{LV} = \left(xp_{y}+yp_{x}\right)^{2}-\frac{c}{2}\left(p_{x}^{2}+p_{y}^{2}\right) + \mathcal{V} + c\,V_\text{LV}$, or equivalently $\mathcal{U} = \mathcal{V} + c\,V_\text{LV}$. For completeness, we state the resulting explicit fourth-order form, i.e.,
\begin{align}
    \mathcal{U}^{(4)} =& \label{U4}
    \frac{\omega_x^2 - \omega_y^2}{2}\left(x^4 - y^4\right)
    - c\left[\frac{\omega_x^2}{2}x^2 + \frac{\omega_y^2}{2}y^2\right]
    \notag\\&
    - c\,\mathcal{C}_4\left[
        \left(x^2 - y^2\right)^2\left(x^2 + y^2\right)
        - c \left(x^4 + y^4\right)
    \right]
    \;.
\end{align}
written again in terms of the frequencies and $\mathcal{C}_4$.

\section{Polynomial class-3 potentials}
\label{app:polynomial-potentials-class-3}

Class 3 in the previous \cref{app:integrable-Hamiltonians} provides an example for an integrable two-particle Hamiltonian which is complex-valued in the PP case but when the $y$-mode is ghostified (via the complex transformation $\mathfrak{C}^{\pm}_{y}$, cf.~\cref{sec:integrable-models}, i.e. taking $y\rightarrow\pm i y$ and $p_y\rightarrow\mp i p_y$) the resulting PG theory is real-valued. In fact, the complexification is real-valued for both signs in $z_{\pm}$ (as part of the model definition in \cref{eq:def-zpm}) and thus, in fact, generates two complex classes. These are distinguished with the occurrence of ``$\pm$'' below.

The same choice for the functions $f(\hat{u})$ and $g(\hat{v})$ as in \cref{app:polynomial-potentials}, \cref{eq:polynomial-f,eq:polynomial-g}, also leads to a polynomial subclass. (Note, however, that $\hat{u}$ and $\hat{v}$ for class 3 differ from $u$ and $v$ for the Liouville class / class 1.)

The polynomial form in terms of $(x,y)$ can be derived in complete analogy to \cref{eq:polynomial-subclass_explicit-potential} and reads
\begin{align}
    V_\text{(class 3)}^{(N)}(x,y) =  
    &\sum_{n=1}^N \sum_{k=1}^{\lceil \frac{n}{2}\rceil-1}\Bigg[
        \frac{\mathcal{C}_n}{2^{n-1}}
        \binom{n}{2k+1}
        \times\notag\\&
        \times 
        r^{2(n-(2k+1))}
        \hat{W}^{2k}
    \Bigg]\;,
    \label{eq:polynomial-subclass_explicit-potential-class3}
\end{align}
where $\lceil \; \rceil$ denotes the ceiling function, $r=\sqrt{x^2-y^2}$, and $\hat{W} = 
\sqrt{r^4 - 2\,c(x \pm y)^2}$ as in the definition of class 3 in \cref{app:integrable-Hamiltonians} (after complexification).

The explicit first few terms of this potential read
\begin{align}
    V_\text{(class 3)}^{(N)}(x,y) = &
    + \mathcal{C}_1 
    + \mathcal{C}_2\left(x^2-y^2\right)
    \notag\\&
    + \mathcal{C}_3\Bigg[
        \left(x^2-y^2\right)^2
        -\frac{1}{2}c\left(x\pm y\right)^2
    \Bigg]
    \notag\\&
    + \mathcal{C}_4\Bigg[
        \left(x^2-y^2\right)^3
        - c\left(x^2-y^2\right)\left(x\pm y\right)^2
    \Bigg]
    \notag\\&
    + \dots\;.
\end{align}
Introducing the frequencies and removing a constant shift, i.e., fixing $\mathcal{C}_1=0$, $\mathcal{C}_2 = \frac{\omega_x^2 + \omega_y^2}{4}$, and $\mathcal{C}_3 = -\frac{\omega_x^2 - \omega_y^2}{2c}$, the $N=4$ case can be written as
\begin{align}
    V_\text{(class 3)}^{(4)}(x,y) &= 
    \frac{\omega_x^2}{2}\,x^2
    - \frac{\omega_y^2}{2}\,y^2
    - \frac{\omega_x^2-\omega_y^2}{4\,c}\Big[
        \notag\\&\quad
        2\left(x^2-y^2\right)^2
        + c\left(x^2+y^2\right)
        - c\left(x\pm y\right)^2
    \Big]
    \notag\\&\quad
    - \mathcal{C}_4\,c\,\left(x^2-y^2\right)(x\pm y)^2
    \notag\\&\quad
    + \mathcal{C}_4\,\left(x^2-y^2\right)^3\;.
\end{align}
In the main text, we specify to $N=4$ and to equal quadratic terms, i.e., to $\omega_x = \omega_y \equiv \omega$.


\bibliographystyle{apsrev4-1}
\bibliography{stableGhosts}

\begin{thebibliography}{55}%
\makeatletter
\providecommand \@ifxundefined [1]{%
 \@ifx{#1\undefined}
}%
\providecommand \@ifnum [1]{%
 \ifnum #1\expandafter \@firstoftwo
 \else \expandafter \@secondoftwo
 \fi
}%
\providecommand \@ifx [1]{%
 \ifx #1\expandafter \@firstoftwo
 \else \expandafter \@secondoftwo
 \fi
}%
\providecommand \natexlab [1]{#1}%
\providecommand \enquote  [1]{``#1''}%
\providecommand \bibnamefont  [1]{#1}%
\providecommand \bibfnamefont [1]{#1}%
\providecommand \citenamefont [1]{#1}%
\providecommand \href@noop [0]{\@secondoftwo}%
\providecommand \href [0]{\begingroup \@sanitize@url \@href}%
\providecommand \@href[1]{\@@startlink{#1}\@@href}%
\providecommand \@@href[1]{\endgroup#1\@@endlink}%
\providecommand \@sanitize@url [0]{\catcode `\\12\catcode `\$12\catcode
  `\&12\catcode `\#12\catcode `\^12\catcode `\_12\catcode `\%12\relax}%
\providecommand \@@startlink[1]{}%
\providecommand \@@endlink[0]{}%
\providecommand \url  [0]{\begingroup\@sanitize@url \@url }%
\providecommand \@url [1]{\endgroup\@href {#1}{\urlprefix }}%
\providecommand \urlprefix  [0]{URL }%
\providecommand \Eprint [0]{\href }%
\providecommand \doibase [0]{http://dx.doi.org/}%
\providecommand \selectlanguage [0]{\@gobble}%
\providecommand \bibinfo  [0]{\@secondoftwo}%
\providecommand \bibfield  [0]{\@secondoftwo}%
\providecommand \translation [1]{[#1]}%
\providecommand \BibitemOpen [0]{}%
\providecommand \bibitemStop [0]{}%
\providecommand \bibitemNoStop [0]{.\EOS\space}%
\providecommand \EOS [0]{\spacefactor3000\relax}%
\providecommand \BibitemShut  [1]{\csname bibitem#1\endcsname}%
\let\auto@bib@innerbib\@empty
\bibitem [{\citenamefont {Ostrogradsky}(1850)}]{Ostrogradsky:1850fid}%
  \BibitemOpen
  \bibfield  {author} {\bibinfo {author} {\bibfnamefont {M.}~\bibnamefont
  {Ostrogradsky}},\ }\href@noop {} {\bibfield  {journal} {\bibinfo  {journal}
  {Mem. Acad. St. Petersbourg}\ }\textbf {\bibinfo {volume} {6}},\ \bibinfo
  {pages} {385} (\bibinfo {year} {1850})}\BibitemShut {NoStop}%
\bibitem [{\citenamefont {Bruneton}\ and\ \citenamefont
  {Esposito-Farese}(2007)}]{Bruneton:2007si}%
  \BibitemOpen
  \bibfield  {author} {\bibinfo {author} {\bibfnamefont {J.-P.}\ \bibnamefont
  {Bruneton}}\ and\ \bibinfo {author} {\bibfnamefont {G.}~\bibnamefont
  {Esposito-Farese}},\ }\href {\doibase 10.1103/PhysRevD.76.129902} {\bibfield
  {journal} {\bibinfo  {journal} {Phys. Rev. D}\ }\textbf {\bibinfo {volume}
  {76}},\ \bibinfo {pages} {124012} (\bibinfo {year} {2007})},\ \bibinfo {note}
  {[Erratum: Phys.Rev.D 76, 129902 (2007)]},\ \Eprint
  {http://arxiv.org/abs/0705.4043} {arXiv:0705.4043 [gr-qc]} \BibitemShut
  {NoStop}%
\bibitem [{\citenamefont {Ketov}\ \emph {et~al.}(2011)\citenamefont {Ketov},
  \citenamefont {Michiaki},\ and\ \citenamefont {Yumibayashi}}]{Ketov:2011re}%
  \BibitemOpen
  \bibfield  {author} {\bibinfo {author} {\bibfnamefont {S.~V.}\ \bibnamefont
  {Ketov}}, \bibinfo {author} {\bibfnamefont {G.}~\bibnamefont {Michiaki}}, \
  and\ \bibinfo {author} {\bibfnamefont {T.}~\bibnamefont {Yumibayashi}}\
  }(\bibinfo {year} {2011})\ \Eprint {http://arxiv.org/abs/1110.1155}
  {arXiv:1110.1155 [hep-th]} \BibitemShut {NoStop}%
\bibitem [{\citenamefont {Woodard}(2015)}]{Woodard:2015zca}%
  \BibitemOpen
  \bibfield  {author} {\bibinfo {author} {\bibfnamefont {R.~P.}\ \bibnamefont
  {Woodard}},\ }\href {\doibase 10.4249/scholarpedia.32243} {\bibfield
  {journal} {\bibinfo  {journal} {Scholarpedia}\ }\textbf {\bibinfo {volume}
  {10}},\ \bibinfo {pages} {32243} (\bibinfo {year} {2015})},\ \Eprint
  {http://arxiv.org/abs/1506.02210} {arXiv:1506.02210 [hep-th]} \BibitemShut
  {NoStop}%
\bibitem [{\citenamefont {Pais}\ and\ \citenamefont
  {Uhlenbeck}(1950)}]{Pais:1950za}%
  \BibitemOpen
  \bibfield  {author} {\bibinfo {author} {\bibfnamefont {A.}~\bibnamefont
  {Pais}}\ and\ \bibinfo {author} {\bibfnamefont {G.~E.}\ \bibnamefont
  {Uhlenbeck}},\ }\href {\doibase 10.1103/PhysRev.79.145} {\bibfield  {journal}
  {\bibinfo  {journal} {Phys. Rev.}\ }\textbf {\bibinfo {volume} {79}},\
  \bibinfo {pages} {145} (\bibinfo {year} {1950})}\BibitemShut {NoStop}%
\bibitem [{\citenamefont {Lee}\ and\ \citenamefont {Wick}(1970)}]{Lee:1970iw}%
  \BibitemOpen
  \bibfield  {author} {\bibinfo {author} {\bibfnamefont {T.~D.}\ \bibnamefont
  {Lee}}\ and\ \bibinfo {author} {\bibfnamefont {G.~C.}\ \bibnamefont {Wick}},\
  }\href {\doibase 10.1103/PhysRevD.2.1033} {\bibfield  {journal} {\bibinfo
  {journal} {Phys. Rev. D}\ }\textbf {\bibinfo {volume} {2}},\ \bibinfo {pages}
  {1033} (\bibinfo {year} {1970})}\BibitemShut {NoStop}%
\bibitem [{\citenamefont {Stelle}(1977)}]{Stelle:1976gc}%
  \BibitemOpen
  \bibfield  {author} {\bibinfo {author} {\bibfnamefont {K.~S.}\ \bibnamefont
  {Stelle}},\ }\href {\doibase 10.1103/PhysRevD.16.953} {\bibfield  {journal}
  {\bibinfo  {journal} {Phys. Rev. D}\ }\textbf {\bibinfo {volume} {16}},\
  \bibinfo {pages} {953} (\bibinfo {year} {1977})}\BibitemShut {NoStop}%
\bibitem [{\citenamefont {Grinstein}\ \emph {et~al.}(2008)\citenamefont
  {Grinstein}, \citenamefont {O'Connell},\ and\ \citenamefont
  {Wise}}]{Grinstein:2007mp}%
  \BibitemOpen
  \bibfield  {author} {\bibinfo {author} {\bibfnamefont {B.}~\bibnamefont
  {Grinstein}}, \bibinfo {author} {\bibfnamefont {D.}~\bibnamefont
  {O'Connell}}, \ and\ \bibinfo {author} {\bibfnamefont {M.~B.}\ \bibnamefont
  {Wise}},\ }\href {\doibase 10.1103/PhysRevD.77.025012} {\bibfield  {journal}
  {\bibinfo  {journal} {Phys. Rev. D}\ }\textbf {\bibinfo {volume} {77}},\
  \bibinfo {pages} {025012} (\bibinfo {year} {2008})},\ \Eprint
  {http://arxiv.org/abs/0704.1845} {arXiv:0704.1845 [hep-ph]} \BibitemShut
  {NoStop}%
\bibitem [{\citenamefont {Linde}(1984)}]{Linde:1984ir}%
  \BibitemOpen
  \bibfield  {author} {\bibinfo {author} {\bibfnamefont {A.~D.}\ \bibnamefont
  {Linde}},\ }\href {\doibase 10.1088/0034-4885/47/8/002} {\bibfield  {journal}
  {\bibinfo  {journal} {Rept. Prog. Phys.}\ }\textbf {\bibinfo {volume} {47}},\
  \bibinfo {pages} {925} (\bibinfo {year} {1984})}\BibitemShut {NoStop}%
\bibitem [{\citenamefont {Linde}(1988)}]{Linde:1988ws}%
  \BibitemOpen
  \bibfield  {author} {\bibinfo {author} {\bibfnamefont {A.~D.}\ \bibnamefont
  {Linde}},\ }\href {\doibase 10.1016/0370-2693(88)90770-8} {\bibfield
  {journal} {\bibinfo  {journal} {Phys. Lett. B}\ }\textbf {\bibinfo {volume}
  {200}},\ \bibinfo {pages} {272} (\bibinfo {year} {1988})}\BibitemShut
  {NoStop}%
\bibitem [{\citenamefont {Kaplan}\ and\ \citenamefont
  {Sundrum}(2006)}]{Kaplan:2005rr}%
  \BibitemOpen
  \bibfield  {author} {\bibinfo {author} {\bibfnamefont {D.~E.}\ \bibnamefont
  {Kaplan}}\ and\ \bibinfo {author} {\bibfnamefont {R.}~\bibnamefont
  {Sundrum}},\ }\href {\doibase 10.1088/1126-6708/2006/07/042} {\bibfield
  {journal} {\bibinfo  {journal} {JHEP}\ }\textbf {\bibinfo {volume} {07}},\
  \bibinfo {pages} {042} (\bibinfo {year} {2006})},\ \Eprint
  {http://arxiv.org/abs/hep-th/0505265} {arXiv:hep-th/0505265} \BibitemShut
  {NoStop}%
\bibitem [{\citenamefont {Brandenberger}\ and\ \citenamefont
  {Peter}(2017)}]{Brandenberger:2016vhg}%
  \BibitemOpen
  \bibfield  {author} {\bibinfo {author} {\bibfnamefont {R.}~\bibnamefont
  {Brandenberger}}\ and\ \bibinfo {author} {\bibfnamefont {P.}~\bibnamefont
  {Peter}},\ }\href {\doibase 10.1007/s10701-016-0057-0} {\bibfield  {journal}
  {\bibinfo  {journal} {Found. Phys.}\ }\textbf {\bibinfo {volume} {47}},\
  \bibinfo {pages} {797} (\bibinfo {year} {2017})},\ \Eprint
  {http://arxiv.org/abs/1603.05834} {arXiv:1603.05834 [hep-th]} \BibitemShut
  {NoStop}%
\bibitem [{\citenamefont {Caldwell}(2002)}]{Caldwell:1999ew}%
  \BibitemOpen
  \bibfield  {author} {\bibinfo {author} {\bibfnamefont {R.~R.}\ \bibnamefont
  {Caldwell}},\ }\href {\doibase 10.1016/S0370-2693(02)02589-3} {\bibfield
  {journal} {\bibinfo  {journal} {Phys. Lett. B}\ }\textbf {\bibinfo {volume}
  {545}},\ \bibinfo {pages} {23} (\bibinfo {year} {2002})},\ \Eprint
  {http://arxiv.org/abs/astro-ph/9908168} {arXiv:astro-ph/9908168} \BibitemShut
  {NoStop}%
\bibitem [{\citenamefont {Di~Valentino}\ \emph {et~al.}(2021)\citenamefont
  {Di~Valentino}, \citenamefont {Mena}, \citenamefont {Pan}, \citenamefont
  {Visinelli}, \citenamefont {Yang}, \citenamefont {Melchiorri}, \citenamefont
  {Mota}, \citenamefont {Riess},\ and\ \citenamefont
  {Silk}}]{DiValentino:2021izs}%
  \BibitemOpen
  \bibfield  {author} {\bibinfo {author} {\bibfnamefont {E.}~\bibnamefont
  {Di~Valentino}}, \bibinfo {author} {\bibfnamefont {O.}~\bibnamefont {Mena}},
  \bibinfo {author} {\bibfnamefont {S.}~\bibnamefont {Pan}}, \bibinfo {author}
  {\bibfnamefont {L.}~\bibnamefont {Visinelli}}, \bibinfo {author}
  {\bibfnamefont {W.}~\bibnamefont {Yang}}, \bibinfo {author} {\bibfnamefont
  {A.}~\bibnamefont {Melchiorri}}, \bibinfo {author} {\bibfnamefont {D.~F.}\
  \bibnamefont {Mota}}, \bibinfo {author} {\bibfnamefont {A.~G.}\ \bibnamefont
  {Riess}}, \ and\ \bibinfo {author} {\bibfnamefont {J.}~\bibnamefont {Silk}},\
  }\href {\doibase 10.1088/1361-6382/ac086d} {\bibfield  {journal} {\bibinfo
  {journal} {Class. Quant. Grav.}\ }\textbf {\bibinfo {volume} {38}},\ \bibinfo
  {pages} {153001} (\bibinfo {year} {2021})},\ \Eprint
  {http://arxiv.org/abs/2103.01183} {arXiv:2103.01183 [astro-ph.CO]}
  \BibitemShut {NoStop}%
\bibitem [{\citenamefont {Cline}\ \emph {et~al.}(2004)\citenamefont {Cline},
  \citenamefont {Jeon},\ and\ \citenamefont {Moore}}]{Cline:2003gs}%
  \BibitemOpen
  \bibfield  {author} {\bibinfo {author} {\bibfnamefont {J.~M.}\ \bibnamefont
  {Cline}}, \bibinfo {author} {\bibfnamefont {S.}~\bibnamefont {Jeon}}, \ and\
  \bibinfo {author} {\bibfnamefont {G.~D.}\ \bibnamefont {Moore}},\ }\href
  {\doibase 10.1103/PhysRevD.70.043543} {\bibfield  {journal} {\bibinfo
  {journal} {Phys. Rev. D}\ }\textbf {\bibinfo {volume} {70}},\ \bibinfo
  {pages} {043543} (\bibinfo {year} {2004})},\ \Eprint
  {http://arxiv.org/abs/hep-ph/0311312} {arXiv:hep-ph/0311312} \BibitemShut
  {NoStop}%
\bibitem [{\citenamefont {Deffayet}\ \emph {et~al.}(2022)\citenamefont
  {Deffayet}, \citenamefont {Mukohyama},\ and\ \citenamefont
  {Vikman}}]{Deffayet:2021nnt}%
  \BibitemOpen
  \bibfield  {author} {\bibinfo {author} {\bibfnamefont {C.}~\bibnamefont
  {Deffayet}}, \bibinfo {author} {\bibfnamefont {S.}~\bibnamefont {Mukohyama}},
  \ and\ \bibinfo {author} {\bibfnamefont {A.}~\bibnamefont {Vikman}},\ }\href
  {\doibase 10.1103/PhysRevLett.128.041301} {\bibfield  {journal} {\bibinfo
  {journal} {Phys. Rev. Lett.}\ }\textbf {\bibinfo {volume} {128}},\ \bibinfo
  {pages} {041301} (\bibinfo {year} {2022})},\ \Eprint
  {http://arxiv.org/abs/2108.06294} {arXiv:2108.06294 [gr-qc]} \BibitemShut
  {NoStop}%
\bibitem [{\citenamefont {Hawking}\ and\ \citenamefont
  {Hertog}(2002)}]{Hawking:2001yt}%
  \BibitemOpen
  \bibfield  {author} {\bibinfo {author} {\bibfnamefont {S.~W.}\ \bibnamefont
  {Hawking}}\ and\ \bibinfo {author} {\bibfnamefont {T.}~\bibnamefont
  {Hertog}},\ }\href {\doibase 10.1103/PhysRevD.65.103515} {\bibfield
  {journal} {\bibinfo  {journal} {Phys. Rev. D}\ }\textbf {\bibinfo {volume}
  {65}},\ \bibinfo {pages} {103515} (\bibinfo {year} {2002})},\ \Eprint
  {http://arxiv.org/abs/hep-th/0107088} {arXiv:hep-th/0107088} \BibitemShut
  {NoStop}%
\bibitem [{\citenamefont {Bender}\ and\ \citenamefont
  {Mannheim}(2008)}]{Bender:2007wu}%
  \BibitemOpen
  \bibfield  {author} {\bibinfo {author} {\bibfnamefont {C.~M.}\ \bibnamefont
  {Bender}}\ and\ \bibinfo {author} {\bibfnamefont {P.~D.}\ \bibnamefont
  {Mannheim}},\ }\href {\doibase 10.1103/PhysRevLett.100.110402} {\bibfield
  {journal} {\bibinfo  {journal} {Phys. Rev. Lett.}\ }\textbf {\bibinfo
  {volume} {100}},\ \bibinfo {pages} {110402} (\bibinfo {year} {2008})},\
  \Eprint {http://arxiv.org/abs/0706.0207} {arXiv:0706.0207 [hep-th]}
  \BibitemShut {NoStop}%
\bibitem [{\citenamefont {Garriga}\ and\ \citenamefont
  {Vilenkin}(2013)}]{Garriga:2012pk}%
  \BibitemOpen
  \bibfield  {author} {\bibinfo {author} {\bibfnamefont {J.}~\bibnamefont
  {Garriga}}\ and\ \bibinfo {author} {\bibfnamefont {A.}~\bibnamefont
  {Vilenkin}},\ }\href {\doibase 10.1088/1475-7516/2013/01/036} {\bibfield
  {journal} {\bibinfo  {journal} {JCAP}\ }\textbf {\bibinfo {volume} {01}},\
  \bibinfo {pages} {036} (\bibinfo {year} {2013})},\ \Eprint
  {http://arxiv.org/abs/1202.1239} {arXiv:1202.1239 [hep-th]} \BibitemShut
  {NoStop}%
\bibitem [{\citenamefont {Salvio}\ and\ \citenamefont
  {Strumia}(2014)}]{Salvio:2014soa}%
  \BibitemOpen
  \bibfield  {author} {\bibinfo {author} {\bibfnamefont {A.}~\bibnamefont
  {Salvio}}\ and\ \bibinfo {author} {\bibfnamefont {A.}~\bibnamefont
  {Strumia}},\ }\href {\doibase 10.1007/JHEP06(2014)080} {\bibfield  {journal}
  {\bibinfo  {journal} {JHEP}\ }\textbf {\bibinfo {volume} {06}},\ \bibinfo
  {pages} {080} (\bibinfo {year} {2014})},\ \Eprint
  {http://arxiv.org/abs/1403.4226} {arXiv:1403.4226 [hep-ph]} \BibitemShut
  {NoStop}%
\bibitem [{\citenamefont {Smilga}(2017)}]{Smilga:2017arl}%
  \BibitemOpen
  \bibfield  {author} {\bibinfo {author} {\bibfnamefont {A.}~\bibnamefont
  {Smilga}},\ }\href {\doibase 10.1142/S0217751X17300253} {\bibfield  {journal}
  {\bibinfo  {journal} {Int. J. Mod. Phys. A}\ }\textbf {\bibinfo {volume}
  {32}},\ \bibinfo {pages} {1730025} (\bibinfo {year} {2017})},\ \Eprint
  {http://arxiv.org/abs/1710.11538} {arXiv:1710.11538 [hep-th]} \BibitemShut
  {NoStop}%
\bibitem [{\citenamefont {Becker}\ \emph {et~al.}(2017)\citenamefont {Becker},
  \citenamefont {Ripken},\ and\ \citenamefont {Saueressig}}]{Becker:2017tcx}%
  \BibitemOpen
  \bibfield  {author} {\bibinfo {author} {\bibfnamefont {D.}~\bibnamefont
  {Becker}}, \bibinfo {author} {\bibfnamefont {C.}~\bibnamefont {Ripken}}, \
  and\ \bibinfo {author} {\bibfnamefont {F.}~\bibnamefont {Saueressig}},\
  }\href {\doibase 10.1007/JHEP12(2017)121} {\bibfield  {journal} {\bibinfo
  {journal} {JHEP}\ }\textbf {\bibinfo {volume} {12}},\ \bibinfo {pages} {121}
  (\bibinfo {year} {2017})},\ \Eprint {http://arxiv.org/abs/1709.09098}
  {arXiv:1709.09098 [hep-th]} \BibitemShut {NoStop}%
\bibitem [{\citenamefont {Anselmi}(2018)}]{Anselmi:2018kgz}%
  \BibitemOpen
  \bibfield  {author} {\bibinfo {author} {\bibfnamefont {D.}~\bibnamefont
  {Anselmi}},\ }\href {\doibase 10.1007/JHEP02(2018)141} {\bibfield  {journal}
  {\bibinfo  {journal} {JHEP}\ }\textbf {\bibinfo {volume} {02}},\ \bibinfo
  {pages} {141} (\bibinfo {year} {2018})},\ \Eprint
  {http://arxiv.org/abs/1801.00915} {arXiv:1801.00915 [hep-th]} \BibitemShut
  {NoStop}%
\bibitem [{\citenamefont {Donoghue}\ and\ \citenamefont
  {Menezes}(2019)}]{Donoghue:2019fcb}%
  \BibitemOpen
  \bibfield  {author} {\bibinfo {author} {\bibfnamefont {J.~F.}\ \bibnamefont
  {Donoghue}}\ and\ \bibinfo {author} {\bibfnamefont {G.}~\bibnamefont
  {Menezes}},\ }\href {\doibase 10.1103/PhysRevD.100.105006} {\bibfield
  {journal} {\bibinfo  {journal} {Phys. Rev. D}\ }\textbf {\bibinfo {volume}
  {100}},\ \bibinfo {pages} {105006} (\bibinfo {year} {2019})},\ \Eprint
  {http://arxiv.org/abs/1908.02416} {arXiv:1908.02416 [hep-th]} \BibitemShut
  {NoStop}%
\bibitem [{\citenamefont {Gross}\ \emph {et~al.}(2021)\citenamefont {Gross},
  \citenamefont {Strumia}, \citenamefont {Teresi},\ and\ \citenamefont
  {Zirilli}}]{Gross:2020tph}%
  \BibitemOpen
  \bibfield  {author} {\bibinfo {author} {\bibfnamefont {C.}~\bibnamefont
  {Gross}}, \bibinfo {author} {\bibfnamefont {A.}~\bibnamefont {Strumia}},
  \bibinfo {author} {\bibfnamefont {D.}~\bibnamefont {Teresi}}, \ and\ \bibinfo
  {author} {\bibfnamefont {M.}~\bibnamefont {Zirilli}},\ }\href {\doibase
  10.1103/PhysRevD.103.115025} {\bibfield  {journal} {\bibinfo  {journal}
  {Phys. Rev. D}\ }\textbf {\bibinfo {volume} {103}},\ \bibinfo {pages}
  {115025} (\bibinfo {year} {2021})},\ \Eprint
  {http://arxiv.org/abs/2007.05541} {arXiv:2007.05541 [hep-th]} \BibitemShut
  {NoStop}%
\bibitem [{\citenamefont {Donoghue}\ and\ \citenamefont
  {Menezes}(2021)}]{Donoghue:2021eto}%
  \BibitemOpen
  \bibfield  {author} {\bibinfo {author} {\bibfnamefont {J.~F.}\ \bibnamefont
  {Donoghue}}\ and\ \bibinfo {author} {\bibfnamefont {G.}~\bibnamefont
  {Menezes}},\ }\href {\doibase 10.1103/PhysRevD.104.045010} {\bibfield
  {journal} {\bibinfo  {journal} {Phys. Rev. D}\ }\textbf {\bibinfo {volume}
  {104}},\ \bibinfo {pages} {045010} (\bibinfo {year} {2021})},\ \Eprint
  {http://arxiv.org/abs/2105.00898} {arXiv:2105.00898 [hep-th]} \BibitemShut
  {NoStop}%
\bibitem [{\citenamefont {Platania}(2022)}]{Platania:2022gtt}%
  \BibitemOpen
  \bibfield  {author} {\bibinfo {author} {\bibfnamefont {A.}~\bibnamefont
  {Platania}},\ }\href {\doibase 10.1007/JHEP09(2022)167} {\bibfield  {journal}
  {\bibinfo  {journal} {JHEP}\ }\textbf {\bibinfo {volume} {09}},\ \bibinfo
  {pages} {167} (\bibinfo {year} {2022})},\ \Eprint
  {http://arxiv.org/abs/2206.04072} {arXiv:2206.04072 [hep-th]} \BibitemShut
  {NoStop}%
\bibitem [{\citenamefont {Pagani}\ \emph {et~al.}(1987)\citenamefont {Pagani},
  \citenamefont {Tecchiolli},\ and\ \citenamefont {Zerbini}}]{Pagani:1987ue}%
  \BibitemOpen
  \bibfield  {author} {\bibinfo {author} {\bibfnamefont {E.}~\bibnamefont
  {Pagani}}, \bibinfo {author} {\bibfnamefont {G.}~\bibnamefont {Tecchiolli}},
  \ and\ \bibinfo {author} {\bibfnamefont {S.}~\bibnamefont {Zerbini}},\ }\href
  {\doibase 10.1007/BF00402140} {\bibfield  {journal} {\bibinfo  {journal}
  {Lett. Math. Phys.}\ }\textbf {\bibinfo {volume} {14}},\ \bibinfo {pages}
  {311} (\bibinfo {year} {1987})}\BibitemShut {NoStop}%
\bibitem [{\citenamefont {Smilga}(2005)}]{Smilga:2004cy}%
  \BibitemOpen
  \bibfield  {author} {\bibinfo {author} {\bibfnamefont {A.~V.}\ \bibnamefont
  {Smilga}},\ }\href {\doibase 10.1016/j.nuclphysb.2004.10.037} {\bibfield
  {journal} {\bibinfo  {journal} {Nucl. Phys. B}\ }\textbf {\bibinfo {volume}
  {706}},\ \bibinfo {pages} {598} (\bibinfo {year} {2005})},\ \Eprint
  {http://arxiv.org/abs/hep-th/0407231} {arXiv:hep-th/0407231} \BibitemShut
  {NoStop}%
\bibitem [{\citenamefont {Boulanger}\ \emph {et~al.}(2019)\citenamefont
  {Boulanger}, \citenamefont {Buisseret}, \citenamefont {Dierick},\ and\
  \citenamefont {White}}]{Boulanger:2018tue}%
  \BibitemOpen
  \bibfield  {author} {\bibinfo {author} {\bibfnamefont {N.}~\bibnamefont
  {Boulanger}}, \bibinfo {author} {\bibfnamefont {F.}~\bibnamefont
  {Buisseret}}, \bibinfo {author} {\bibfnamefont {F.}~\bibnamefont {Dierick}},
  \ and\ \bibinfo {author} {\bibfnamefont {O.}~\bibnamefont {White}},\ }\href
  {\doibase 10.1140/epjc/s10052-019-6569-y} {\bibfield  {journal} {\bibinfo
  {journal} {Eur. Phys. J. C}\ }\textbf {\bibinfo {volume} {79}},\ \bibinfo
  {pages} {60} (\bibinfo {year} {2019})},\ \Eprint
  {http://arxiv.org/abs/1811.07733} {arXiv:1811.07733 [physics.class-ph]}
  \BibitemShut {NoStop}%
\bibitem [{\citenamefont {Damour}\ and\ \citenamefont
  {Smilga}(2022)}]{Damour:2021fva}%
  \BibitemOpen
  \bibfield  {author} {\bibinfo {author} {\bibfnamefont {T.}~\bibnamefont
  {Damour}}\ and\ \bibinfo {author} {\bibfnamefont {A.}~\bibnamefont
  {Smilga}},\ }\href {\doibase 10.1103/PhysRevD.105.045018} {\bibfield
  {journal} {\bibinfo  {journal} {Phys. Rev. D}\ }\textbf {\bibinfo {volume}
  {105}},\ \bibinfo {pages} {045018} (\bibinfo {year} {2022})},\ \Eprint
  {http://arxiv.org/abs/2110.11175} {arXiv:2110.11175 [hep-th]} \BibitemShut
  {NoStop}%
\bibitem [{\citenamefont {Carroll}\ \emph {et~al.}(2003)\citenamefont
  {Carroll}, \citenamefont {Hoffman},\ and\ \citenamefont
  {Trodden}}]{Carroll:2003st}%
  \BibitemOpen
  \bibfield  {author} {\bibinfo {author} {\bibfnamefont {S.~M.}\ \bibnamefont
  {Carroll}}, \bibinfo {author} {\bibfnamefont {M.}~\bibnamefont {Hoffman}}, \
  and\ \bibinfo {author} {\bibfnamefont {M.}~\bibnamefont {Trodden}},\ }\href
  {\doibase 10.1103/PhysRevD.68.023509} {\bibfield  {journal} {\bibinfo
  {journal} {Phys. Rev. D}\ }\textbf {\bibinfo {volume} {68}},\ \bibinfo
  {pages} {023509} (\bibinfo {year} {2003})},\ \Eprint
  {http://arxiv.org/abs/astro-ph/0301273} {arXiv:astro-ph/0301273} \BibitemShut
  {NoStop}%
\bibitem [{\citenamefont {Ilhan}\ and\ \citenamefont
  {Kovner}(2013)}]{Ilhan:2013xe}%
  \BibitemOpen
  \bibfield  {author} {\bibinfo {author} {\bibfnamefont {I.~B.}\ \bibnamefont
  {Ilhan}}\ and\ \bibinfo {author} {\bibfnamefont {A.}~\bibnamefont {Kovner}},\
  }\href {\doibase 10.1103/PhysRevD.88.044045} {\bibfield  {journal} {\bibinfo
  {journal} {Phys. Rev. D}\ }\textbf {\bibinfo {volume} {88}},\ \bibinfo
  {pages} {044045} (\bibinfo {year} {2013})},\ \Eprint
  {http://arxiv.org/abs/1301.4879} {arXiv:1301.4879 [hep-th]} \BibitemShut
  {NoStop}%
\bibitem [{\citenamefont {Pav\v{s}i\v{c}}(2016)}]{Pavsic:2016ykq}%
  \BibitemOpen
  \bibfield  {author} {\bibinfo {author} {\bibfnamefont {M.}~\bibnamefont
  {Pav\v{s}i\v{c}}},\ }\href {\doibase 10.1142/S0219887816300154} {\bibfield
  {journal} {\bibinfo  {journal} {Int. J. Geom. Meth. Mod. Phys.}\ }\textbf
  {\bibinfo {volume} {13}},\ \bibinfo {pages} {1630015} (\bibinfo {year}
  {2016})},\ \Eprint {http://arxiv.org/abs/1607.06589} {arXiv:1607.06589
  [gr-qc]} \BibitemShut {NoStop}%
\bibitem [{\citenamefont {Pav\v{s}i\v{c}}(2013)}]{Pavsic:2013noa}%
  \BibitemOpen
  \bibfield  {author} {\bibinfo {author} {\bibfnamefont {M.}~\bibnamefont
  {Pav\v{s}i\v{c}}},\ }\href {\doibase 10.1142/S0217732313501654} {\bibfield
  {journal} {\bibinfo  {journal} {Mod. Phys. Lett. A}\ }\textbf {\bibinfo
  {volume} {28}},\ \bibinfo {pages} {1350165} (\bibinfo {year} {2013})},\
  \Eprint {http://arxiv.org/abs/1302.5257} {arXiv:1302.5257 [gr-qc]}
  \BibitemShut {NoStop}%
\bibitem [{\citenamefont {Pav\v{s}i\v{c}}(2020)}]{Pavsic:2020aqi}%
  \BibitemOpen
  \bibfield  {author} {\bibinfo {author} {\bibfnamefont {M.}~\bibnamefont
  {Pav\v{s}i\v{c}}},\ }\href {\doibase 10.1142/S0217751X20300203} {\bibfield
  {journal} {\bibinfo  {journal} {Int. J. Mod. Phys. A}\ }\textbf {\bibinfo
  {volume} {35}},\ \bibinfo {pages} {2030020} (\bibinfo {year} {2020})},\
  \Eprint {http://arxiv.org/abs/2012.04976} {arXiv:2012.04976 [hep-th]}
  \BibitemShut {NoStop}%
\bibitem [{\citenamefont {Salle}\ and\ \citenamefont
  {Lefschetz}(1961)}]{Lefschetz}%
  \BibitemOpen
  \bibfield  {author} {\bibinfo {author} {\bibfnamefont {J.~L.}\ \bibnamefont
  {Salle}}\ and\ \bibinfo {author} {\bibfnamefont {S.}~\bibnamefont
  {Lefschetz}},\ }\href@noop {} {\emph {\bibinfo {title} {{Stability By
  Liapunov's Direct Methods with Applications}}}}\ (\bibinfo  {publisher}
  {Academic Press},\ \bibinfo {year} {1961})\BibitemShut {NoStop}%
\bibitem [{\citenamefont {Liouville}(1855)}]{Liouville1855note}%
  \BibitemOpen
  \bibfield  {author} {\bibinfo {author} {\bibfnamefont {J.}~\bibnamefont
  {Liouville}},\ }\href@noop {} {\bibfield  {journal} {\bibinfo  {journal}
  {Journal de Math{\'e}matiques pures et appliqu{\'e}es}\ }\textbf {\bibinfo
  {volume} {20}},\ \bibinfo {pages} {137} (\bibinfo {year} {1855})}\BibitemShut
  {NoStop}%
\bibitem [{\citenamefont {Arnold}(2013)}]{Arnold2013book}%
  \BibitemOpen
  \bibfield  {author} {\bibinfo {author} {\bibfnamefont {V.~I.}\ \bibnamefont
  {Arnold}},\ }\href@noop {} {\emph {\bibinfo {title} {Mathematical methods of
  classical mechanics}}},\ Vol.~\bibinfo {volume} {60}\ (\bibinfo  {publisher}
  {Springer Science \& Business Media},\ \bibinfo {year} {2013})\BibitemShut
  {NoStop}%
\bibitem [{\citenamefont {Darboux}(1901)}]{darboux1901archives}%
  \BibitemOpen
  \bibfield  {author} {\bibinfo {author} {\bibfnamefont {G.}~\bibnamefont
  {Darboux}},\ }in\ \href@noop {} {\emph {\bibinfo {booktitle} {Ser}}},\
  Vol.~\bibinfo {volume} {2}\ (\bibinfo {year} {1901})\ p.\ \bibinfo {pages}
  {371}\BibitemShut {NoStop}%
\bibitem [{\citenamefont {Whittaker}(1964)}]{Whittaker1964book}%
  \BibitemOpen
  \bibfield  {author} {\bibinfo {author} {\bibfnamefont {E.~T.}\ \bibnamefont
  {Whittaker}},\ }\href@noop {} {\emph {\bibinfo {title} {A treatise on the
  analytical dynamics of particles and rigid bodies}}}\ (\bibinfo  {publisher}
  {CUP Archive},\ \bibinfo {year} {1964})\BibitemShut {NoStop}%
\bibitem [{\citenamefont {Fri{\v{s}}}\ \emph {et~al.}(1967)\citenamefont
  {Fri{\v{s}}}, \citenamefont {Smorodinskii}, \citenamefont {Uhl{\i}r},\ and\
  \citenamefont {Winternitz}}]{Frivs1967symmetry}%
  \BibitemOpen
  \bibfield  {author} {\bibinfo {author} {\bibfnamefont {J.}~\bibnamefont
  {Fri{\v{s}}}}, \bibinfo {author} {\bibfnamefont {Y.~A.}\ \bibnamefont
  {Smorodinskii}}, \bibinfo {author} {\bibfnamefont {M.}~\bibnamefont
  {Uhl{\i}r}}, \ and\ \bibinfo {author} {\bibfnamefont {P.}~\bibnamefont
  {Winternitz}},\ }\href@noop {} {\bibfield  {journal} {\bibinfo  {journal}
  {Sov. J. Nucl. Phys}\ }\textbf {\bibinfo {volume} {4}},\ \bibinfo {pages}
  {444} (\bibinfo {year} {1967})}\BibitemShut {NoStop}%
\bibitem [{\citenamefont {Holt}(1982)}]{Holt1982construction}%
  \BibitemOpen
  \bibfield  {author} {\bibinfo {author} {\bibfnamefont {C.~R.}\ \bibnamefont
  {Holt}},\ }\href@noop {} {\bibfield  {journal} {\bibinfo  {journal} {Journal
  of Mathematical Physics}\ }\textbf {\bibinfo {volume} {23}},\ \bibinfo
  {pages} {1037} (\bibinfo {year} {1982})}\BibitemShut {NoStop}%
\bibitem [{\citenamefont {Ankiewicz}\ and\ \citenamefont
  {Pask}(1983)}]{Ankiewicz1983complete}%
  \BibitemOpen
  \bibfield  {author} {\bibinfo {author} {\bibfnamefont {A.}~\bibnamefont
  {Ankiewicz}}\ and\ \bibinfo {author} {\bibfnamefont {C.}~\bibnamefont
  {Pask}},\ }\href@noop {} {\bibfield  {journal} {\bibinfo  {journal} {Journal
  of Physics A: Mathematical and General}\ }\textbf {\bibinfo {volume} {16}},\
  \bibinfo {pages} {4203} (\bibinfo {year} {1983})}\BibitemShut {NoStop}%
\bibitem [{\citenamefont {Dorizzi}\ \emph {et~al.}(1983)\citenamefont
  {Dorizzi}, \citenamefont {Grammaticos},\ and\ \citenamefont
  {Ramani}}]{Ramani:1983}%
  \BibitemOpen
  \bibfield  {author} {\bibinfo {author} {\bibfnamefont {B.}~\bibnamefont
  {Dorizzi}}, \bibinfo {author} {\bibfnamefont {B.}~\bibnamefont
  {Grammaticos}}, \ and\ \bibinfo {author} {\bibfnamefont {A.}~\bibnamefont
  {Ramani}},\ }\href {\doibase 10.1063/1.525975} {\bibfield  {journal}
  {\bibinfo  {journal} {Journal of Mathematical Physics}\ }\textbf {\bibinfo
  {volume} {24}},\ \bibinfo {pages} {2282} (\bibinfo {year} {1983})},\ \Eprint
  {http://arxiv.org/abs/https://doi.org/10.1063/1.525975}
  {https://doi.org/10.1063/1.525975} \BibitemShut {NoStop}%
\bibitem [{\citenamefont {Grammaticos}\ \emph {et~al.}(1983)\citenamefont
  {Grammaticos}, \citenamefont {Dorizzi},\ and\ \citenamefont
  {Ramani}}]{Ramani:1983a}%
  \BibitemOpen
  \bibfield  {author} {\bibinfo {author} {\bibfnamefont {B.}~\bibnamefont
  {Grammaticos}}, \bibinfo {author} {\bibfnamefont {B.}~\bibnamefont
  {Dorizzi}}, \ and\ \bibinfo {author} {\bibfnamefont {A.}~\bibnamefont
  {Ramani}},\ }\href {\doibase 10.1063/1.525976} {\bibfield  {journal}
  {\bibinfo  {journal} {Journal of Mathematical Physics}\ }\textbf {\bibinfo
  {volume} {24}},\ \bibinfo {pages} {2289} (\bibinfo {year} {1983})},\ \Eprint
  {http://arxiv.org/abs/https://doi.org/10.1063/1.525976}
  {https://doi.org/10.1063/1.525976} \BibitemShut {NoStop}%
\bibitem [{\citenamefont {Thompson}(1984)}]{Thompson1984darboux}%
  \BibitemOpen
  \bibfield  {author} {\bibinfo {author} {\bibfnamefont {G.}~\bibnamefont
  {Thompson}},\ }\href@noop {} {\bibfield  {journal} {\bibinfo  {journal}
  {Journal of Physics A: Mathematical and General}\ }\textbf {\bibinfo {volume}
  {17}},\ \bibinfo {pages} {985} (\bibinfo {year} {1984})}\BibitemShut
  {NoStop}%
\bibitem [{\citenamefont {Sen}(1985)}]{Sen1985integrable}%
  \BibitemOpen
  \bibfield  {author} {\bibinfo {author} {\bibfnamefont {T.}~\bibnamefont
  {Sen}},\ }\href@noop {} {\bibfield  {journal} {\bibinfo  {journal} {Physics
  Letters A}\ }\textbf {\bibinfo {volume} {111}},\ \bibinfo {pages} {96}
  (\bibinfo {year} {1985})}\BibitemShut {NoStop}%
\bibitem [{\citenamefont {Hietarinta}(1987)}]{Hietarinta:1986tw}%
  \BibitemOpen
  \bibfield  {author} {\bibinfo {author} {\bibfnamefont {J.}~\bibnamefont
  {Hietarinta}},\ }\href {\doibase 10.1016/0370-1573(87)90089-5} {\bibfield
  {journal} {\bibinfo  {journal} {Phys. Rept.}\ }\textbf {\bibinfo {volume}
  {147}},\ \bibinfo {pages} {87} (\bibinfo {year} {1987})}\BibitemShut
  {NoStop}%
\bibitem [{\citenamefont {Liouville}(1846)}]{Liouville1846}%
  \BibitemOpen
  \bibfield  {author} {\bibinfo {author} {\bibfnamefont {J.}~\bibnamefont
  {Liouville}},\ }\href@noop {} {\bibfield  {journal} {\bibinfo  {journal}
  {Journal de math{\'e}matiques pures et appliqu{\'e}es}\ ,\ \bibinfo {pages}
  {345}} (\bibinfo {year} {1846})}\BibitemShut {NoStop}%
\bibitem [{\citenamefont {Robert}\ and\ \citenamefont
  {Smilga}(2008)}]{Robert:2006nj}%
  \BibitemOpen
  \bibfield  {author} {\bibinfo {author} {\bibfnamefont {D.}~\bibnamefont
  {Robert}}\ and\ \bibinfo {author} {\bibfnamefont {A.~V.}\ \bibnamefont
  {Smilga}},\ }\href {\doibase 10.1063/1.2904474} {\bibfield  {journal}
  {\bibinfo  {journal} {J. Math. Phys.}\ }\textbf {\bibinfo {volume} {49}},\
  \bibinfo {pages} {042104} (\bibinfo {year} {2008})},\ \Eprint
  {http://arxiv.org/abs/math-ph/0611023} {arXiv:math-ph/0611023} \BibitemShut
  {NoStop}%
\bibitem [{\citenamefont {Inc.}(2022{\natexlab{a}})}]{Mathematica:RK}%
  \BibitemOpen
  \bibfield  {author} {\bibinfo {author} {\bibfnamefont {W.~R.}\ \bibnamefont
  {Inc.}},\ }\href
  {https://reference.wolfram.com/language/tutorial/NDSolveExplicitRungeKutta.html}
  {\enquote {\bibinfo {title} {Mathematica, {V}ersion 13.1,
  {E}xplicit{R}unge{K}utta {M}ethod for {NDS}olve},}\ } (\bibinfo {year}
  {2022}{\natexlab{a}}),\ \bibinfo {note} {{C}hampaign, {IL}}\BibitemShut
  {NoStop}%
\bibitem [{\citenamefont {Shampine}(2018)}]{Shampine:1994numerical}%
  \BibitemOpen
  \bibfield  {author} {\bibinfo {author} {\bibfnamefont {L.~F.}\ \bibnamefont
  {Shampine}},\ }\href@noop {} {\emph {\bibinfo {title} {Numerical solution of
  ordinary differential equations}}}\ (\bibinfo  {publisher} {Routledge},\
  \bibinfo {year} {2018})\BibitemShut {NoStop}%
\bibitem [{\citenamefont {Inc.}(2022{\natexlab{b}})}]{Mathematica:Stiffness}%
  \BibitemOpen
  \bibfield  {author} {\bibinfo {author} {\bibfnamefont {W.~R.}\ \bibnamefont
  {Inc.}},\ }\href
  {https://reference.wolfram.com/language/tutorial/NDSolveStiffnessTest.html}
  {\enquote {\bibinfo {title} {Mathematica, {V}ersion 13.1, {S}tiffness
  {D}etection},}\ } (\bibinfo {year} {2022}{\natexlab{b}}),\ \bibinfo {note}
  {{C}hampaign, {IL}}\BibitemShut {NoStop}%
\bibitem [{\citenamefont {Mitsopoulos}\ \emph {et~al.}(2020)\citenamefont
  {Mitsopoulos}, \citenamefont {Tsamparlis},\ and\ \citenamefont
  {Paliathanasis}}]{Mitsopoulos_2020}%
  \BibitemOpen
  \bibfield  {author} {\bibinfo {author} {\bibfnamefont {A.}~\bibnamefont
  {Mitsopoulos}}, \bibinfo {author} {\bibfnamefont {M.}~\bibnamefont
  {Tsamparlis}}, \ and\ \bibinfo {author} {\bibfnamefont {A.}~\bibnamefont
  {Paliathanasis}},\ }\href {\doibase 10.3390/sym12101655} {\bibfield
  {journal} {\bibinfo  {journal} {Symmetry}\ }\textbf {\bibinfo {volume}
  {12}},\ \bibinfo {pages} {1655} (\bibinfo {year} {2020})}\BibitemShut
  {NoStop}%
\end{thebibliography}%


\end{document}